\documentclass[twocolumn,tighten]{aastex62}

\pdfoutput=1

\usepackage{graphicx}
\usepackage{amsmath}
\usepackage{color}
\usepackage{ulem}
\usepackage{float}
\usepackage{threeparttable}
\usepackage{subdepth}
\usepackage{ulem}

\newcommand{\kms}{km~s$^{-1}$}
\newcommand{\kkmspc}{K~km~s$^{-1}$ pc$^2$}
\newcommand{\angstrom}{\textup{\AA}}
\newcommand{\jybeamkms}{Jy~beam$^{-1}$~km~s$^{-1}$}
\newcommand{\mjybeam}{mJy~beam$^{-1}$}

\shortauthors{Tan et al.}
\begin{document}

%\title{Resolving the Molecular Gas and Dust in Ultraluminous Infrared QSO Hosts with ALMA}
%at $z$ = 0.06-0.19 

\title{Resolving the Interstellar Medium in Ultraluminous Infrared QSO Hosts with ALMA}

\author[0000-0003-3032-0948]{Qing-Hua Tan}
\affil{Purple Mountain Observatory \& Key Laboratory for Radio Astronomy, Chinese Academy of Sciences, 10 Yuanhua Road, Nanjing 210033, People's Republic of China}

\author{Yu Gao}
\affil{Purple Mountain Observatory \& Key Laboratory for Radio Astronomy, Chinese Academy of Sciences, 10 Yuanhua Road, Nanjing 210033, People's Republic of China}

\author{Kotaro Kohno}
\affil{Institute of Astronomy, Graduate School of Science, The University of Tokyo, Osawa, Mitaka, Tokyo 181-0015, Japan}
\affil{Research Center for the Early Universe, Graduate School of Science, The University of Tokyo, 7-3-1 Hongo, Bunkyo-ku, Tokyo 113-0033, Japan}

\author{Xiao-Yang Xia}
\affil{Tianjin Astrophysics Center, Tianjin Normal University, Tianjin 300387, People’s Republic of China}

\author{Alain Omont}
\affil{Sorbonne Universit$\acute{e}$, UPMC Universit$\acute{e}$ Paris 6 and CNRS, UMR 7095, Institut d'Astrophysique de Paris, 98bis boulevard Arago, 75014 Paris, France}

\author{Cai-Na Hao}
\affil{Tianjin Astrophysics Center, Tianjin Normal University, Tianjin 300387, People’s Republic of China}

\author{Shu-De Mao}
\affil{Physics Department and Tsinghua Centre for Astrophysics, Tsinghua University, Beijing, 100084, China}
\affil{National Astronomical Observatories, Chinese Academy of Sciences, 20A Datun Rd, Chaoyang District, Beijing 100012, China}

\author{Emanuele Daddi}
\affil{CEA, IRFU, DAp, AIM, Universit$\acute{e}$ Paris-Saclay, Universite Paris Diderot,
Sorbonne Paris Cit$\acute{e}$, CNRS, F-91191 Gif-sur-Yvette, France}

\author{Yong Shi}
\affil{School of Astronomy and Space Science, Nanjing University, Nanjing 210093, Peopleʼs Republic of China}
\affil{Key Laboratory of Modern Astronomy and Astrophysics (Nanjing University), Ministry of Education, Nanjing 210093, Peopleʼs Republic of China}

\author{Ying-He Zhao}
\affil{Yunnan Observatories, Chinese Academy of Sciences, Kunming 650011, People’s Republic of China}

\author{Pierre Cox}
\affil{Institut d'Astrophysique de Paris, CNRS and Universite Pierre et Marie Curie, 98bis boulevard Arago, 75014 Paris, France}
%\affil{Joint ALMA Observatory, 3107 Alonso de C$\acute{o}$rdova, Vitacura, Santiago, Chile}
%\affil{CNRS, UMR 7095, \\
%Institut d’Astrophysique de Paris, F-75014 Paris, France}

%\end{CJK*}

%\author{August Muench}
%\affiliation{American Astronomical Society \\
%2000 Florida Ave., NW, Suite 300 \\
%Washington, DC 20009-1231, USA}
%\collaboration{(AAS Journals Data Scientists collaboration)}

%\author{Butler Burton}
%\affiliation{National Radio Astronomy Observatory}
%\affiliation{AAS Journals Associate Editor-in-Chief}
%\nocollaboration

%% Note that the \and command from previous versions of AASTeX is now
%% depreciated in this version as it is no longer necessary. AASTeX 
%% automatically takes care of all commas and "and"s between authors names.

%% AASTeX 6.2 has the new \collaboration and \nocollaboration commands to
%% provide the collaboration status of a group of authors. These commands 
%% can be used either before or after the list of corresponding authors. The
%% argument for \collaboration is the collaboration identifier. Authors are
%% encouraged to surround collaboration identifiers with ()s. The 
%% \nocollaboration command takes no argument and exists to indicate that
%% the nearby authors are not part of surrounding collaborations.

%% Mark off the abstract in the ``abstract'' environment. 
\begin{abstract}

We present ALMA observations of the CO(1$-$0) line and 3-mm continuum emission in eight ultraluminous infrared (IR) quasi-stellar objects (QSOs) at $z = 0.06-0.19$. All eight IR QSO hosts are clearly resolved in their CO molecular gas emission with a median source size of 3.2 kpc, and seven out of eight sources are detected in 3-mm continuum, which is found to be more centrally concentrated with respect to molecular gas with sizes of 0.4$-$1.0 kpc. Our observations reveal a diversity of CO morphology and kinematics for the IR QSO systems which can be roughly classified into three categories, rotating gas disk with ordered velocity gradient, compact CO peak with disturbed velocity, and multiple CO distinct sources undergoing a merger between luminous QSO and a companion galaxy separated by a few kpc. The molecular gas in three of IR QSO hosts are found to be rotation-dominated with the ratio of the maximum rotation velocity to the local velocity dispersion of $V_{\rm rot}/\sigma=4-6$. Basic estimates of the dynamical masses within the CO-emitting regions give masses between $7.4\times10^9$ and $6.9\times10^{10}$ $M_\odot$. We find an increasing trend between BH mass accretion rate and star formation rate (SFR) over three orders of magnitude in far-IR luminosity/SFR, in line with the correlation between QSO bolometric luminosity and SF activity, indicative of a likely direct connection between AGN and SF activity over galaxy evolution timescales.

%The correlation between $M_{\rm BH}$ and bulge velocity dispersion is found to be slightly tighter than that with $M_{\rm dyn}$ for all types of QSOs in this study.

\end{abstract}

%% Keywords should appear after the \end{abstract} command. 
%% See the online documentation for the full list of available subject
%% keywords and the rules for their use.
%\keywords{editorials, notices --- miscellaneous --- catalogs --- surveys}


%% From the front matter, we move on to the body of the paper.
%% Sections are demarcated by \section and \subsection, respectively.
%% Observe the use of the LaTeX \label
%% command after the \subsection to give a symbolic KEY to the
%% subsection for cross-referencing in a \ref command.
%% You can use LaTeX's \ref and \label commands to keep track of
%% cross-references to sections, equations, tables, and figures.
%% That way, if you change the order of any elements, LaTeX will
%% automatically renumber them.
%%
%% We recommend that authors also use the natbib \citep
%% and \citet commands to identify citations.  The citations are
%% tied to the reference list via symbolic KEYs. The KEY corresponds
%% to the KEY in the \bibitem in the reference list below. 

%\latex\ \footnote{\url{http://www.latex-project.org/}} is a document markup
%in the online documentation at \url{http://journals.aas.org/authors/aastex.html}.
\section{Introduction} \label{sec:intro}

There is growing evidence that the growth histories of the central supermassive black hole (SMBH) are closely related to that of their host galaxies \citep[see][and references therein]{kormendy13}. Tight correlations have been observed between SMBH masses and global properties of their host galaxies, such as the bulge mass and the velocity dispersion of host galaxy \citep[e.g.,][]{magorrian98,ferrarese00,tremaine02,mcconnell13}, for massive elliptical galaxies and galaxies with classical bulges, likely point to a symbiotic connection between bulge formation and BH growth. The possible starburst-active galactic nuclei (AGN) connection is relevant to our understanding of the galaxy formation and its evolution across cosmic time, and the origin of nuclear activity and associated BH growth \citep[][and references therein]{hickox18}. 

According to the most popular models of AGN-galaxy coevolution, an obscured AGN is formed and fed by an accretion disk of material after the nuclei of gas-rich merging galaxies coalesce. The most luminous sources that reach a quasi-stellar object (QSO)-like luminosity experience the so-called feedback or blow-out phase, which expels and/or heats up the gas of the host galaxy, in the form of outflowing winds, manifesting itself as an optically bright QSO \citep{sanders88,sanders96,silk98,dimatteo05,springel05,hopkins06,lonsdale06,fabian12,moreno19}. Obtaining a comprehensive picture of the evolution of massive galaxies from a buried ``starburst phase" to an optically shining ``QSO phase" is one of the central motivation of QSOs and ultraluminous infrared galaxies (ULIRGs; $L_{\rm IR}\geqslant 10^{12}\ L_\odot$) studies \citep[e.g.,][]{shi14,zhang16,shangguan19}.

Cold molecular gas in galaxies provides the fuel for both star formation and BH accretion \citep[see e.g.,][]{carilli13,vito14}. Studying the distribution and conditions of the molecular interstellar medium (ISM) in star-forming galaxies with actively accreting SMBHs is therefore crucial to understanding the exact physical processes involved in the coevolution. The rotational transitions of CO lines are good tracers for probing the physical properties of the molecular gas in galaxies. To date, the CO line emission has been widely detected in about 100 QSOs from local universe out to redshift of $z>6$ \citep{carilli13,banerji17}. In addition, the [CII] 158 $\mu$m line emission is another main coolant of the ISM and has been increasingly detected at redshift $z>2$ in recent studies with the advent of Atacama Large Millimeter/submillimeter Array (ALMA) \citep[e.g.,][]{decarli18,smit18,zanella18}. Both the CO and [CII] imaging of infrared(IR)-luminous QSOs provide information on the spatial distribution of gas and the dynamical masses of the QSO host galaxies \citep[e.g.,][]{yun04,wang10,wang13,aravena11,brusa15,banerji17,trakhtenbrot17,willott17,decarli18,feruglio18,lu18,hill19}. However, the majority of high-$z$ QSO hosts are spatially unresolved due to the limited angular resolution and sensitivity of current data (the typical scale is $\gtrsim$ 3 kpc at high-$z$), making the characterization of the gas distribution and kinematics, and the estimate of dynamical mass uncertain. 

\citet{hao05} studied a sample of type 1 AGNs selected from the local ULIRGs. These galaxies are referred to as IR QSOs and were originally compiled from the ULIRGs in the QDOT redshift survey \citep{lawrence99}, the 1 Jy ULIRG survey \citep{kim98}, and the cross-correlation of the $IRAS$ Point Source Catalog with the $ROSAT$ All-sky Survey Catalog. By comparing the properties of IR versus optical luminosities for IR QSOs with optically selected PG QSOs and narrow-line Seyfert 1 galaxies, they found that IR QSOs are significantly different from the other two classes of AGNs with IR excess in the far-IR, implying that massive starbursts may play a significant role in the energy output of IR QSOs. The optical spectroscopic and X-ray observations also show evidence for young, growing QSOs with high accretion rates to their central black holes \citep{zheng02,hao05}. Moreover, fast ionized outflows with velocity up to $\sim$1000 \kms\ have been observed in some of IR QSOs \citep{zheng02}. The study of mid-IR spectroscopic properties showed that the slope of MIR continua for IR QSOs is intermediate between those of classical PG QSOs and ULIRGs \citep{cao08}. All of these properties suggest that IR QSOs are likely the objects caught in the short-lived ``transition" phase between ultraluminous starburst and QSO stages, which is expected to be characterized by heavily obscured AGN with large reservoirs of gas not yet fully consumed and complex kinematics, including strong winds and outflows produced in the feedback process \citep{sanders88,hopkins08,fabian12}.

To verify the postulate that the far-IR excess in IR QSOs originate primarily from starbursts, \citet{xia12} performed the first CO survey in 19 IR QSOs ($z<0.4$) selected from \citet{hao05} with the IRAM 30m. The CO observations show that the gas masses are a few times $10^9-10^{10}\ M_\odot$, one order of magnitude higher than that of PG QSOs \citep{evans01,evans06,scoville03}. This supports the scenario that the IR QSO hosts are indeed equipped with a large reservoir of molecular gas, providing fuel for both the star formation and the accretion of AGN. In addition, the CO-detected QSOs at {\bf $z>2$} are found to be gas-rich and follow a similar $L^\prime_{\rm CO}-L_{\rm FIR}$ relation to local IR QSOs \citep[e.g.,][]{riechers06,wang10,sharon16,fan18}. This suggests that the far-IR emission from IR-luminous, high-$z$ QSOs is also powered mainly by starbursts. A large amount of gas is also found in a few WISE-selected local IR bright QSOs \citep{zhao16b}. With the addition of six local IR QSOs that have been published in CO detection previously, we have a sample of 25 IR QSOs in total, which we consider as a representative IR QSO sample in the local universe \citep[see][]{xia12}. Many local (U)LIRGs have well-resolved CO images \citep[e.g.,][]{wilson08,ueda14,xu14,xu15,zhao16a,zhao17,cao18}, but up to now the only IR QSO that has been well-studied is the nearest one, Mrk~231, which shows CO concentration in the central 1 kpc region with massive molecular outflow \citep[e.g.,][]{downes98,cicone12,feruglio15}. To fully characterize the molecular gas properties for IR QSOs, i.e., gas distribution, excitation, and kinematics, CO imaging observations of a large sample of IR QSOs with high spatial resolution are needed. These will be crucial to disentangling the interplay between star formation and BH feeding and unveiling the details of the co-evolution of BH and their host galaxies.

Studies of luminosity function for large samples of $IRAS$ galaxies reveal that the space densities of ULIRGs and QSOs are comparable in the local universe \citep{soifer86,sanders88}, while the fraction of IR QSOs is less than 10\% of ULIRGs \citep{zheng02}. These facts indicate that the IR QSO phase may last only a few times 10$^7$ yr if the space density of objects is simply related to the timescale of different phases \citep{hao05}. There is increasing evidence that the comoving space density of ULIRGs evolves with redshift, and their contribution to the star formation history rises with redshift and likely dominates at $z>1$ \citep{elbaz02,daddi04,lefloch05,magnelli11}. This may imply that the comoving space density of IR QSOs is correspondingly higher at higher redshift and the objects at the IR QSO phase tend to be more common in the early universe. As the IR QSOs we selected are the most luminous IR QSOs in the local universe, they should be considered as the nearest templates to their more extreme counterparts at high-$z$ discovered by deep surveys. Therefore, a systematic investigation of local IR QSOs will provide the best local analogues and valuable information for our understanding of galaxy formation and evolution in the early universe.

In this study, we report our ALMA observations of CO(1$-$0) line and 3-mm continuum emission (project 2015.1.01147.S) of eight IR QSOs at $z=0.06-0.19$, aimed at characterizing the distribution and kinematics of molecular ISM of local IR QSO host galaxies. We selected sources that were previously detected with bright CO line emission \citep{xia12} and have declination of $\delta < +40^\circ$. The ALMA observations are described in Section~\ref{sec:obs}. In Section~\ref{sec:res}, we present results on CO line and 3-mm continuum morphologies, CO kinematics, gas masses, source sizes, and dynamical masses. In Section~\ref{sec:discussion}, we discuss the properties of IR QSO hosts based on the gas morphology and kinematics, and attempt to investigate the growth of SMBHs and their host galaxies in IR-luminous QSO phase by comparison with QSOs observed at high-$z$, and the evolutionary status of IR QSOs. We summarize our main findings in Section~\ref{sec:summary}. Throughout this work we assume a standard $\Lambda$CDM cosmology with $H_{\rm 0}$ = 70 \kms\ Mpc$^{-1}$, $\Omega_{\rm m}$ = 0.3, and $\Omega_\Lambda$ = 0.7.

\section{ALMA Observations} \label{sec:obs}

The ALMA observations of IR QSOs in our sample were carried out between 2016 August and 2017 August with 36$-$44 12m diameter antennas in the extended configuration with baselines 15$-$3637 m, under good weather conditions with precipitable water vapor (PWV) ranging between 0.4 and 2.2~mm. We used the ALMA band 3 receiver and configured the correlator with spectral window centered at the CO(1$-$0) line frequency (Table~\ref{tab:obs}). The spectral resolution is 0.488 MHz, or 1.3$-$1.5 \kms\ at the observed line frequency. 

Observations of the science targets were interleaved with nearby phase calibrators (e.g., bright quasars) that were observed every 6$-$7 minutes. Bandpass calibration was performed through observations of J0006$-$0623, J0750+1231, J0854+2006, J1229+0203, J1550+0527, and J2258$-$2758, which were also used as flux calibrators. The typical uncertainty of flux calibration is estimated to be 5$-$10\%. The pointing was checked on bandpass and phase calibrators. The average on-source integration time on the target was $\sim$17 minutes per source (see Table~\ref{tab:obs}).

The ALMA data were calibrated with CASA\footnote{\url{http://casa.nrao.edu/}} \citep{mcmullin07} 4.7.0 and 4.7.2 (for IRAS F11119+3257 and IRAS F23411+0228 that were observed in 2017 July and August) in pipeline mode, by executing the calibration scripts corresponding to the release data of the observations. Dust continuum images were produced for each target from the calibrated visibilities, by combining the line-free channels from the spectral window in multi-frequency synthesis mode using the CASA task \texttt{clean}. The continuum emission was subtracted in the $uv$-plane before making the line images. The continuum-subtracted line visibilities were obtained by fitting a $uv$-plane model of the continuum emission with a zeroth-order polynomial that was then subtracted using the \texttt{uvcontsub} task. The typical full width at half maximum (FWHM) synthesized beam size is $\sim$0.\arcsec45 by using Briggs weighting with a robust parameter of 0.5 (see Table~\ref{tab:obs}), corresponding to $\sim$ 0.8 kpc at $z=0.1$. By adopting a Briggs weighting with robust = 2.0 (close to natural weighting), we obtain a $\sim$20\% larger beam size but deeper sensitivity. The achieved 1$\sigma$ rms noise levels in the spectral velocity resolution of 25~\kms\ are 0.4$-$0.9~\mjybeam. To maximize the signal-to-noise ratio (SNR) of our observations, we adopt the cubes of both line and continuum produced by Briggs cleaning with robust parameter of 2.0 for data analysis in the following, unless otherwise specified.

%%%%%%%%%%%%%%- Table-1 - %%%%%%%%%%%%%%

\begin{deluxetable*}{lrrcccccCChC}
\tablenum{1}
\centering
\tabletypesize{\scriptsize}
\addtolength{\tabcolsep}{-2.pt}
\tablecaption{Description of the ALMA CO(1-0) observations}\label{tab:obs}
\tablewidth{0pt}
\tablehead{
\colhead{Source} & \colhead{RA$\tablenotemark{a}$} & \colhead{DEC$\tablenotemark{a}$} & \colhead{Date} & \colhead{Obs. freq} & \colhead{Min, Max} & \colhead{Antennas} & \colhead{On-source} & \colhead{FoV} & \colhead{Synt. beam$\tablenotemark{b}$} & \nocolhead{Physical Scale} & \colhead{Scale$\tablenotemark{c}$}\\
\colhead{} & \colhead{(J2000)} & \colhead{(J2000)} & \colhead{} & \colhead{(GHz)} & \colhead{baseline (m)} & \colhead{} & \colhead{time (min)} & \colhead{(arcsec)} & \colhead{(arcsec)} & \nocolhead{(kpc/arcsec)} & \colhead{(kpc)}
}
\startdata
I~ZW~1 & 00:53:34.9 & 12:41:36.2 & Aug. 2016 & 108.644 & (15, 1545) & 42 &  4.02 & 46.4\times 46.4 & 0.57\times 0.46 & 1.18 & 0.67 \\
IRAS 06269$-$0543 & 06:29:24.7 & $-$05:45:26.0 & Oct. 2016 & 103.197 & (18, 1808) & 44 & 15.62 & 48.8\times 48.8 & 0.44\times 0.40 & 2.12 & 0.93 \\
IRAS F11119+3257 & 11:14:38.9 & 32:41:33.0 & Aug. 2017 & 96.866 & (21, 3637) & 40 & 36.28 & 52.0\times 52.0 & 0.40\times 0.23  & 3.17 & 1.27 \\
IRAS Z11598$-$0112 & 12:02:26.6 & $-$01:29:15.3 & Aug. 2016 & 100.149 & (15, 1813) & 38 & 10.58 & 50.3\times 50.3 & 0.47\times 0.38 & 2.63 & 1.24 \\
IRAS F15069+1808 & 15:09:13.7 & 17:57:11.0 & Sep. 2016 & 98.439 & (15, 2483) & 38 & 11.58 & 51.2\times 51.2 & 0.58\times 0.33 & 2.91 & 1.69 \\
IRAS F15462$-$0450 & 15:48:56.8 & $-$04:59:33.5 & Sep. 2016 & 104.792 & (15, 3143) & 40 & 18.13 & 48.1\times 48.1 & 0.35\times 0.29 & 1.84 & 0.64 \\
IRAS F22454$-$1744 & 22:48:04.1 & $-$17:28:28.5 & Aug. 2016 & 103.105 & (15, 1462) & 38 & 12.08 & 48.9\times 48.9 & 0.57\times 0.47 & 2.13 & 1.21 \\
IRAS F23411+0228 & 23:43:39.7 & 02:45:05.7 & Aug. 2016 & 105.562 & (15, 1462) & 38 & 17.13 & 47.7\times 47.7 & 0.56\times 0.44$\tablenotemark{d}$ & 1.71 & 0.96 \\
                 &            &            & Jul. 2017 & 105.562 & (21, 2196) & 36 & 17.13 & 47.7\times 47.7 & - & - & - \\
\enddata
%\tablecomments{}
\tablenotetext{a}{Observing phase center}
\tablenotetext{b}{Using Briggs robust weighting with a robustness parameter of 0.5.}
\tablenotetext{c}{Spatial physical scale corresponding to the major axis FWHM of the synthesized beam.}
\tablenotetext{d}{Combination of the two configuration data.}
\end{deluxetable*}

\section{Results and analysis} \label{sec:res}

\subsection{Distribution of the CO Line and 3-mm Continuum Emission}\label{subsec:distribution}

All eight IR QSOs observed in our sample are clearly resolved in the CO(1$-$0) line emission (i.e., the observed sizes of CO emission are more than 2 times the synthesized beams; see Figure~\ref{fig:moment}). We detect 3-mm continuum emission in seven (non-detection in IRAS F15069+1808) out of eight IR QSOs, with four of which are marginally resolved. The 1$\sigma$ rms noise level of 3-mm continuum emission in the eight IR QSOs observed is 30$-$78 $\mu$Jy, and the total flux of the 3-mm continuum emission measured for each object is summarized in Table~\ref{tab:meas}. The velocity-integrated CO(1$-$0) line maps and 3-mm continuum maps of the eight IR QSO hosts are presented in Figure~\ref{fig:moment}. The molecular gas in the IR QSO hosts mapped in CO emission are found to distribute in the nuclear regions and the peaks of CO emission are coincident with the 3-mm continuum emission peaks (the exception is IRAS 06269$-$0543, where the peak of continuum emission is shifted by $\sim$0.\arcsec19, equivalent to $\sim$0.4 kpc, with respect to the CO emission\footnote{The 1$\sigma$ positional uncertainties of CO and continuum peaks are 0.\arcsec014 and 0.\arcsec020, respectively. The positional uncertainty is estimated based on the formula $\Delta \alpha = \Delta \beta$ = 0.6$\theta$(S/N)$^{-1}$ \citep{ivison07}, where $\Delta \alpha$ and $\Delta \beta$ are the rms errors in right ascension and declination, respectively, $\theta$ is the FWHM of the beam and S/N is the signal-to-noise ratio of the detection.}). The CO(1$-$0) spectra of the QSO hosts (together with the companions we discuss below) are presented in Figure~\ref{fig:spec}.

High-resolution CO maps reveal a diversity of morphologies for the molecular gas in our sample of IR QSO hosts for the first time. Five of our galaxies show a predominantly compact morphology, while the rest three of galaxies (IRAS~F15069+1808, IRAS~F22454$-$1744, and IRAS~F23411+0228) are resolved into multiple distinct objects in the CO emission, show evidence for merger with a galaxy accompanying each of the IR QSOs. 

Two (I~ZW~1 and IRAS~F11119+3257) of the eight IR QSOs in our sample have been mapped in CO emission previously. The CO(1$-$0) imaging observations of I~ZW~1 were obtained with the IRAM Plateau de Bure Interferometer (PdBI) and the BIMA millimeter interferometer at 1\arcsec .9 and 0\arcsec .7 angular resolution \citep{schinnerer98, staguhn04}, respectively. Compared with these observations, our ALMA observations at 0\arcsec .6 resolution resolve the emission in the nuclear region with improved $uv$-coverage and angular resolution. In contrast to the BIMA observations, the ALMA observations do not show any particular spatial structure in the nuclear region. This is similar to the spatial distribution of CO(2$-$1) emission obtained with PdBI at 0\arcsec .9 resolution \citep{staguhn04}. We detect two spiral arm-like structures to the west and the east of the nucleus in the ALMA CO map respectively, consistent with that of the PdBI CO(1$-$0) observations at 1\arcsec .9 resolution \citep{schinnerer98}. A kinematic analysis of the PdBI CO(1$-$0) data shows evidence for a circumnuclear molecular ring with a diameter of 1\arcsec .5 \citep[$\sim$1.8 kpc;][]{schinnerer98}. However, our ALMA observations do not identify this nuclear starburst ring structure. In addition, both optical and near-IR observations reveal that I~ZW~1 is likely undergoing a minor merger with a companion galaxy 15\arcsec .6 west of the QSO \citep{hutchings90, schinnerer98, canalizo01}. This companion is not detected in our ALMA observations. 

IRAS~F11119+3257 has been observed with ALMA in CO(1$-$0) emission at about $2\arcsec .8$ resolution \citep{veilleux17}, with a main objective of detecting the molecular outflow. Compared to these observations that do not spatially resolve the CO morphology, our ALMA observations at 0\arcsec .4 resolution clearly resolve the distribution of both CO gas and 3-mm continuum emission in the host galaxy of IRAS F11119+3257 on $\sim$ 1 kpc scale. However, our data are not deep enough for the detection of broad wings in the CO emission.

The CO map of IRAS~F15462$-$0450 exhibits clumpy structures outside the central bright compact source (F15462$-$0450C). The integrated CO line flux suggests that the detection is significant at the $\sim$ 5$\sigma$ and $\sim$ 7$\sigma$ level for the clumps to the northeast (NE) and southwest (SW), respectively. It is worth noting that these two clumps are almost symmetrically located in a straight line across the CO peak of the central source with a separation of 1.\arcsec6 ($\sim$ 2.9 kpc) between the central peak and the clump in each side. One possibility for the symmetric structure may be related to a massive molecular outflow in the QSO host. Deeper CO observations and higher resolution imaging in multi-wavelength (e.g., radio continuum, optical) are needed to reveal the nature of this feature. The spectra extracted for these two clumps, if associated with a redshifted CO(1$-$0) transition, show a velocity shift of $\sim -60$ \kms\ and $\sim$ 40 \kms\ from the CO line of F15462$-$0450C for NE and SW, respectively. We detect significant 3-mm continuum emission from the main body of IRAS~F15462$-$0450 at the $\sim$ 10$\sigma$ level with a flux density of 0.645$\pm$0.066 mJy  (measured from a two-dimensional (2D) Gaussian fit to the continuum image, see Table~\ref{tab:meas}). No 3-mm continuum emission was detected in the clumpy structures NE and SW. 

For IRAS~F15069+1808, one of the IR QSO systems resolved into multiple distinct CO sources, we identify two bright sources in the CO map which we denote as NW and SE (northwest and southeast) and refer the overlap region as Overlap (see Figure~\ref{fig:moment}). Both bright sources show an elongated morphology along the north-east-south-west axis. IRAS~F15069+1808 is likely in the mid-stage of a merger when the galaxies have collided and begun to merge. The projected separation between the NW and SE sources is 2.\arcsec48, equivalent to a projected physical distance of $\sim$ 7.2 kpc. In the continuum image, we do not detect any signal above 3$\sigma$, placing a 3$\sigma$ upper limit on 3-mm continuum emission $S_{\rm 3mm}<0.15$ mJy.

%%%%%%%%%%%%%%%- Fig-1 -%%%%%%%%%%%%%%%%

\begin{figure*}[htbp]
\centering
\includegraphics[width=0.23\linewidth]{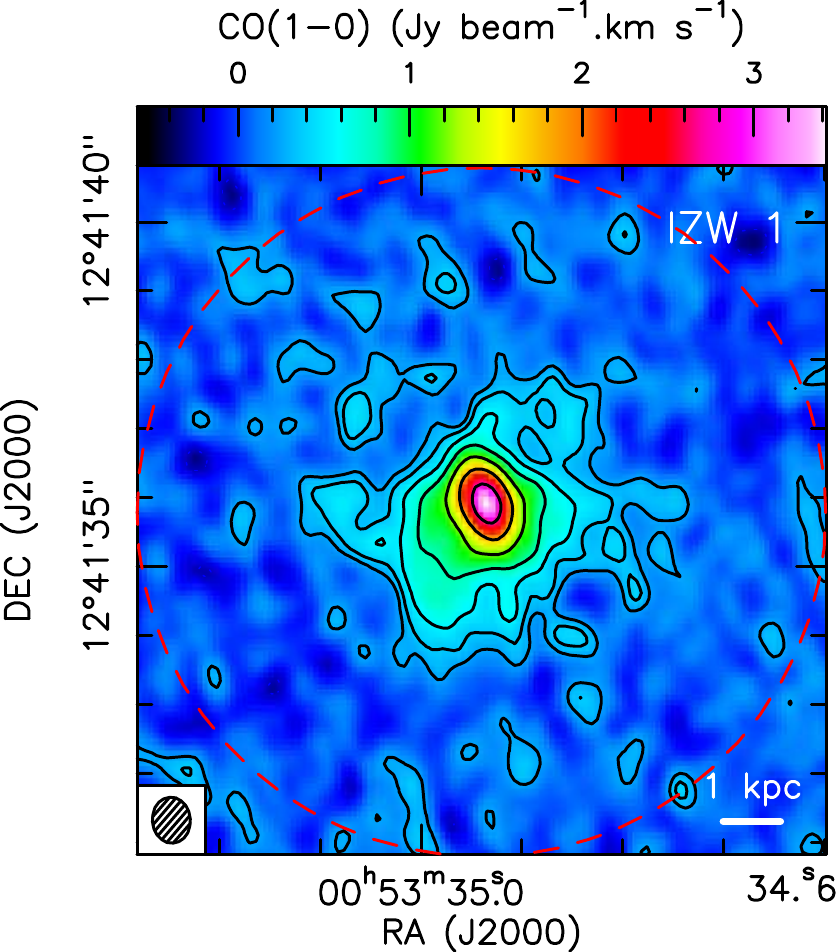}
\includegraphics[width=0.23\linewidth]{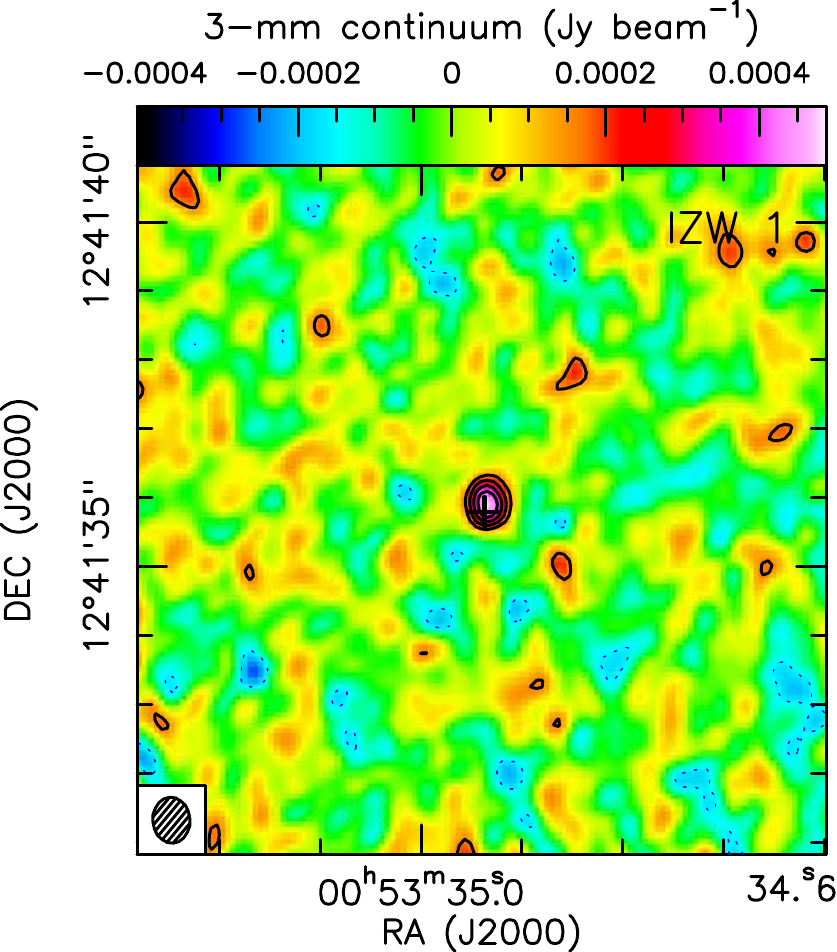}
\includegraphics[width=0.23\linewidth]{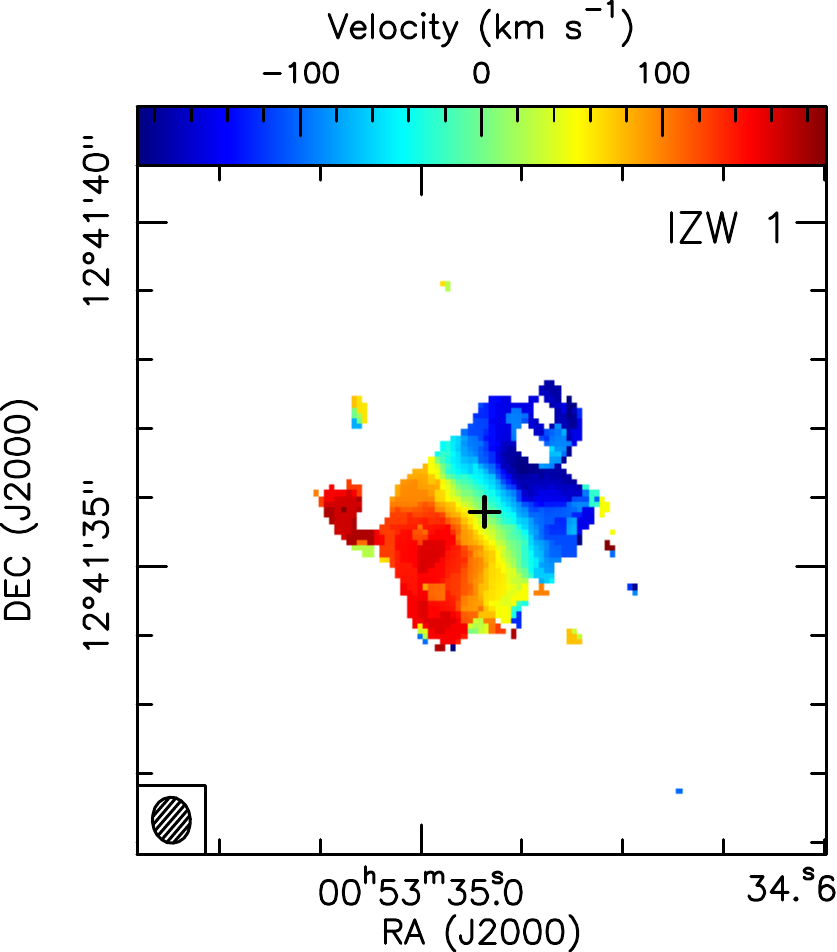}
\includegraphics[width=0.23\linewidth]{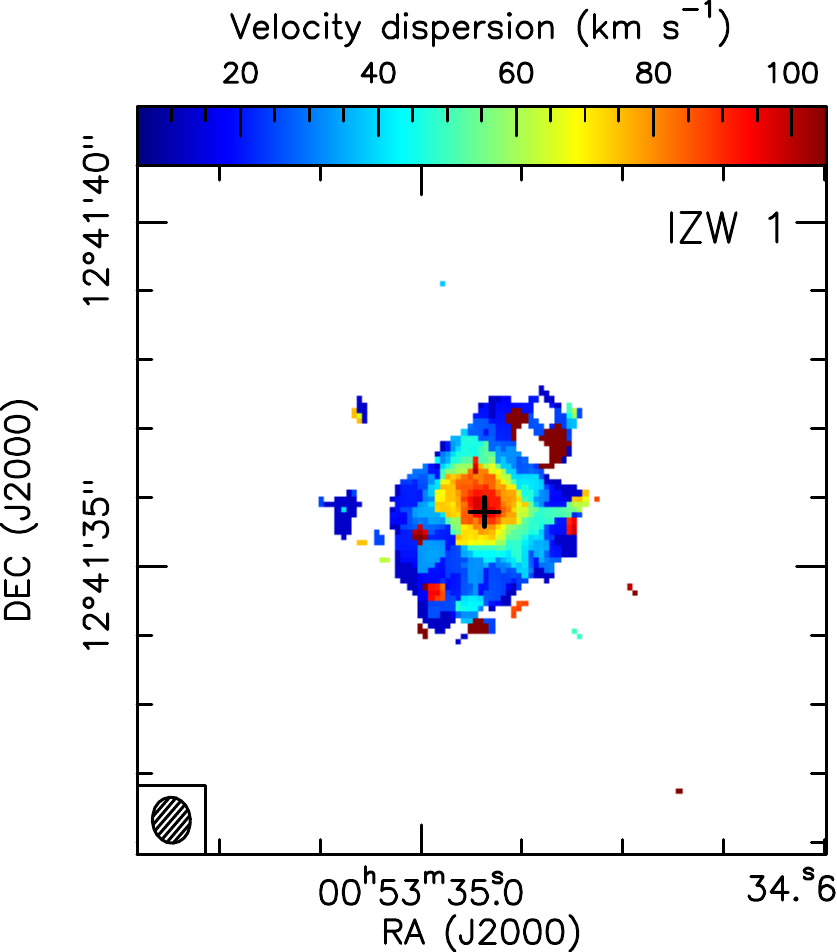}\\
\vspace{2pt}
\includegraphics[width=0.23\linewidth]{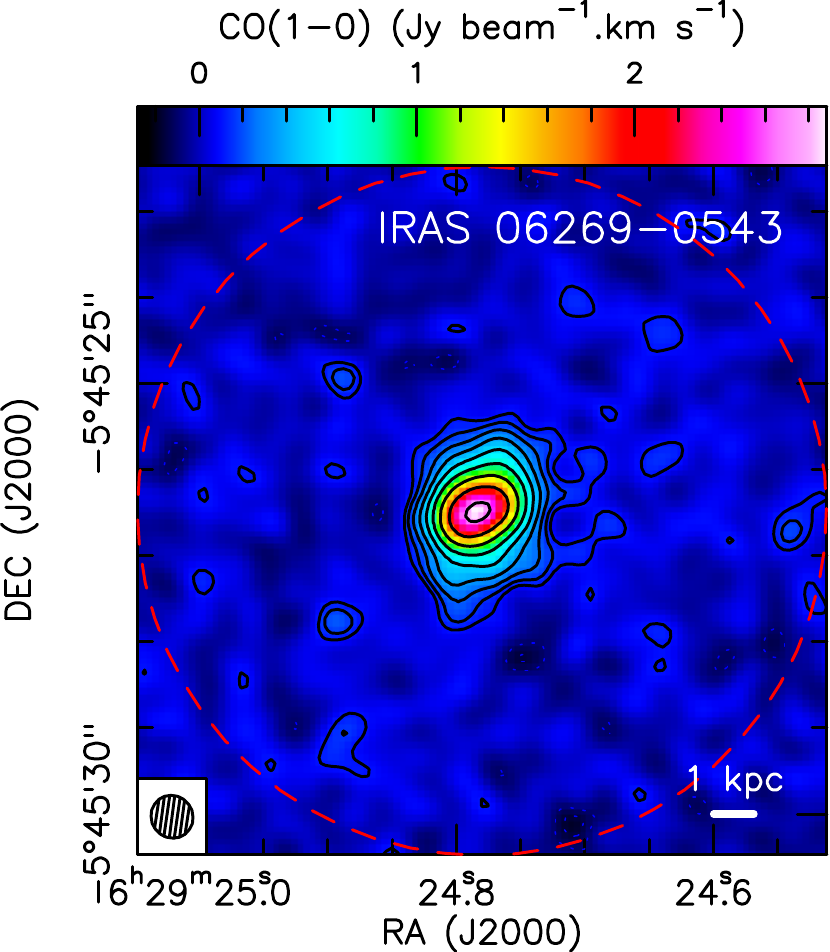}
\includegraphics[width=0.23\linewidth]{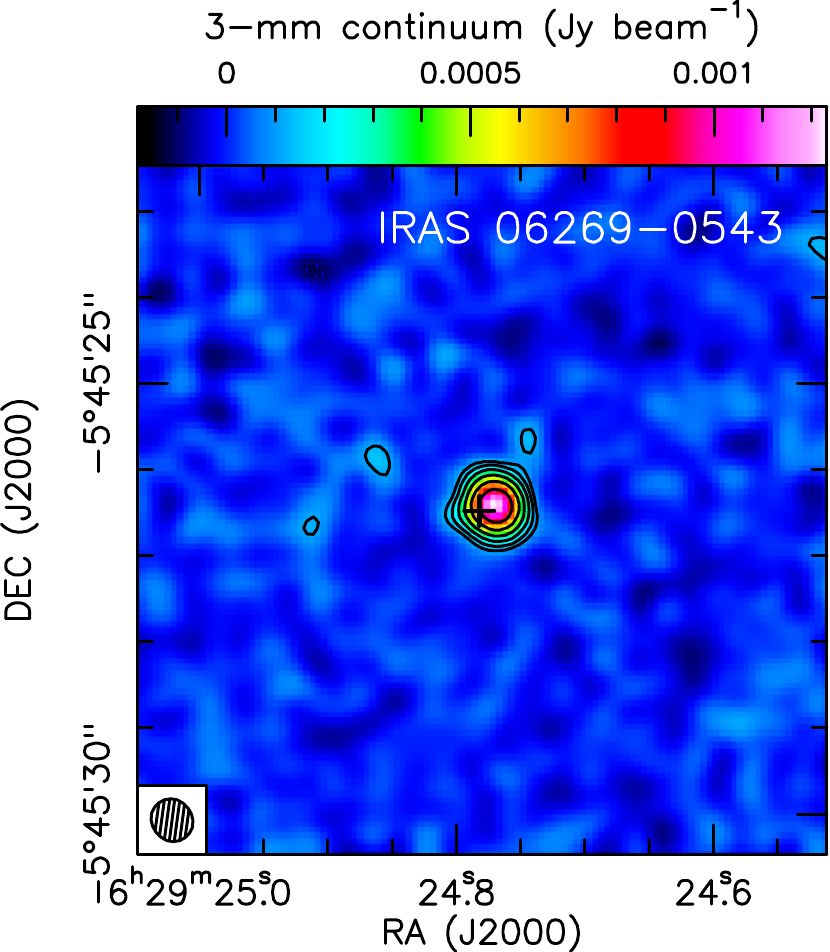}
\includegraphics[width=0.23\linewidth]{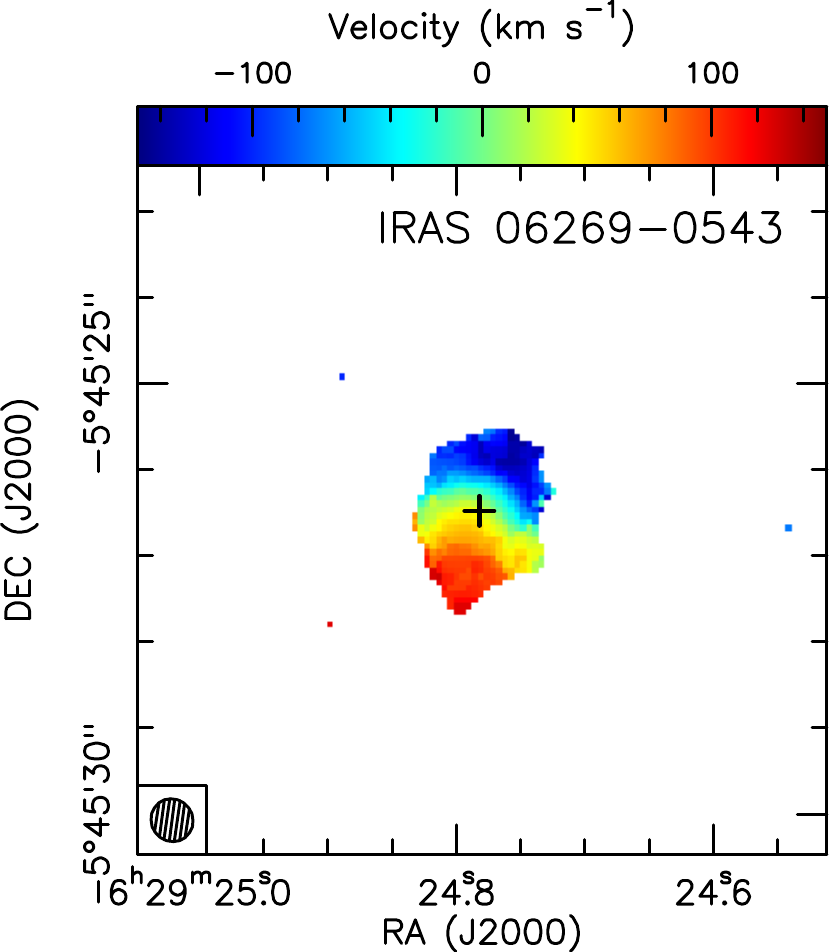}
\includegraphics[width=0.23\linewidth]{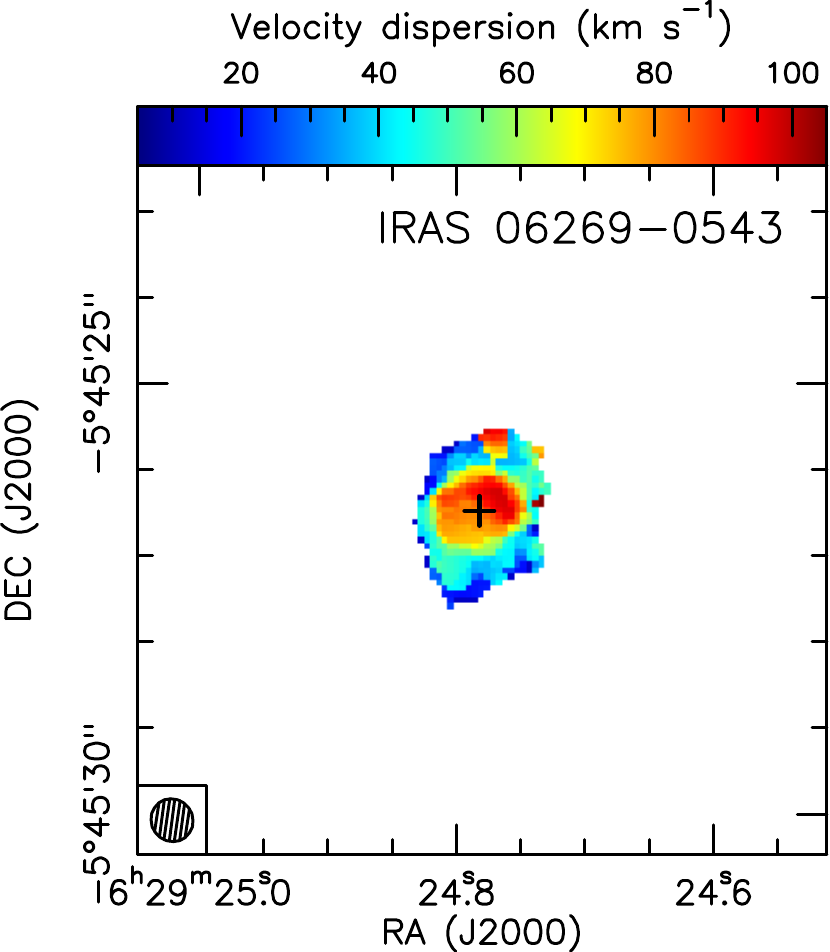}\\
\vspace{2pt}
\includegraphics[width=0.23\linewidth]{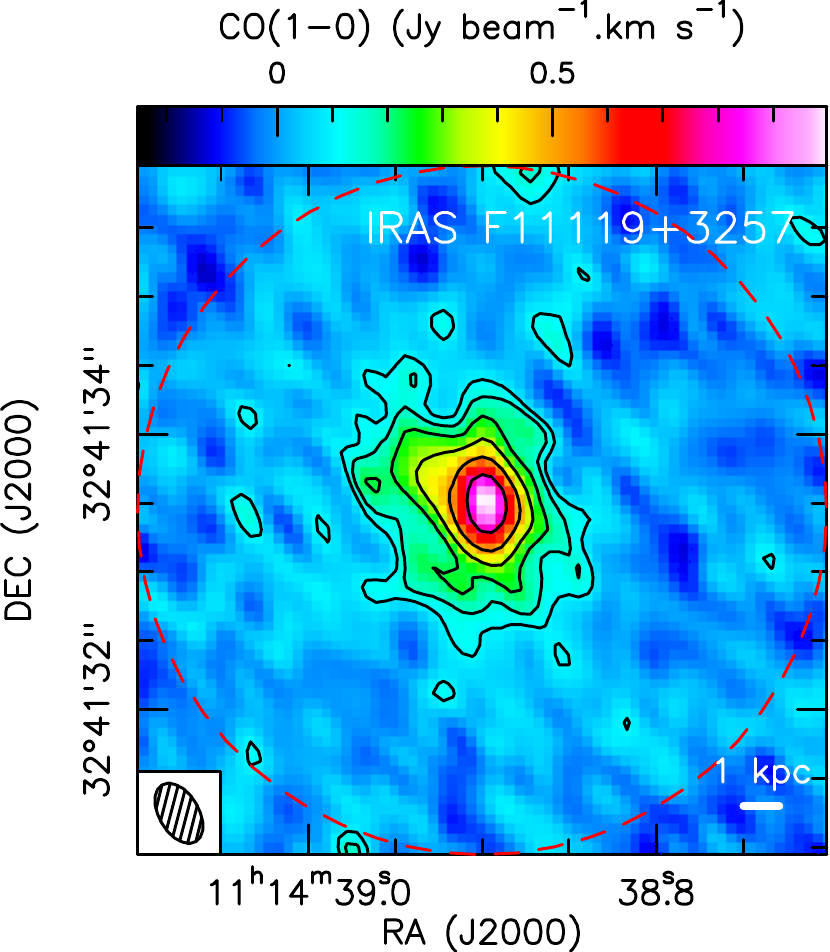}
\includegraphics[width=0.23\linewidth]{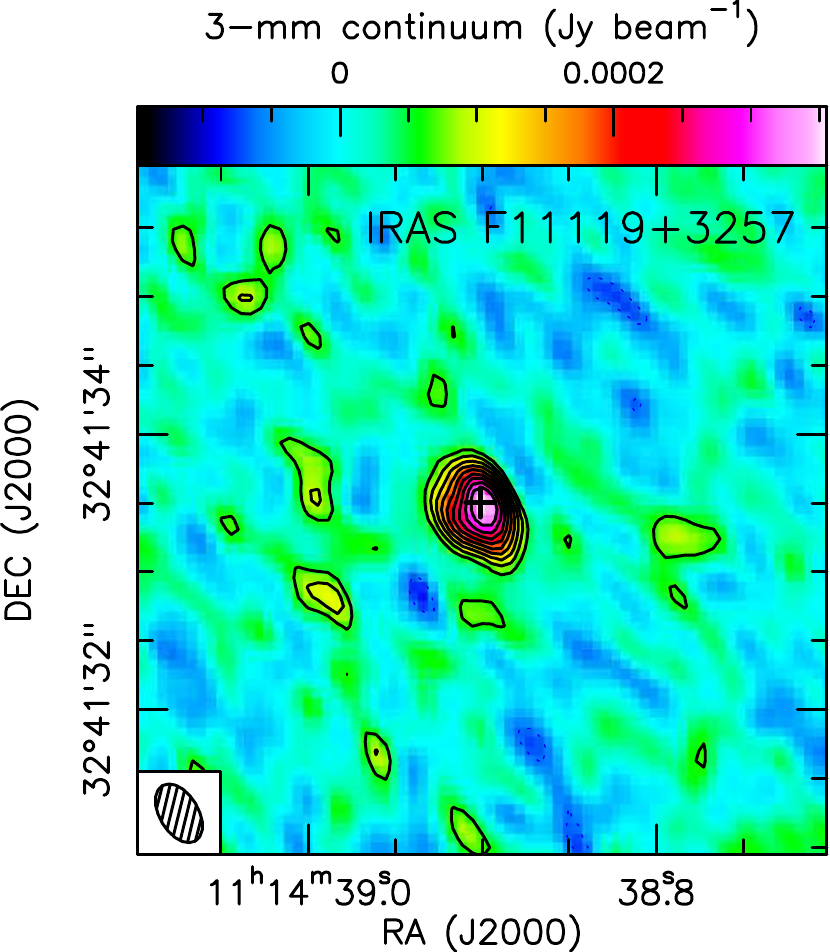}
\includegraphics[width=0.23\linewidth]{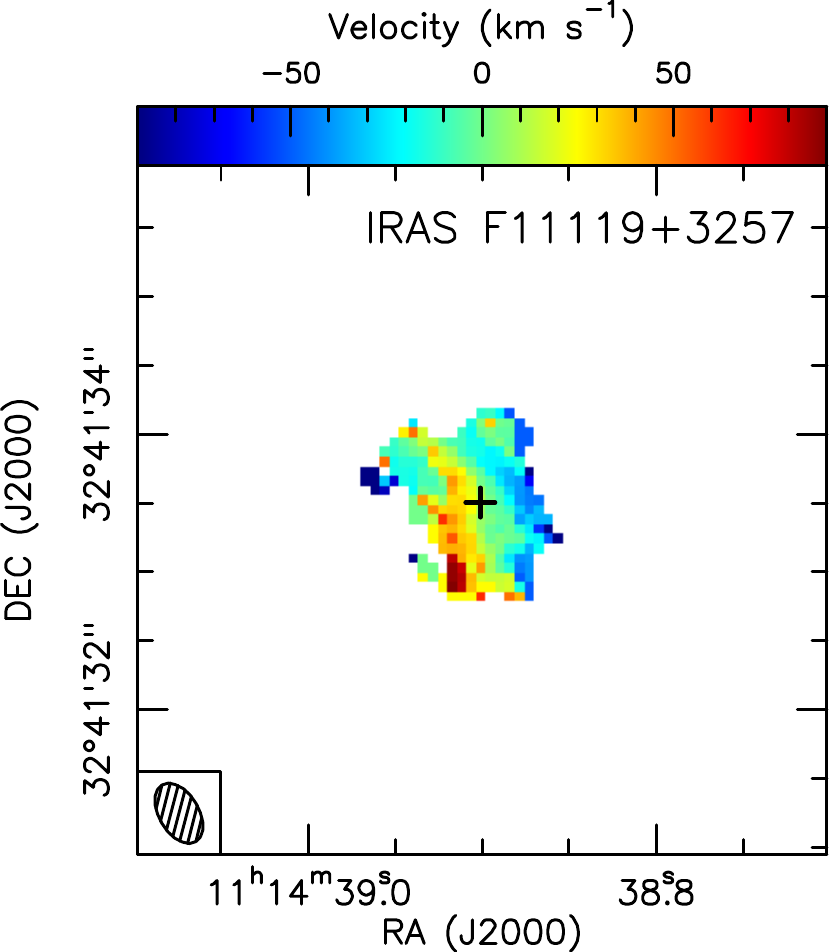}
\includegraphics[width=0.23\linewidth]{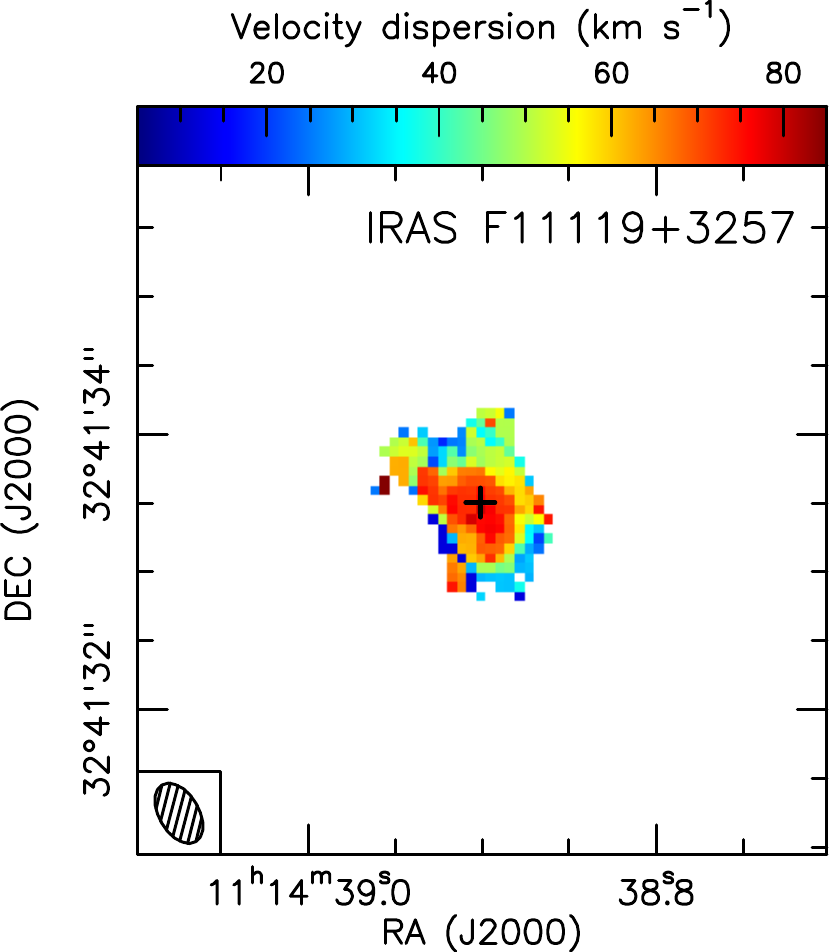}\\
\vspace{2pt}
\includegraphics[width=0.23\linewidth]{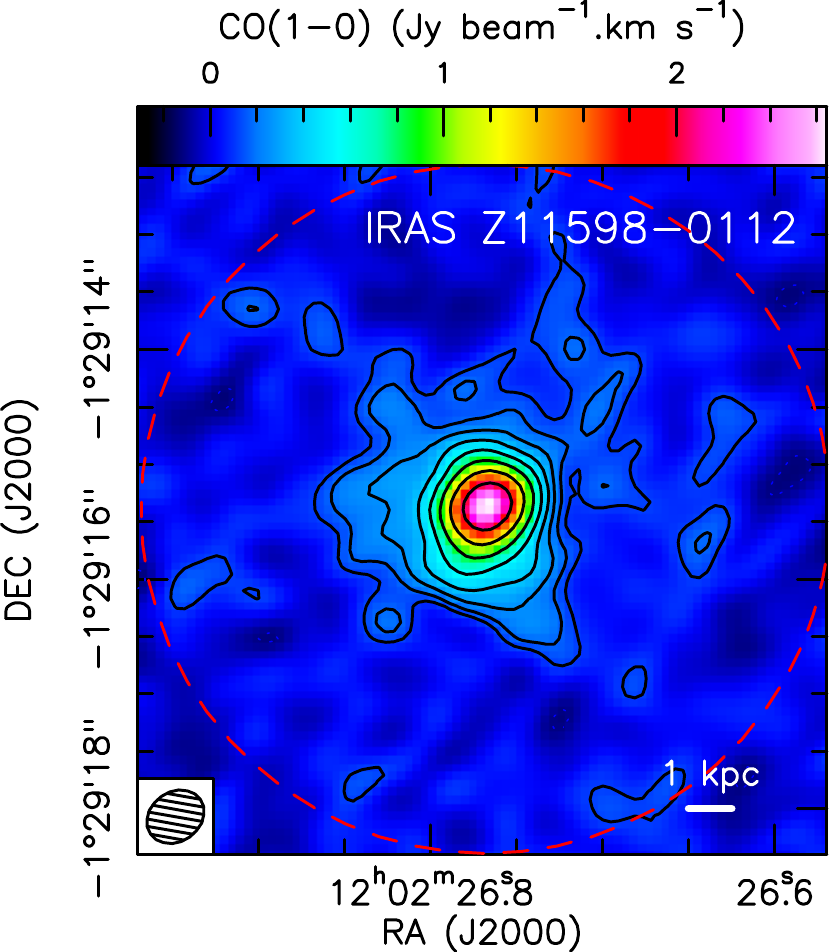}
\includegraphics[width=0.23\linewidth]{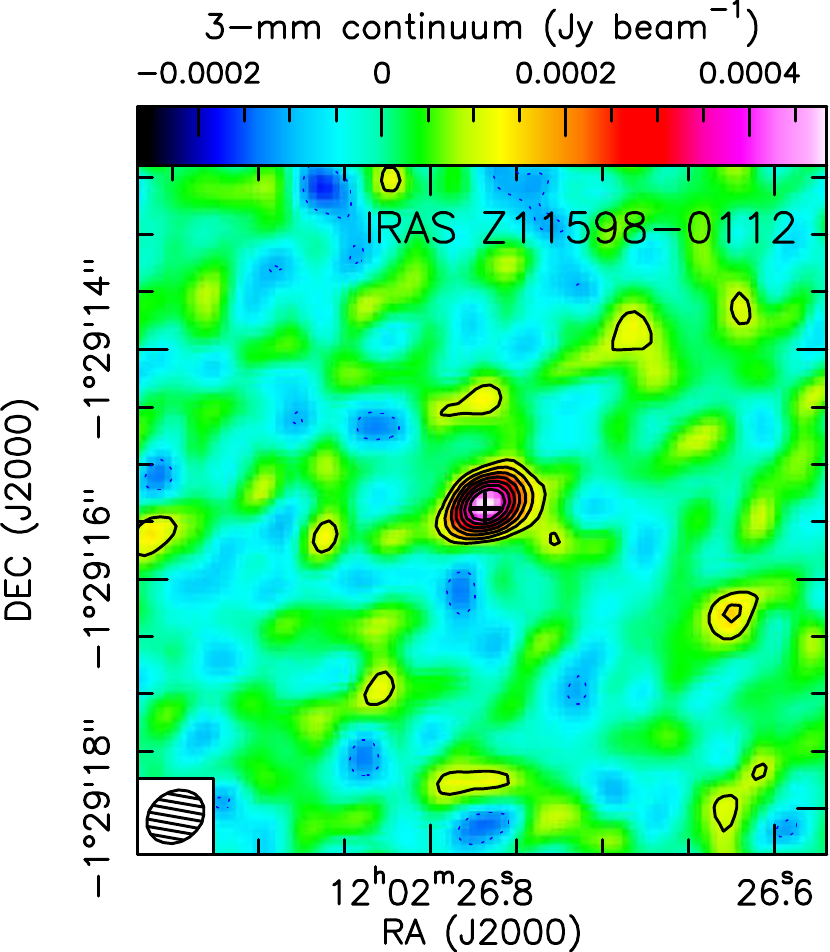}
\includegraphics[width=0.23\linewidth]{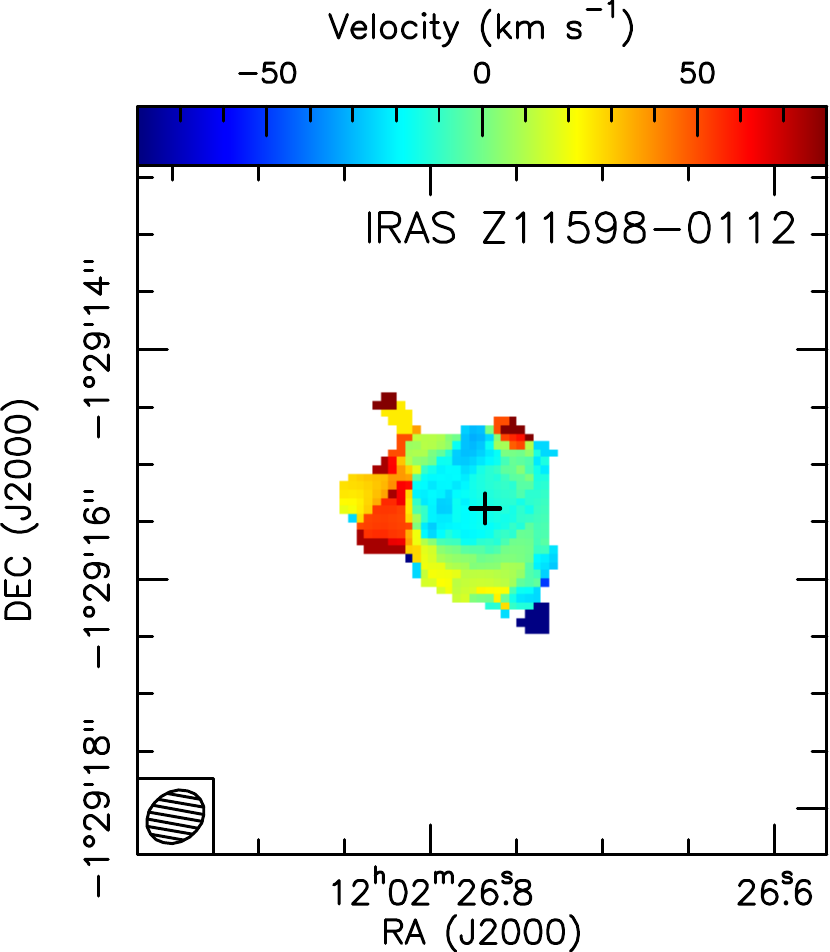}
\includegraphics[width=0.23\linewidth]{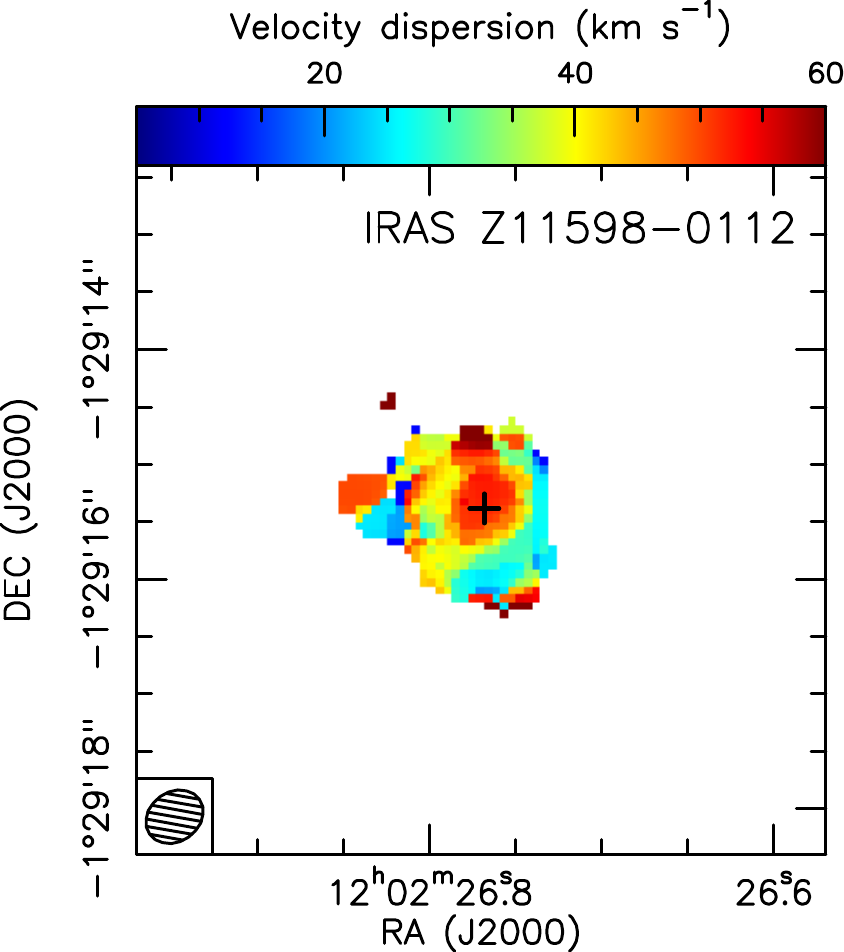}\\
\caption{The velocity-integrated CO(1-0) line maps (left), 3-mm continuum maps (middle-left), CO velocity fields (middle-right), and velocity dispersion maps (right) of the eight CO-mapping IR QSOs. Resolution of $0.\arcsec4-0.\arcsec6$ are achieved through Briggs weighting with robustness=2.0 and the beam size is shown in the bottom-left of each panel. The line intensity-weighted velocity and velocity dispersion maps are calculated using pixels detected at $\geqslant$ 3$\sigma$ and are masked where the integrated intensity is $<$ 3$\sigma$ in each object. The black cross symbol shown in the continuum, velocity, and velocity dispersion maps indicates the CO position derived from a 2D Gaussian fit to the integrated line map. The lowest contours in CO velocity integrated map and continuum map are $\pm$2$\sigma$, and contours increase by factors of 1.5 for the line image while the contours increase in steps of 1$\sigma$ for the continuum (except for IRAS~06269$-$0543, where the continuum contours increase by factors of 1.5). The 1$\sigma$ rms noise for CO line and continuum maps are: 0.119 and 0.078, 0.052 and 0.037, 0.050 and 0.031, 0.056 and 0.048, 0.079 and 0.050, 0.050 and 0.037, 0.058 and 0.047, 0.039 and 0.030, \jybeamkms\ and \mjybeam\ for I~ZW~1, IRAS~06269$-$0543, IRAS~F11119+3257, IRAS~Z11598$-$0112, IRAS~F15069+1808, IRAS~F15462$-$0450, IRAS~F22454$-$1744, and IRAS~F23411+0228, respectively. The red-dashed circles represent the circular apertures used to extract CO spectra for comparison (see Section~\ref{subsec:spect}). For the objects that are resolved into multiple CO sources, IRAS~F15069+1808, IRAS~F15462$-$0450, IRAS~F22454$-$1744, and IRAS~F23411+0228, each CO component is labelled in C-Center, N-North, E-East, NW, SE, NE, and SW, respectively. \label{fig:moment}}
\end{figure*}

%\clearpage

\setcounter{figure}{0}
\begin{figure*}[htbp]
\centering
\includegraphics[width=0.23\linewidth]{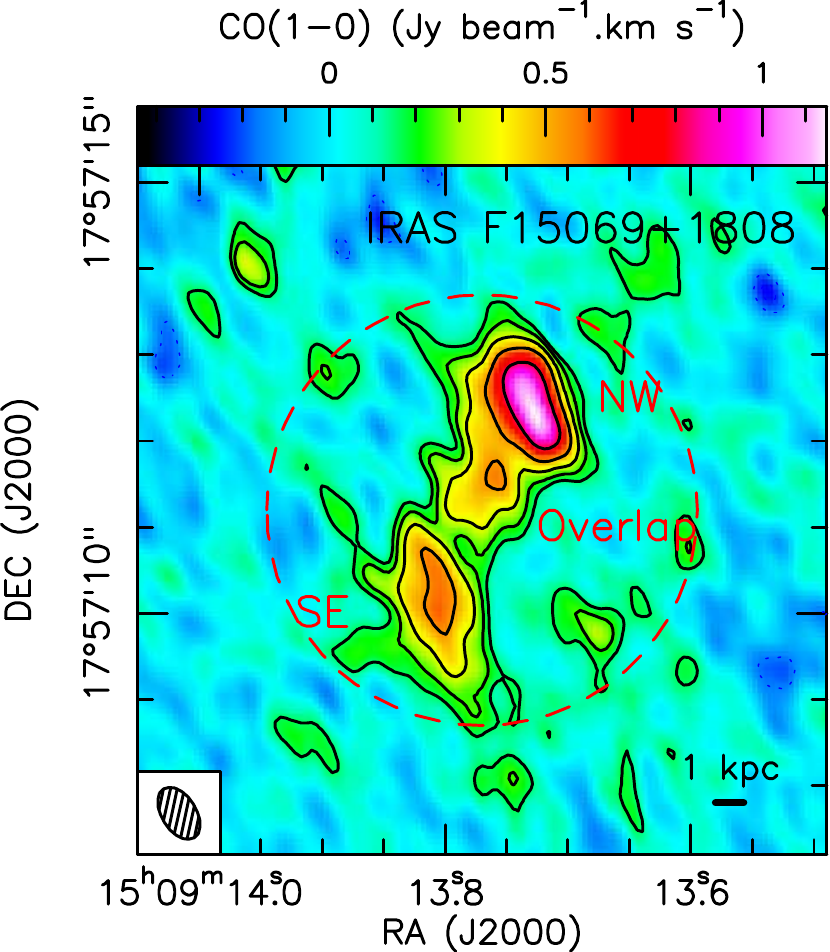}
\includegraphics[width=0.23\linewidth]{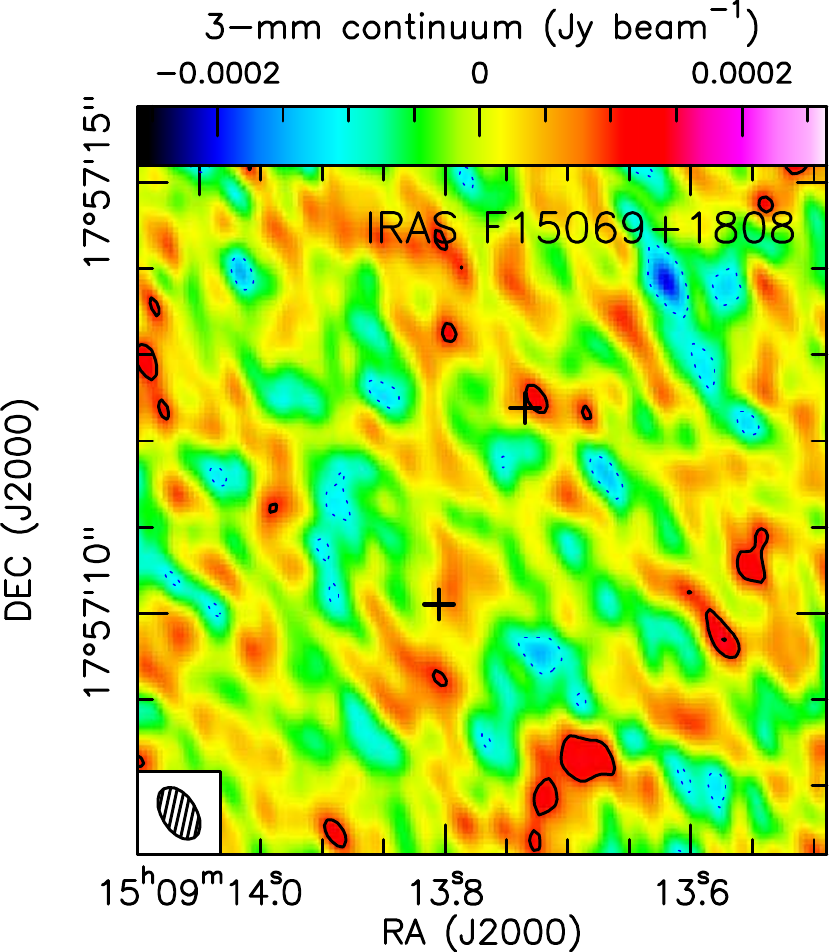}
\includegraphics[width=0.23\linewidth]{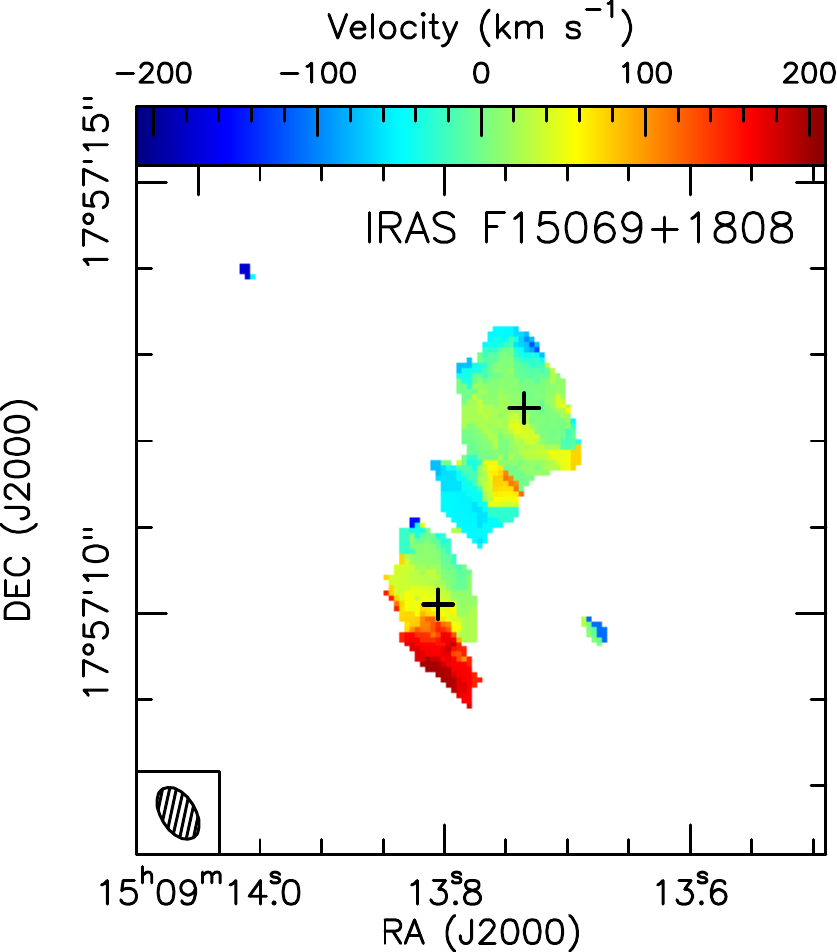}
\includegraphics[width=0.23\linewidth]{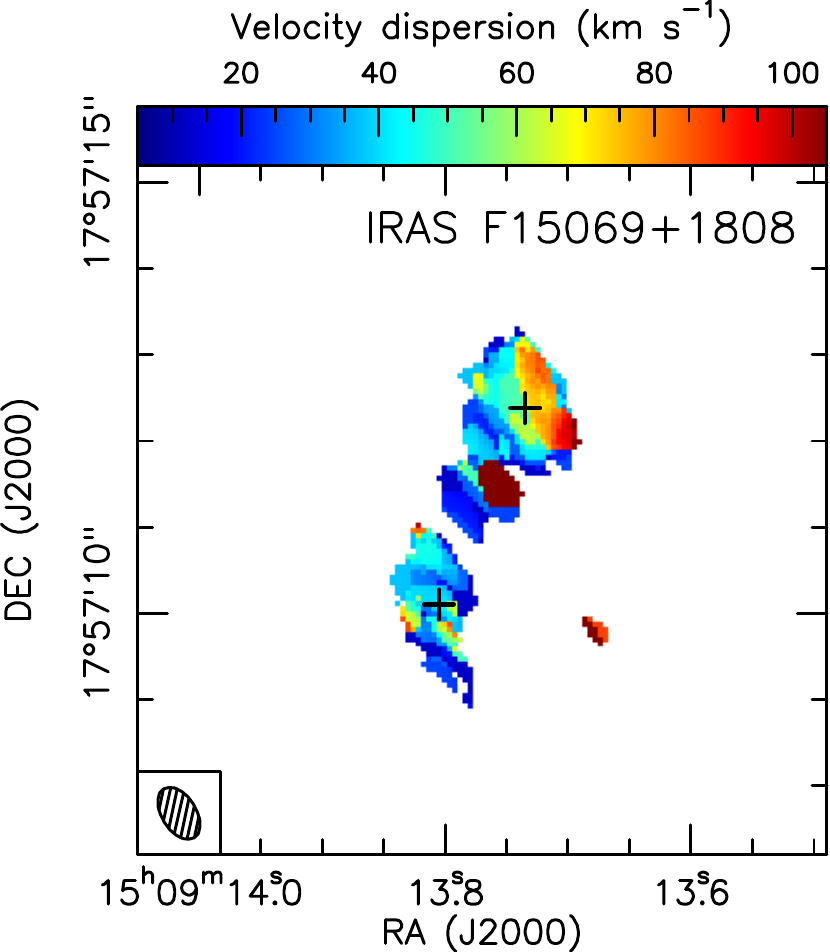}\\
\vspace{2pt}
\includegraphics[width=0.23\linewidth]{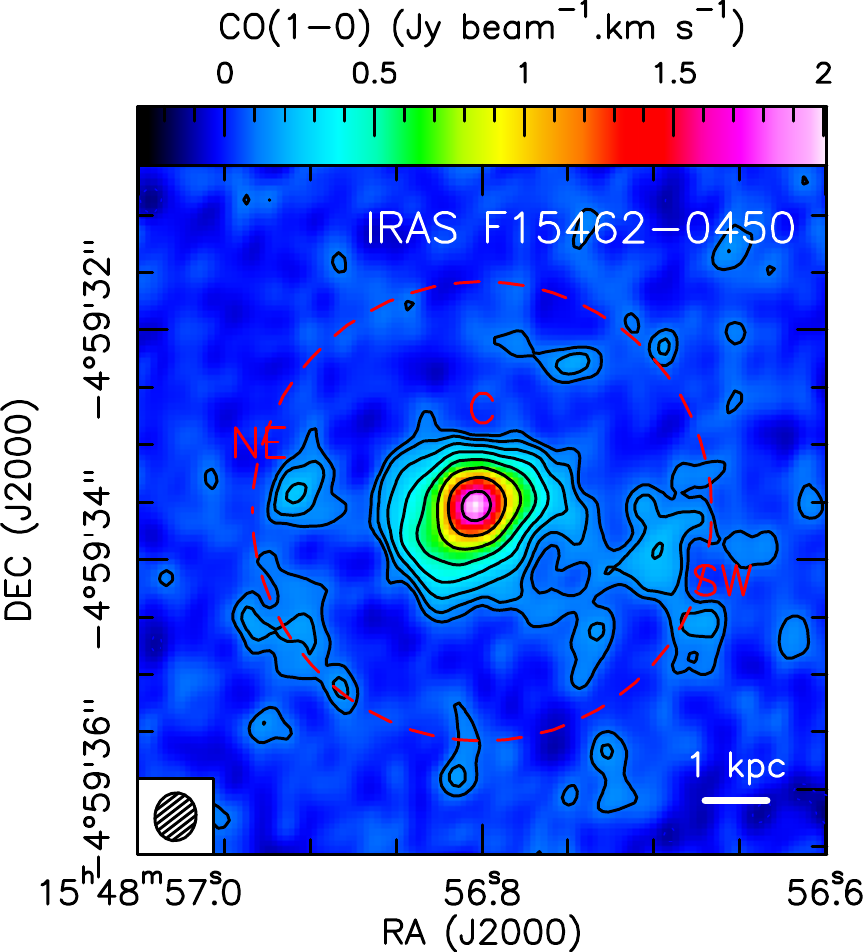}
\includegraphics[width=0.23\linewidth]{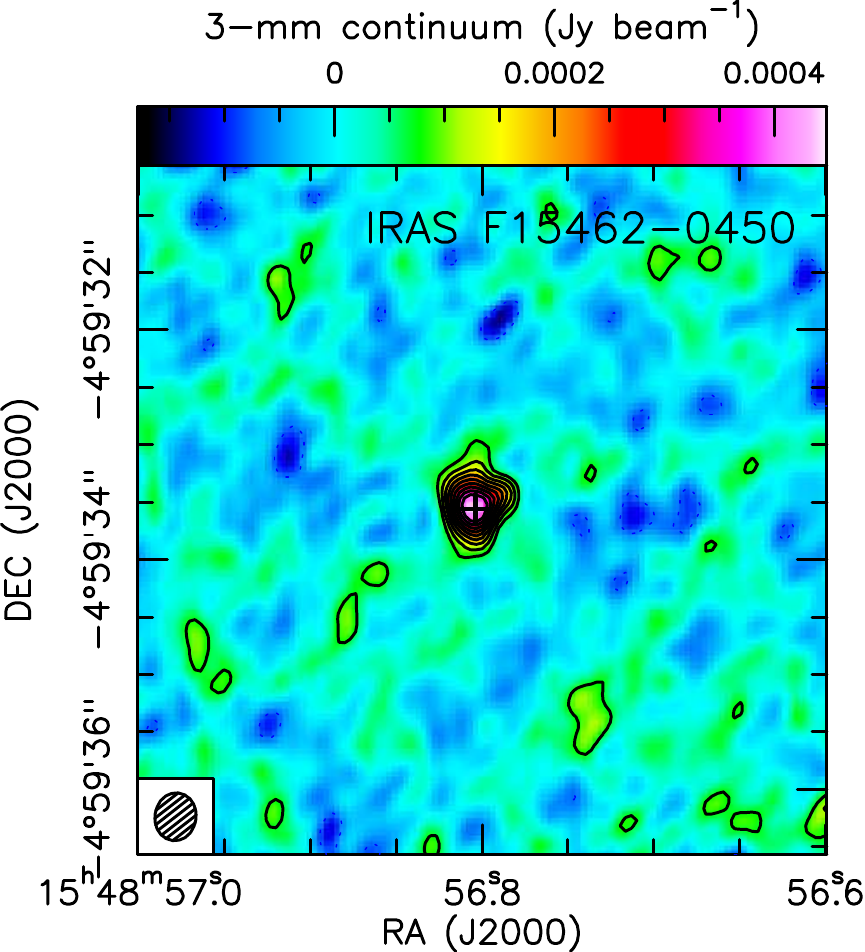}
\includegraphics[width=0.23\linewidth]{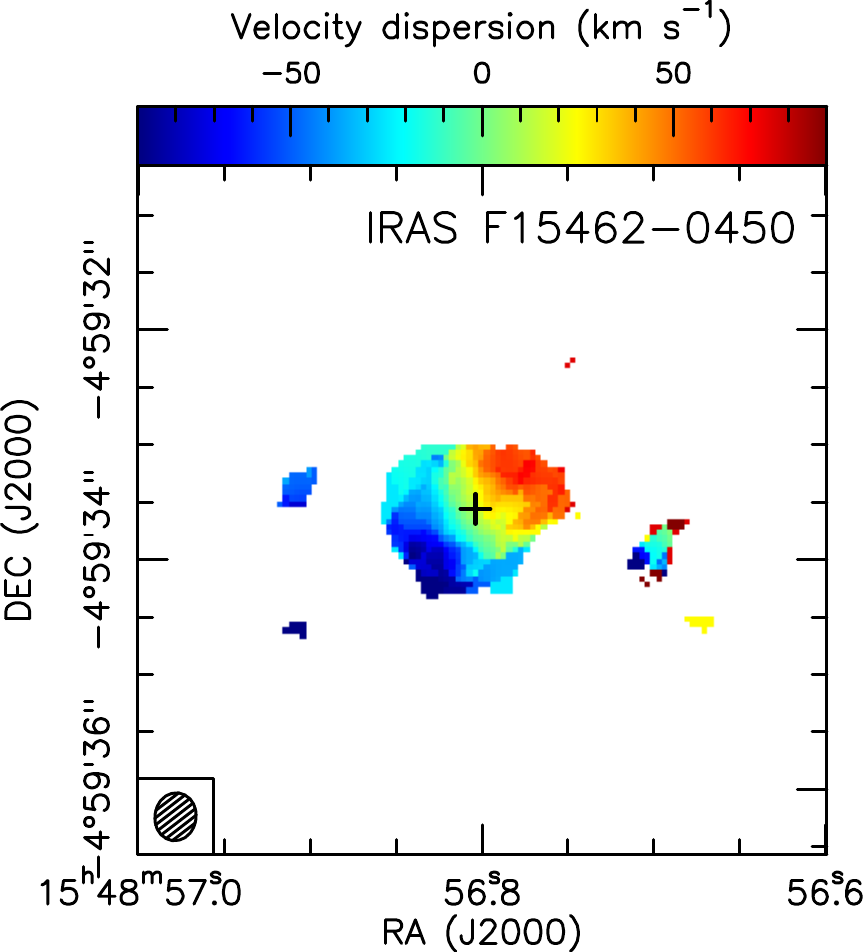}
\includegraphics[width=0.23\linewidth]{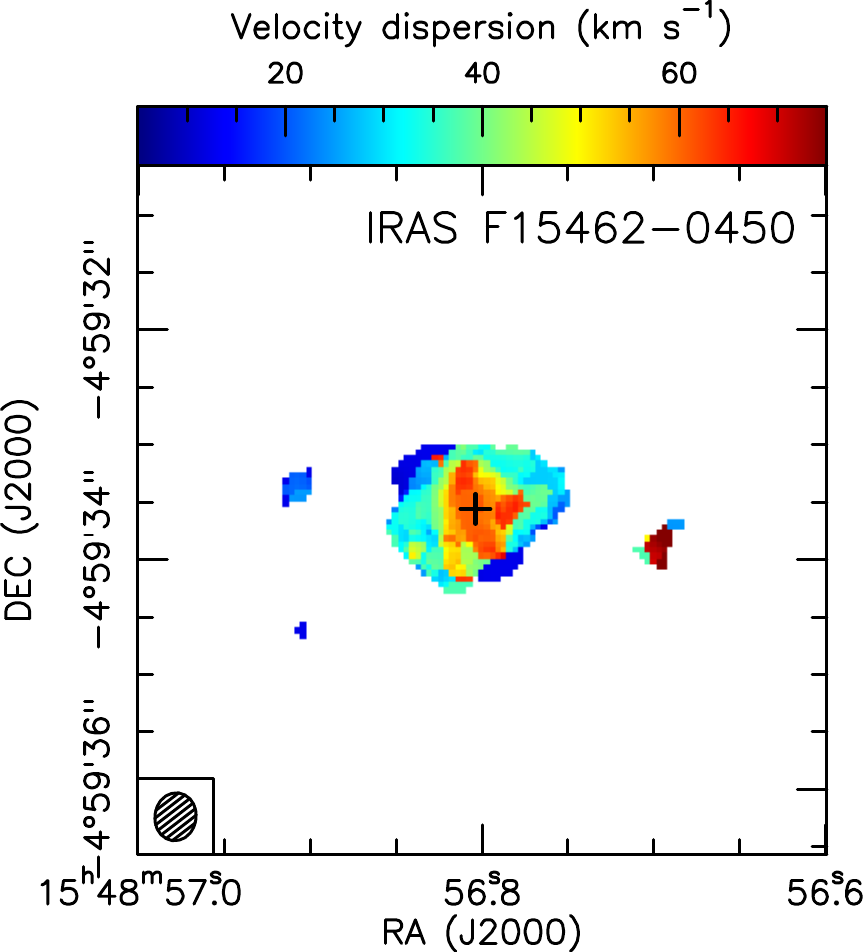}\\
\vspace{2pt}
\includegraphics[width=0.23\linewidth]{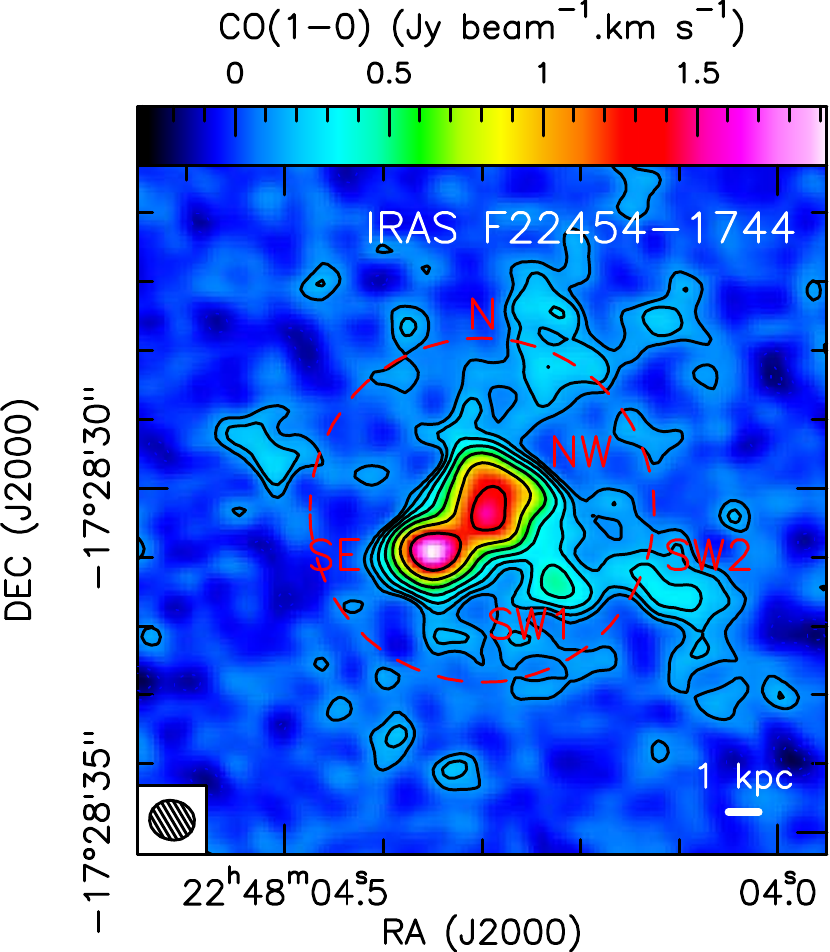}
\includegraphics[width=0.23\linewidth]{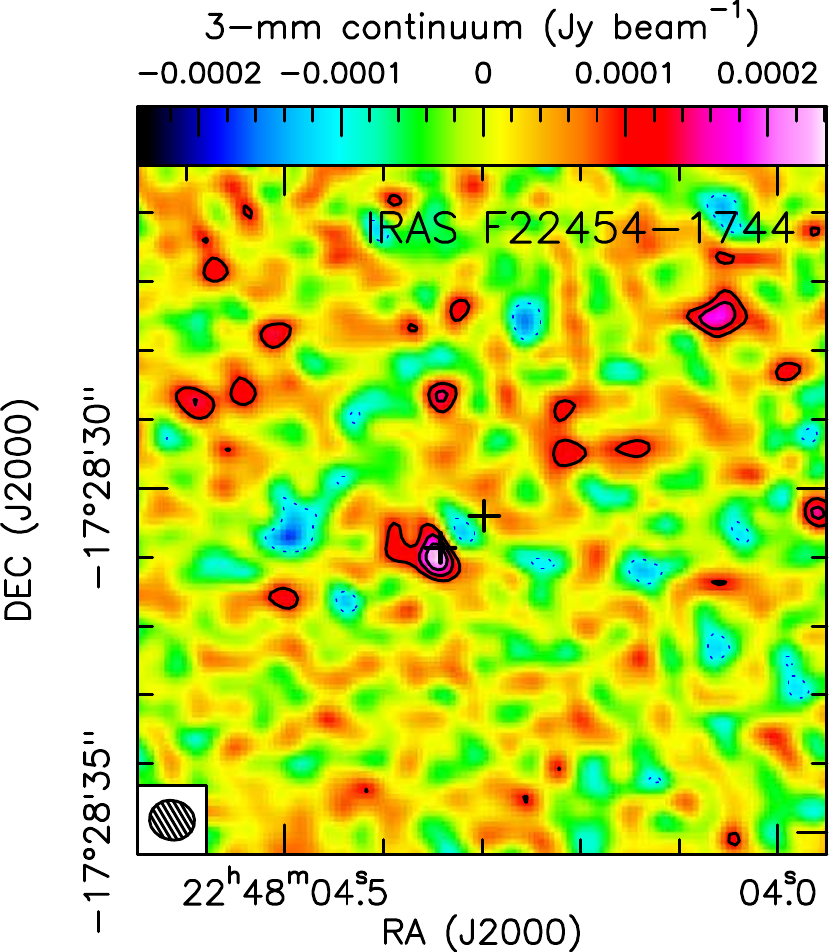}
\includegraphics[width=0.23\linewidth]{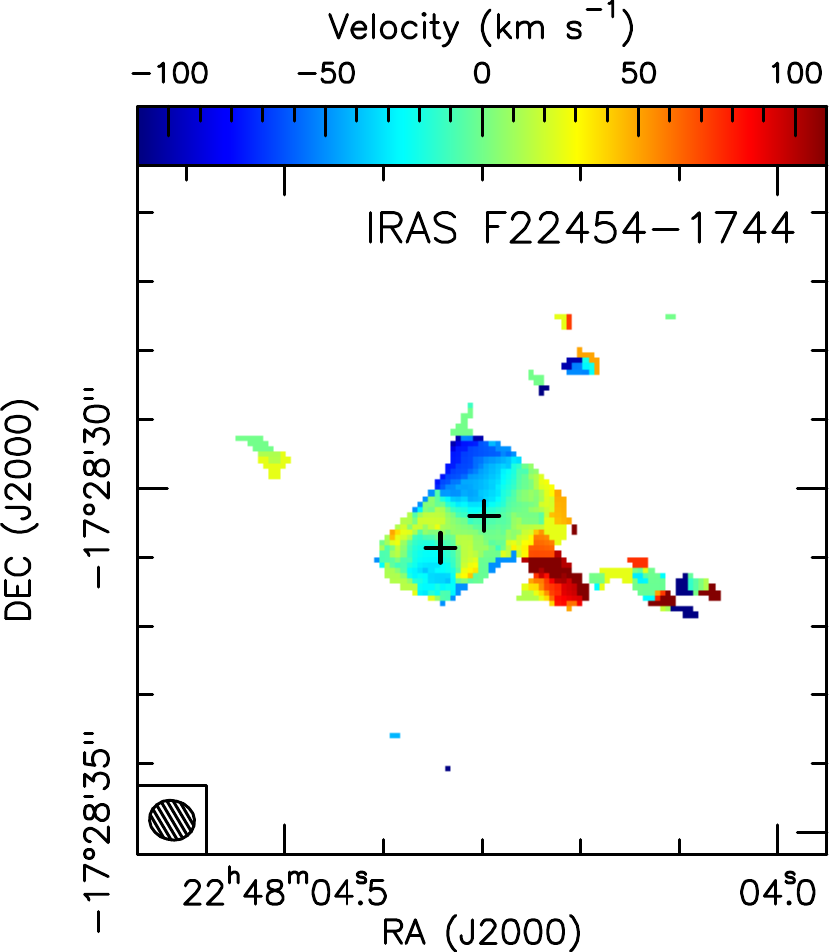}
\includegraphics[width=0.23\linewidth]{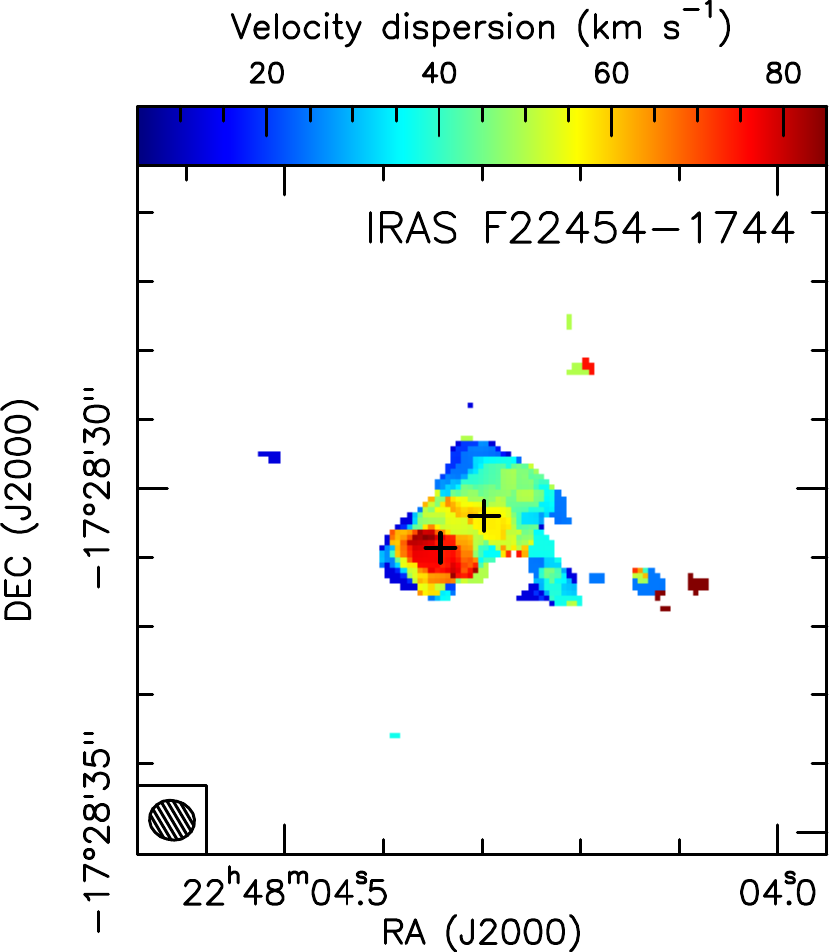}\\
\vspace{2pt}
\includegraphics[width=0.23\linewidth]{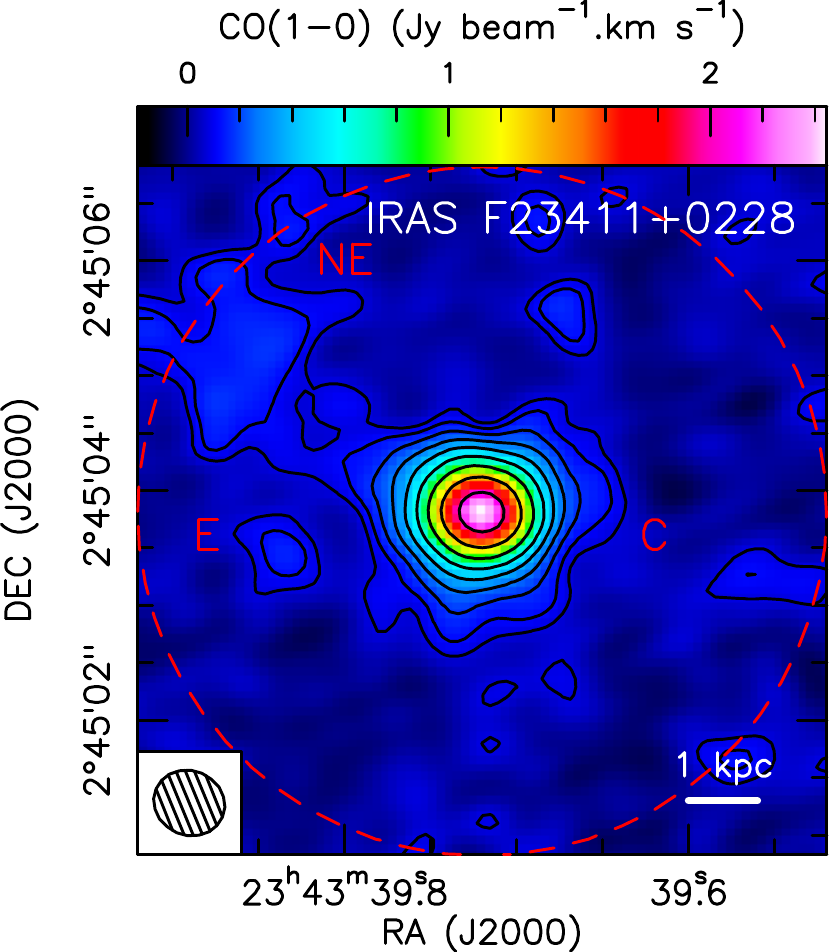}
\includegraphics[width=0.23\linewidth]{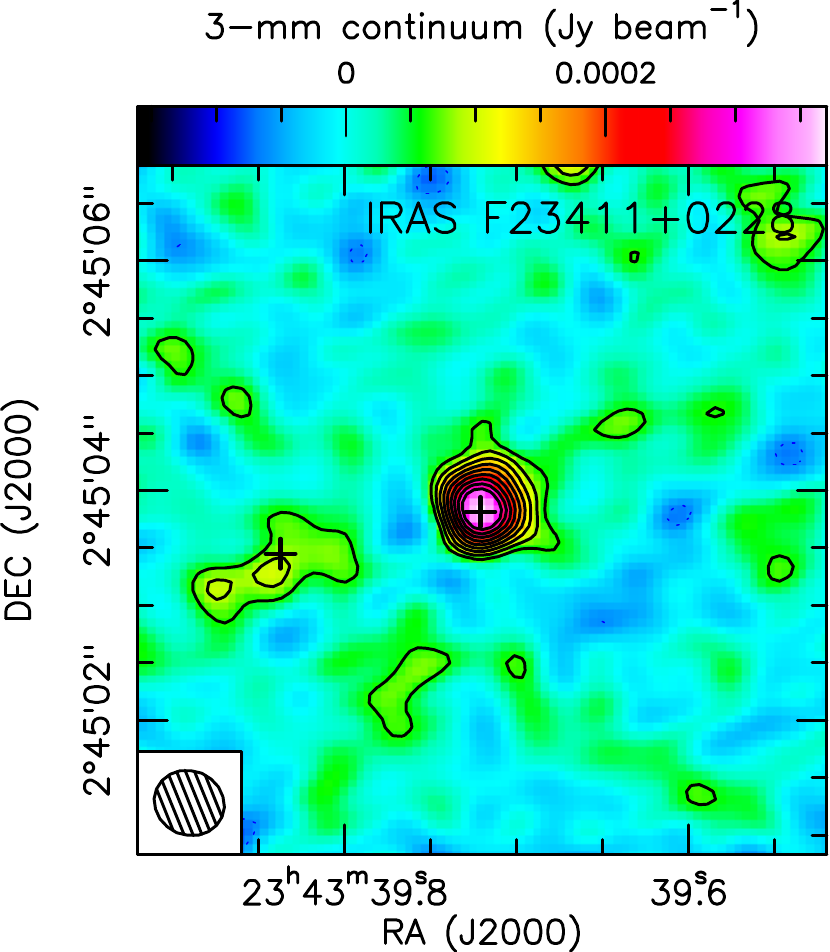}
\includegraphics[width=0.23\linewidth]{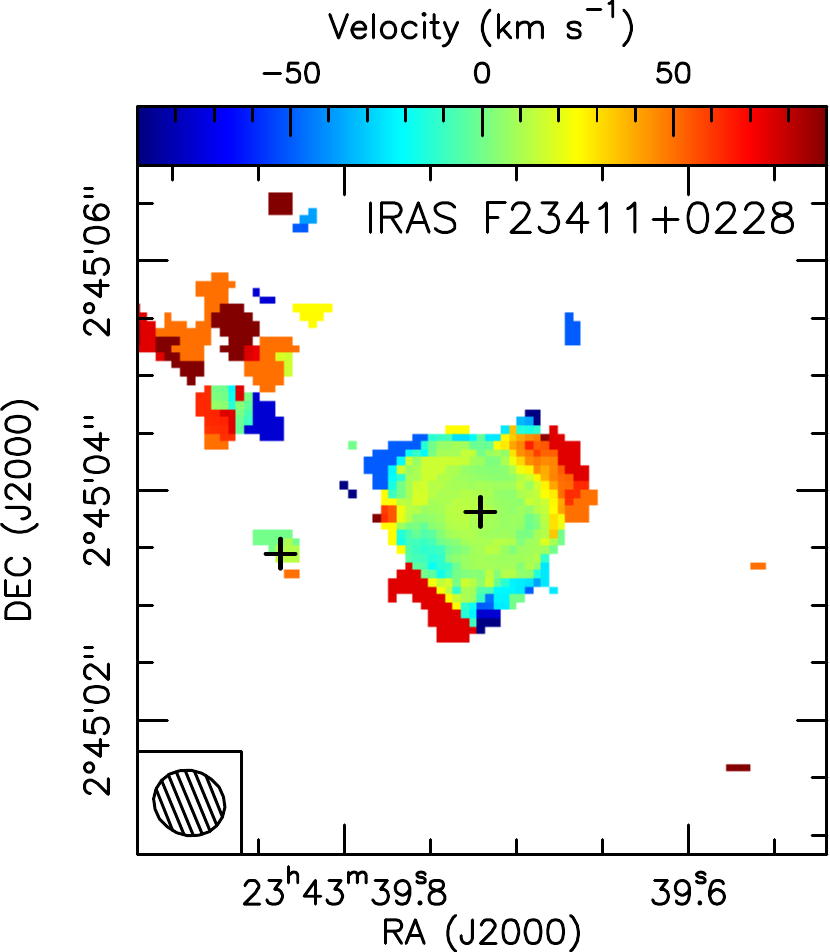}
\includegraphics[width=0.23\linewidth]{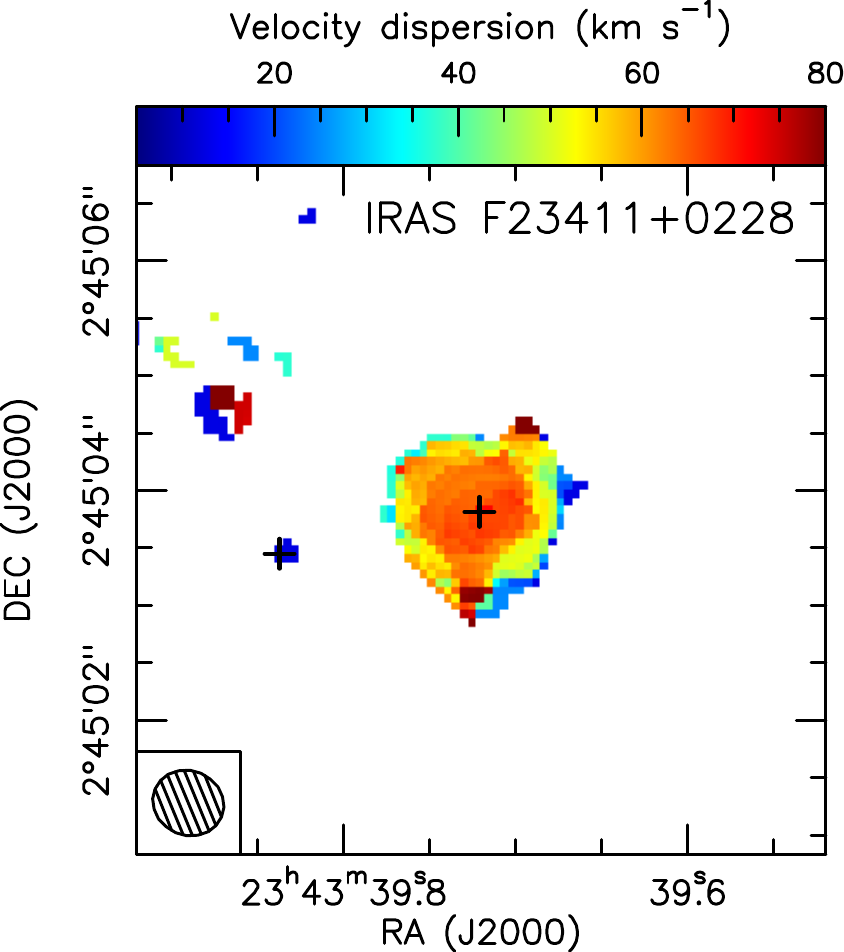}\\
\caption{(Continued)}
\end{figure*}
%%%%%%%%%%%%%%%%%%%%%%%%%%%%%%%%%%%%%%%

%%%%%%%%%%%%%%%- Fig-2 -%%%%%%%%%%%%%%%%

\begin{figure*}[htbp]
\centering
\includegraphics[width=0.24\linewidth]{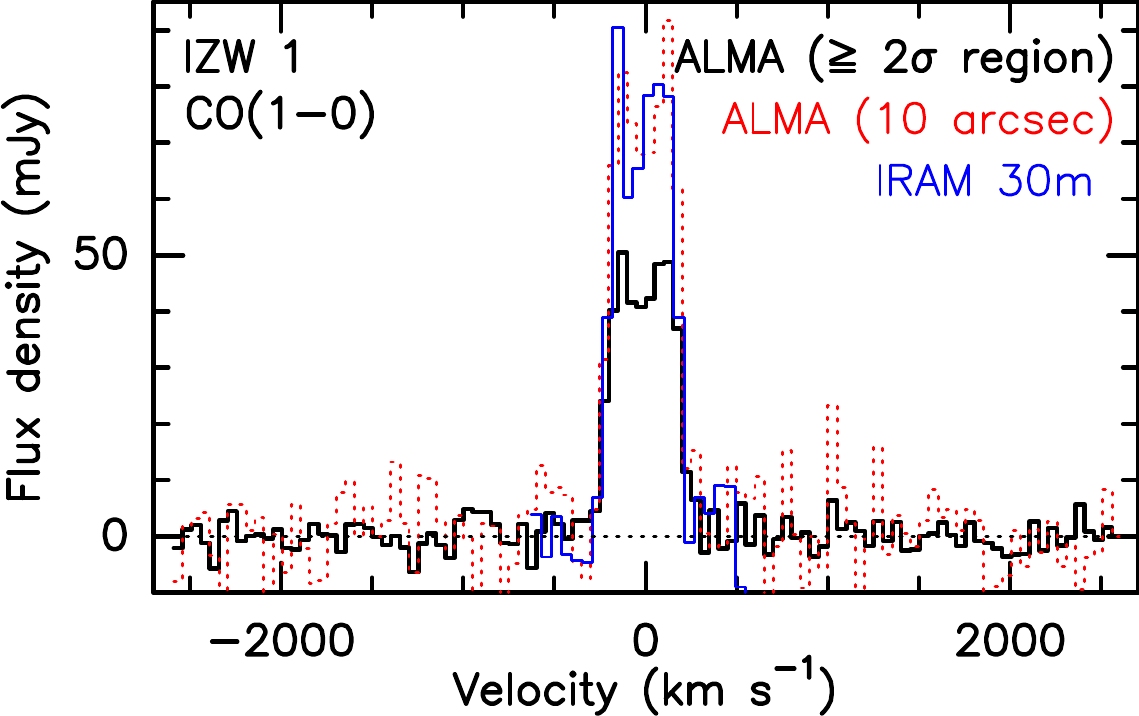}
\includegraphics[width=0.24\linewidth]{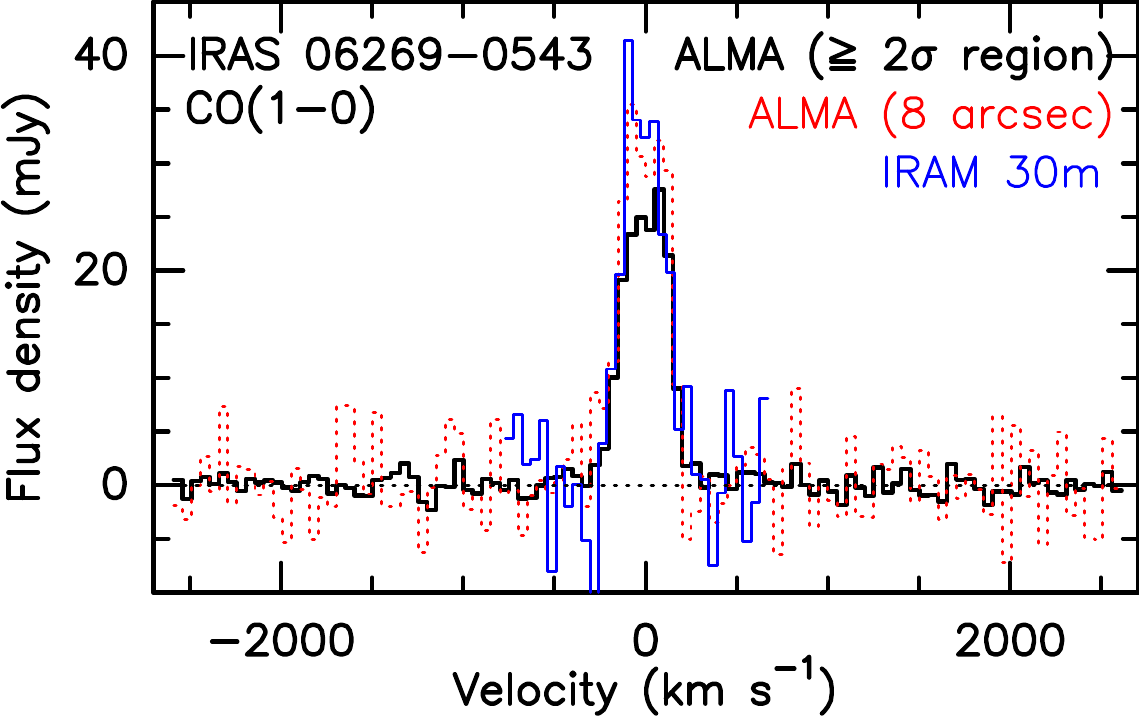}
\includegraphics[width=0.24\linewidth]{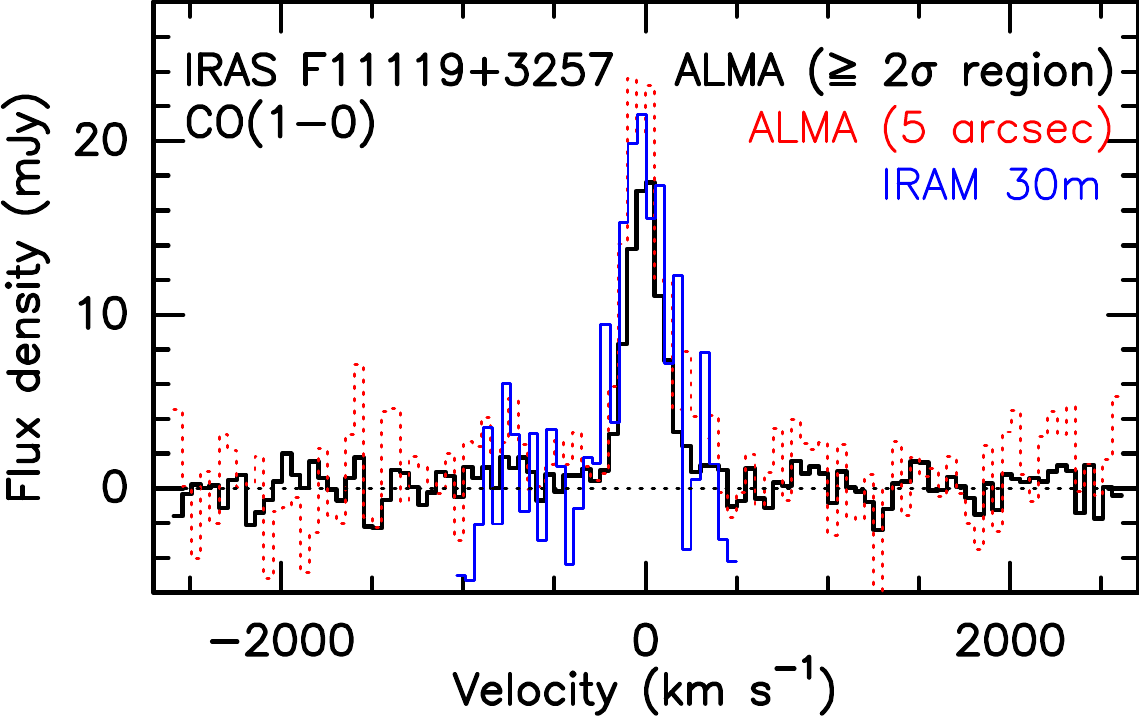}
\includegraphics[width=0.24\linewidth]{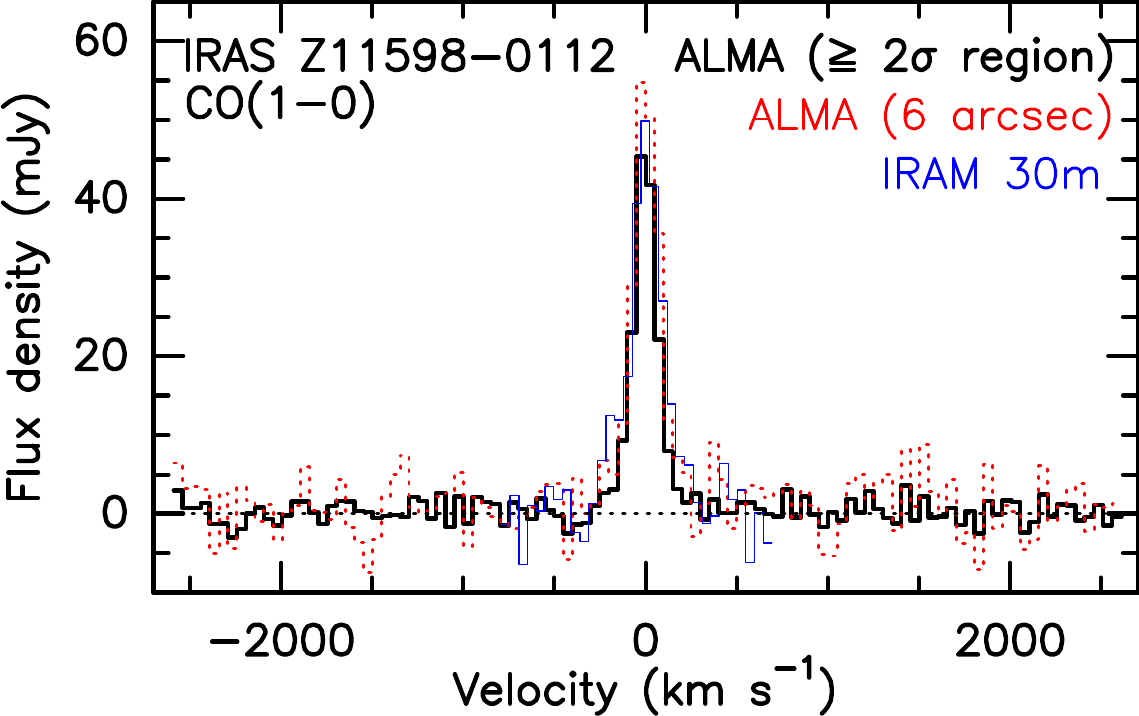}\\
\vspace{5pt}
\includegraphics[width=0.24\linewidth]{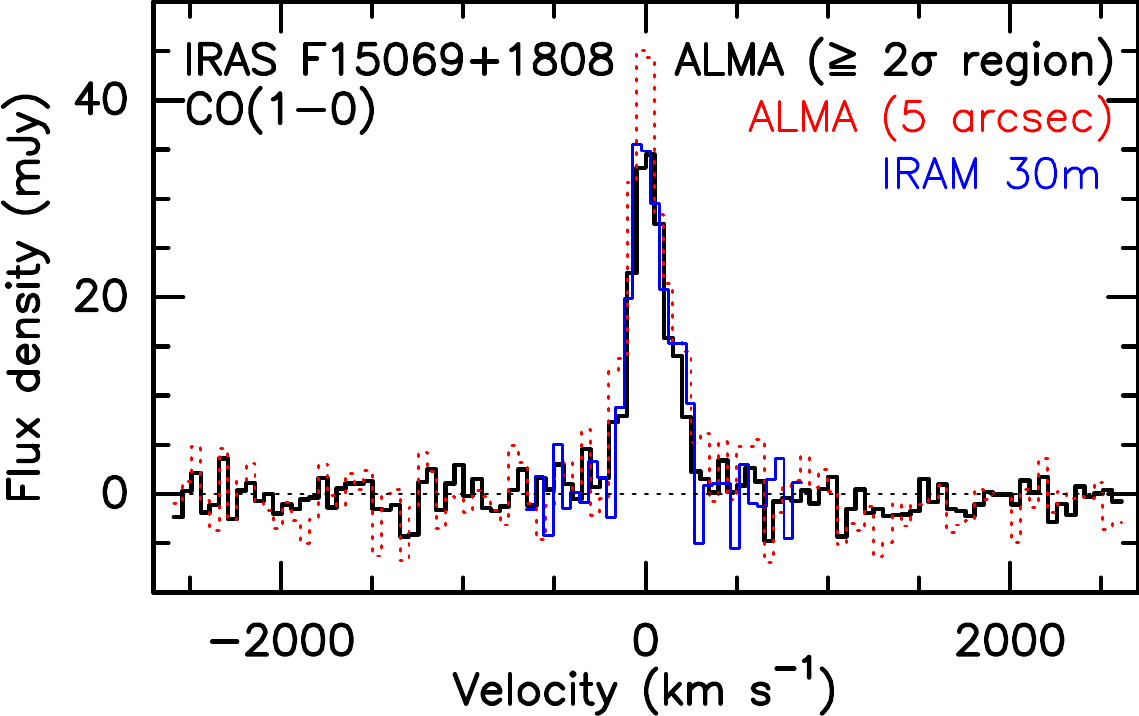}
\includegraphics[width=0.24\linewidth]{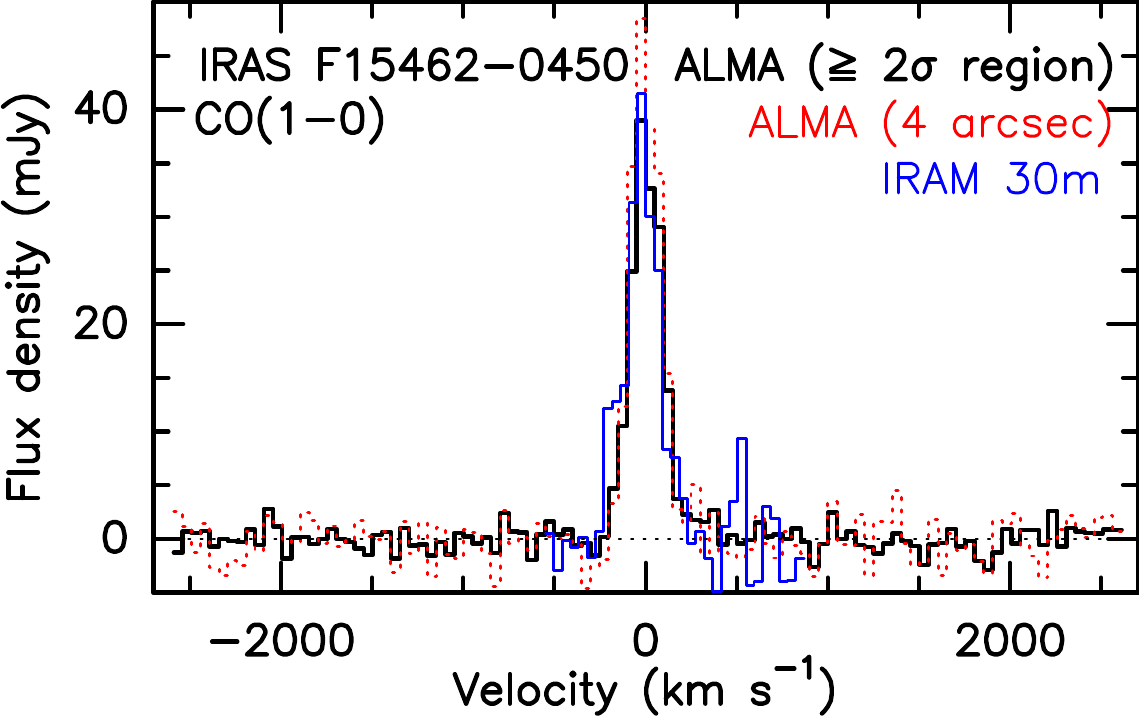}
\includegraphics[width=0.25\linewidth]{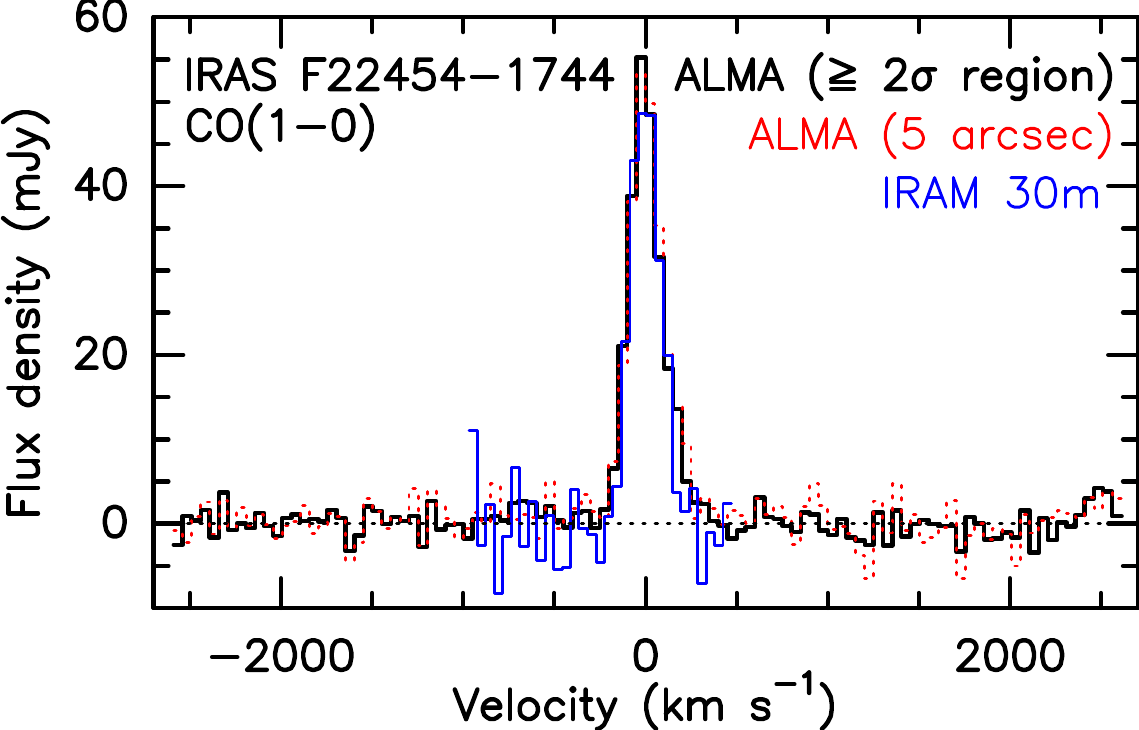}
\includegraphics[width=0.25\linewidth]{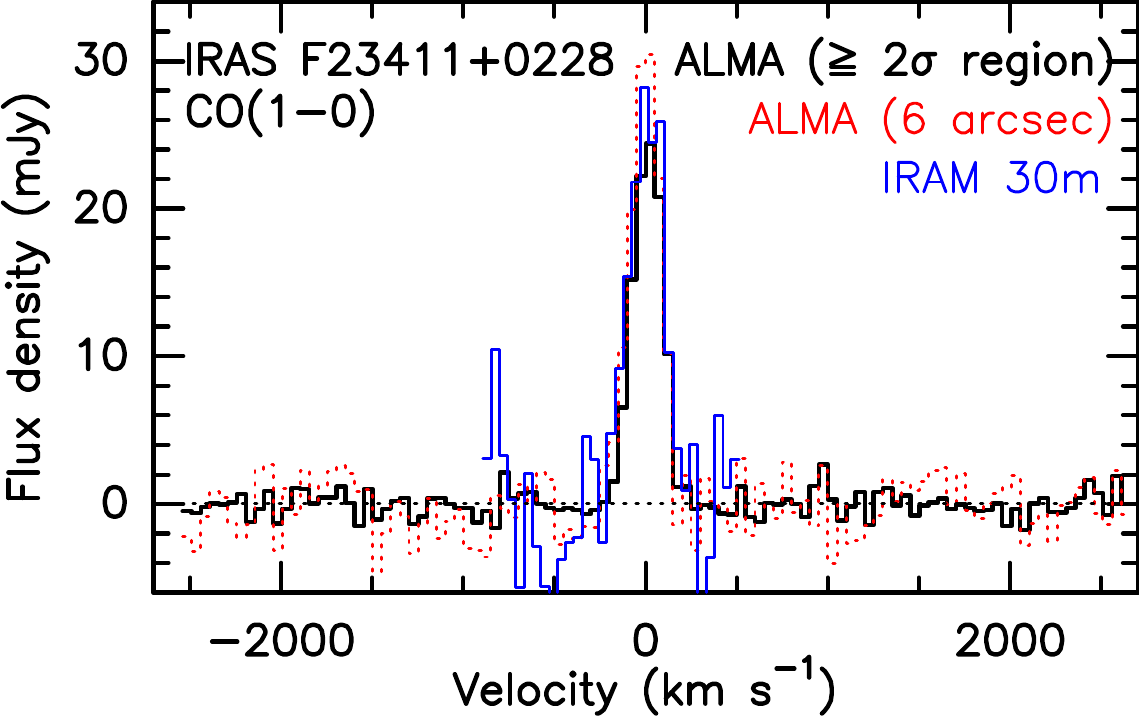}\\
\caption{ALMA continuum-subtracted CO(1-0) emission line spectra of IR QSO systems extracted from the region within the $\geqslant$2$\sigma$ signal in the velocity-integrated map (black) and from a circular aperture (red-dotted; see the aperture shown in Figure~\ref{fig:moment}) with diameter listed in the top right, compared with the CO(1-0) line observed with the IRAM 30m (blue). The zero velocity corresponds to the redshifts listed in Table~\ref{tab:meas}. All spectra are binned to a channel width of $\sim$50~\kms\ with Hanning smoothing.\label{fig:spec}}
\end{figure*}

%%%%%%%%%%%%%%%%%%%%%%%%%%%%%%%%%%%%%%%

Similar to the interacting system IRAS~F15069+1808 described above, the CO gas in IRAS~F22454$-$1744 is also resolved into multiple sources with two arm-like structures (north and southwest arms, denoted as N and SW) originating from the two bright nuclear sources (denoted as NW and SE, Figure~\ref{fig:moment}), which are undergoing merging with a projected separation of 0.\arcsec78 ($\sim$ 1.7 kpc) between the two central CO peaks. The 3-mm continuum emission is detected in the nuclear region of F22454$-$1744SE at the $4-5\sigma$ level with a flux density of 0.253$\pm$0.078 mJy, whereas non-detected in the source NW. The SW arm is resolved into two clumps (SW1 and SW2), which form a filamentary structure. 

The CO distribution of IRAS~F23411+0228 shows that the molecular gas is strongly concentrated in the central region (F23411+0228C) accompanied by an arm-like feature to the NE and a clumpy structure outside the main body of the galaxy to the east (F23411+0228E). We detect 3-mm continuum emission in F23411+0228C at $\sim$ 8$\sigma$ level with flux of 0.433$\pm$0.055 mJy. Most importantly, we also tentatively detect 3-mm continuum emission with flux of 86$\pm$30 $\mu$Jy from F23411+0228E, supporting the reality of such a feature. The angular separation between C and E components in IRAS~F23411+0228 system is 1.\arcsec77 ($\sim$ 3.0 kpc). The spectrum extracted for F23411+0228E demonstrates that the line emission (see Figure~\ref{fig:spec}), if associated with a redshifted CO(1$-$0) transition, comes from the same redshift as the main galaxy, suggesting that the source E is likely physically related to the IRAS~F23411+0228 system. However, it is not clear whether this is due to an ongoing merger or interaction, or whether it is a clumpy star-forming region at large radii.  

\begin{figure*}[htbp]
\centering
\includegraphics[width=0.25\linewidth]{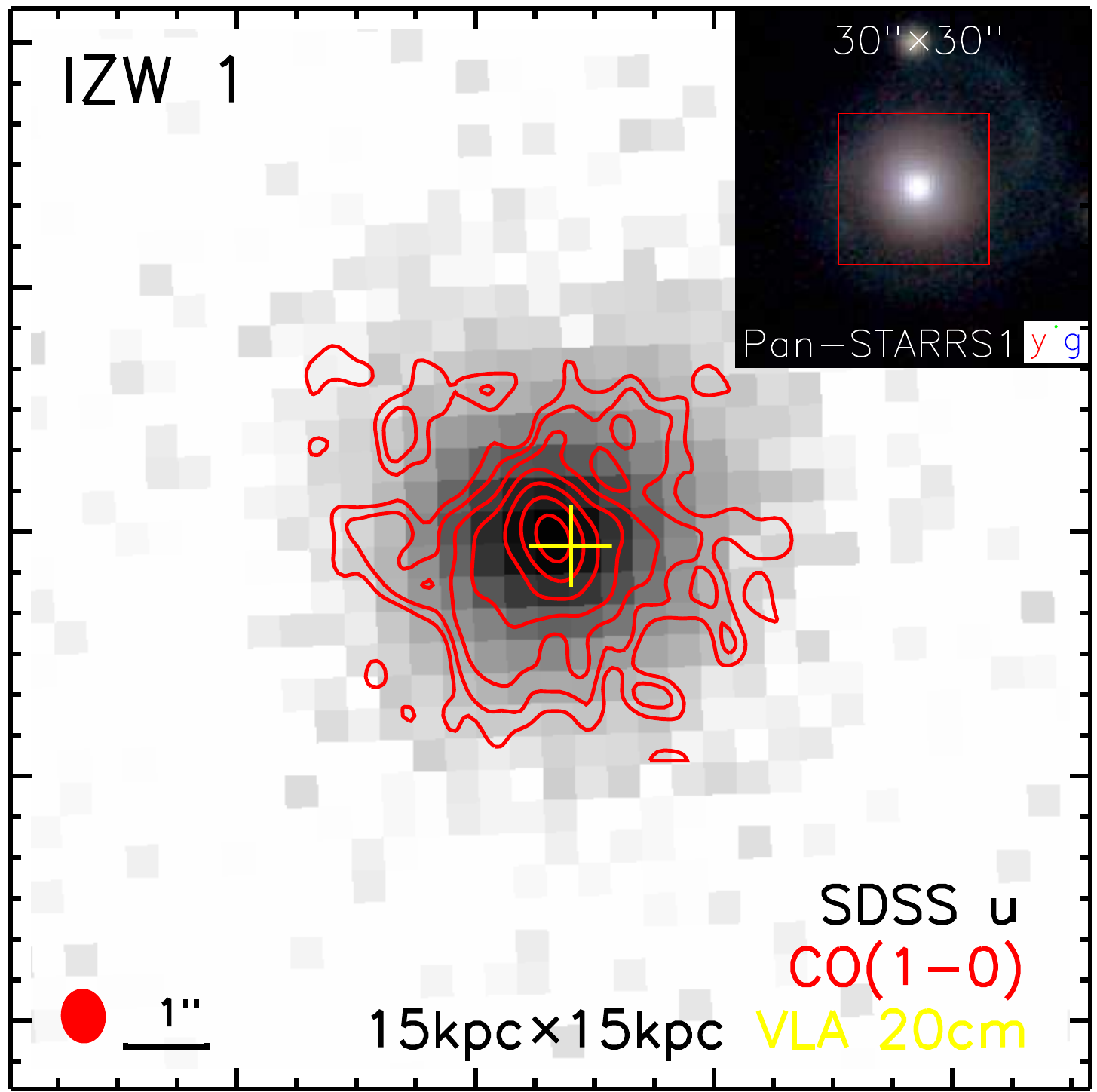}
\hspace{-6pt}
\includegraphics[width=0.25\linewidth]{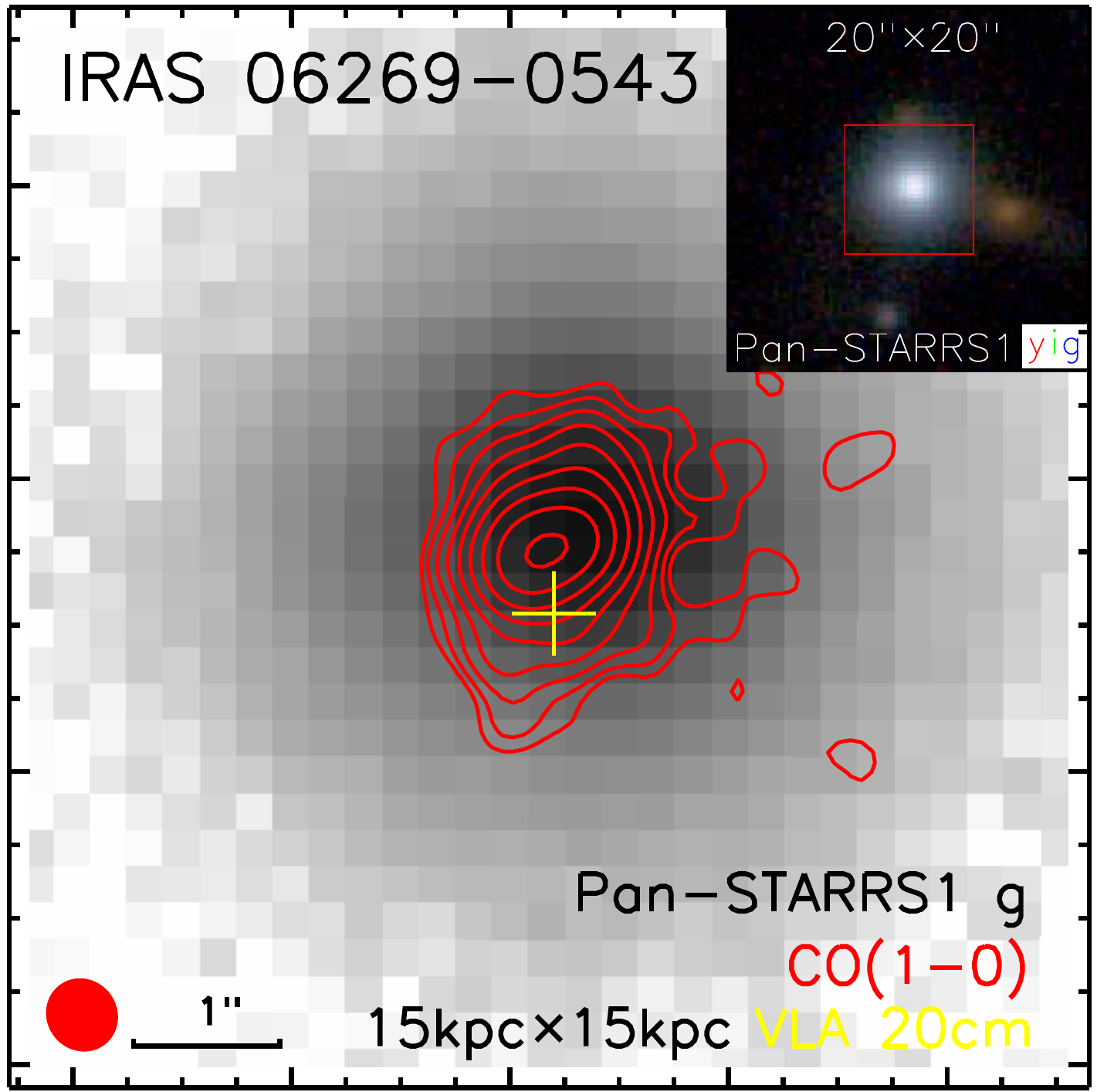}
\hspace{-6pt}
\includegraphics[width=0.25\linewidth]{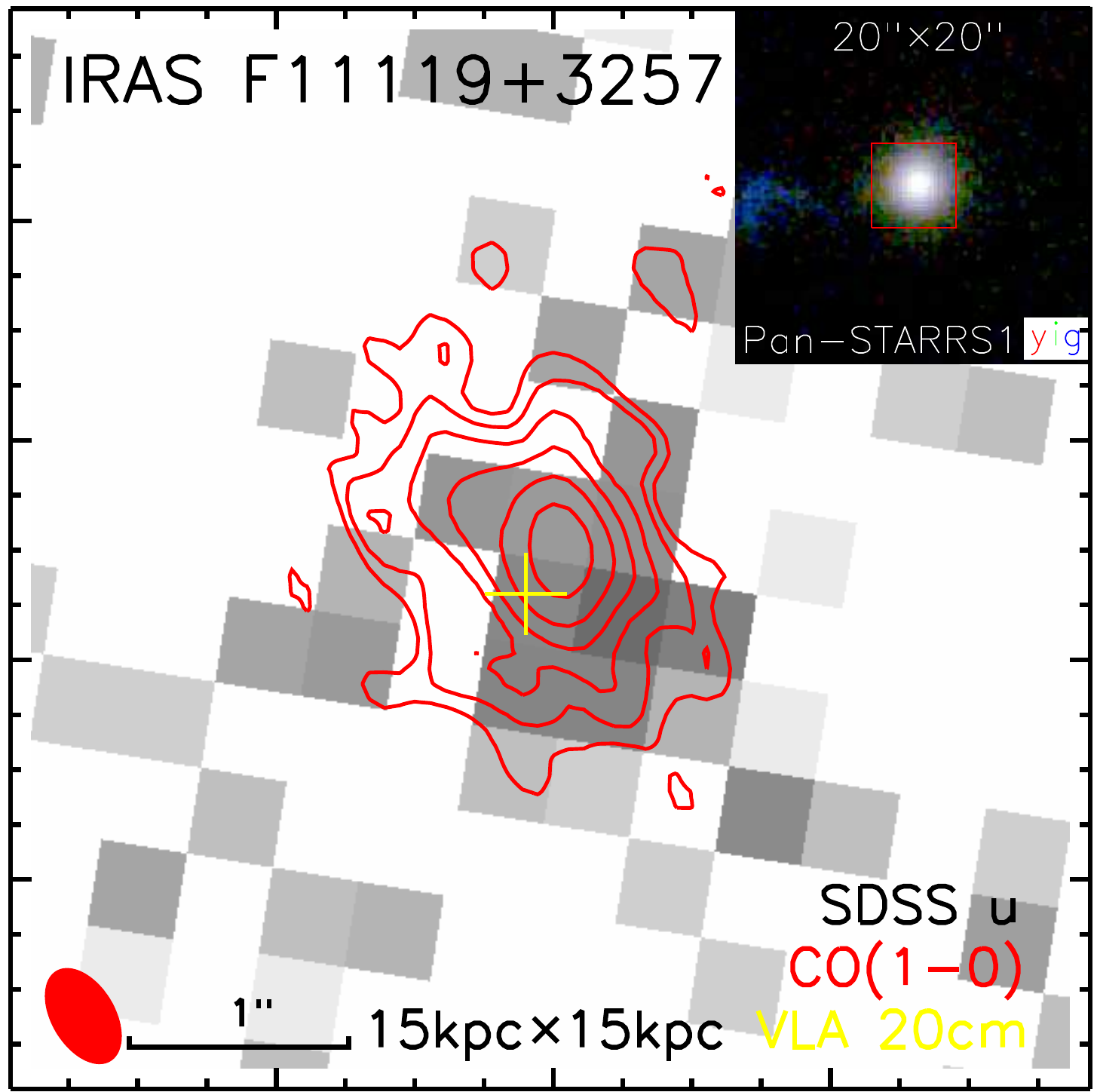}\\
%\vspace{-3pt}
\includegraphics[width=0.25\linewidth]{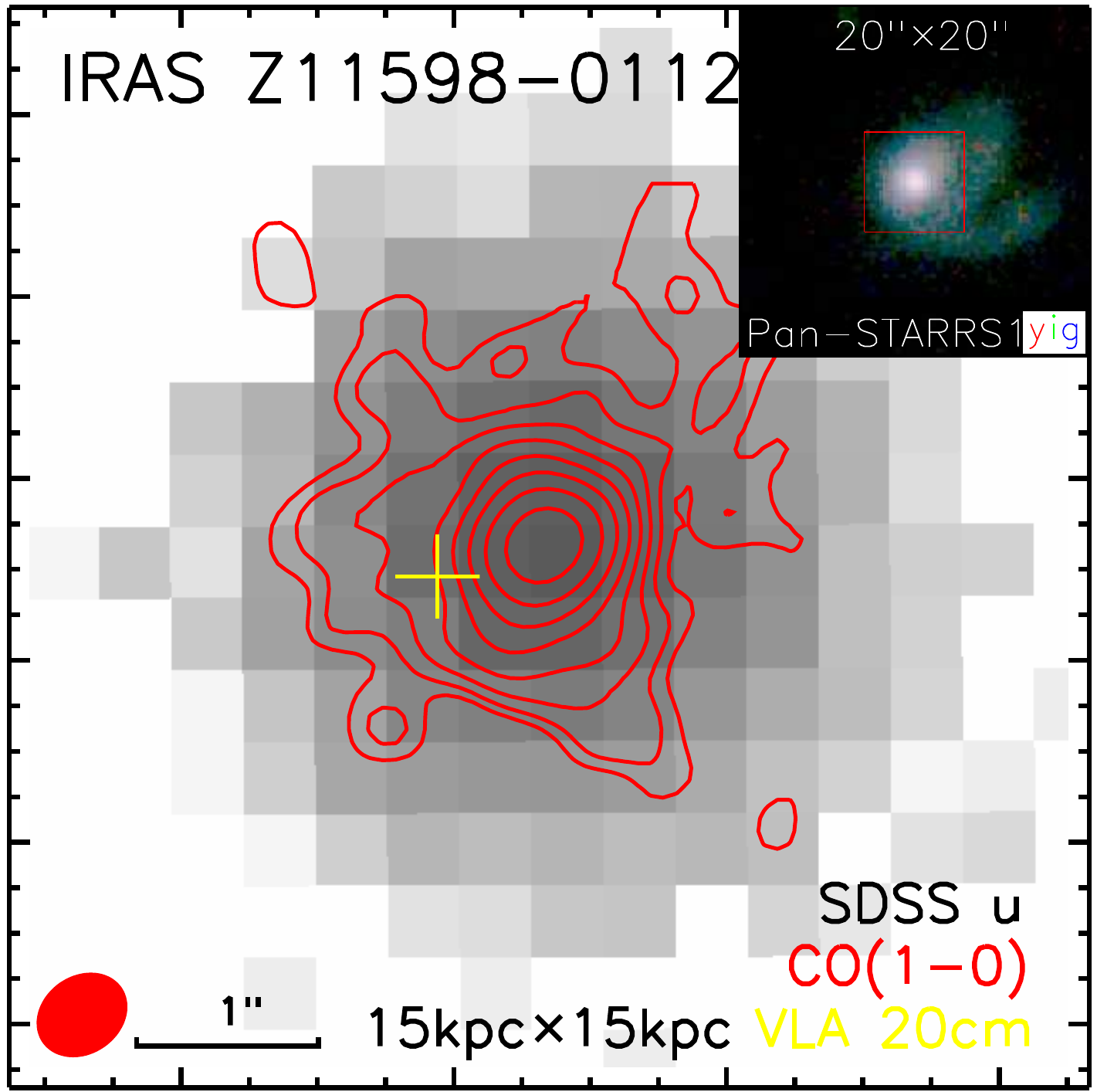}
\hspace{-6pt}
\includegraphics[width=0.25\linewidth]{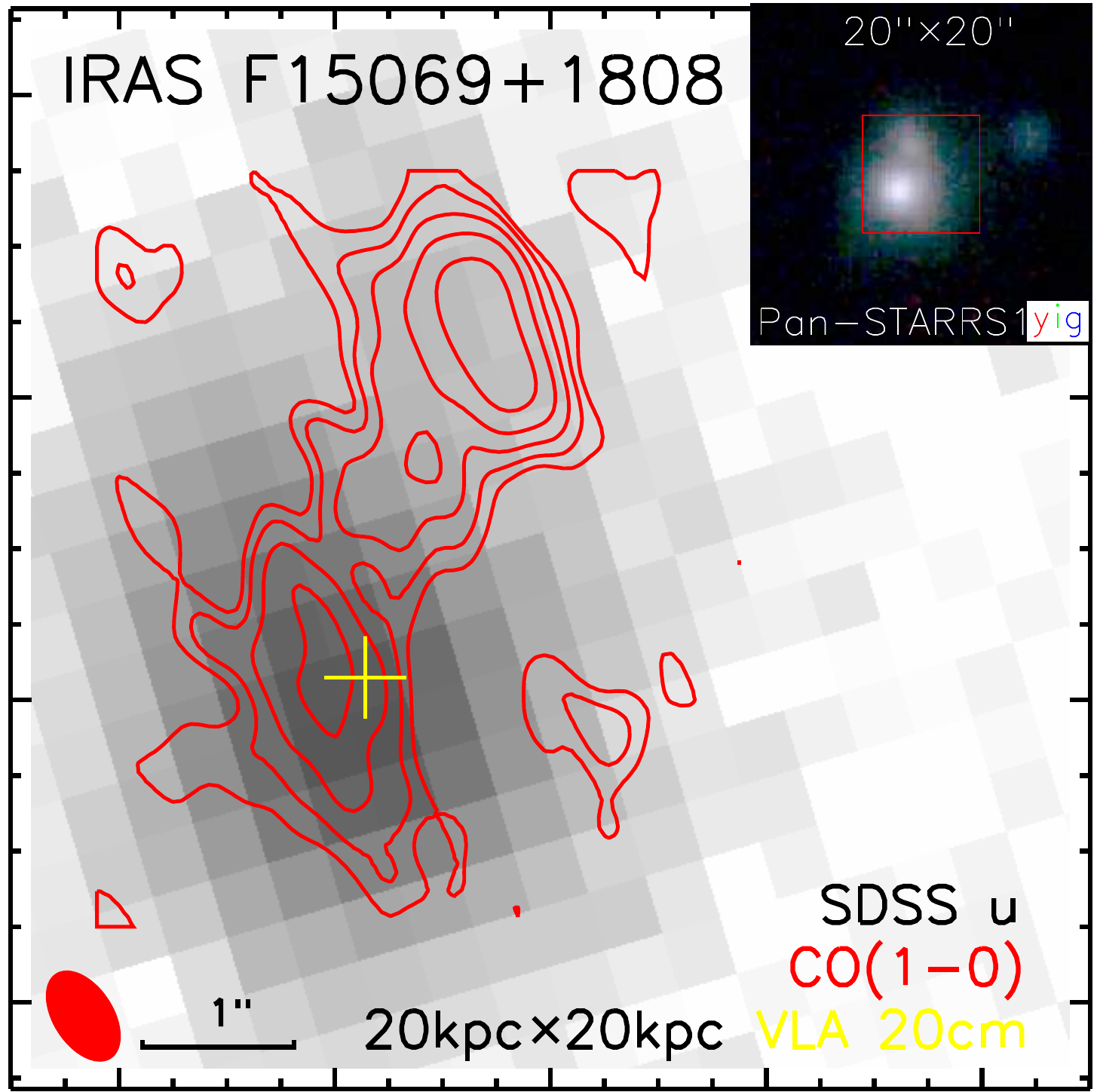}
\hspace{-6pt}
\includegraphics[width=0.25\linewidth]{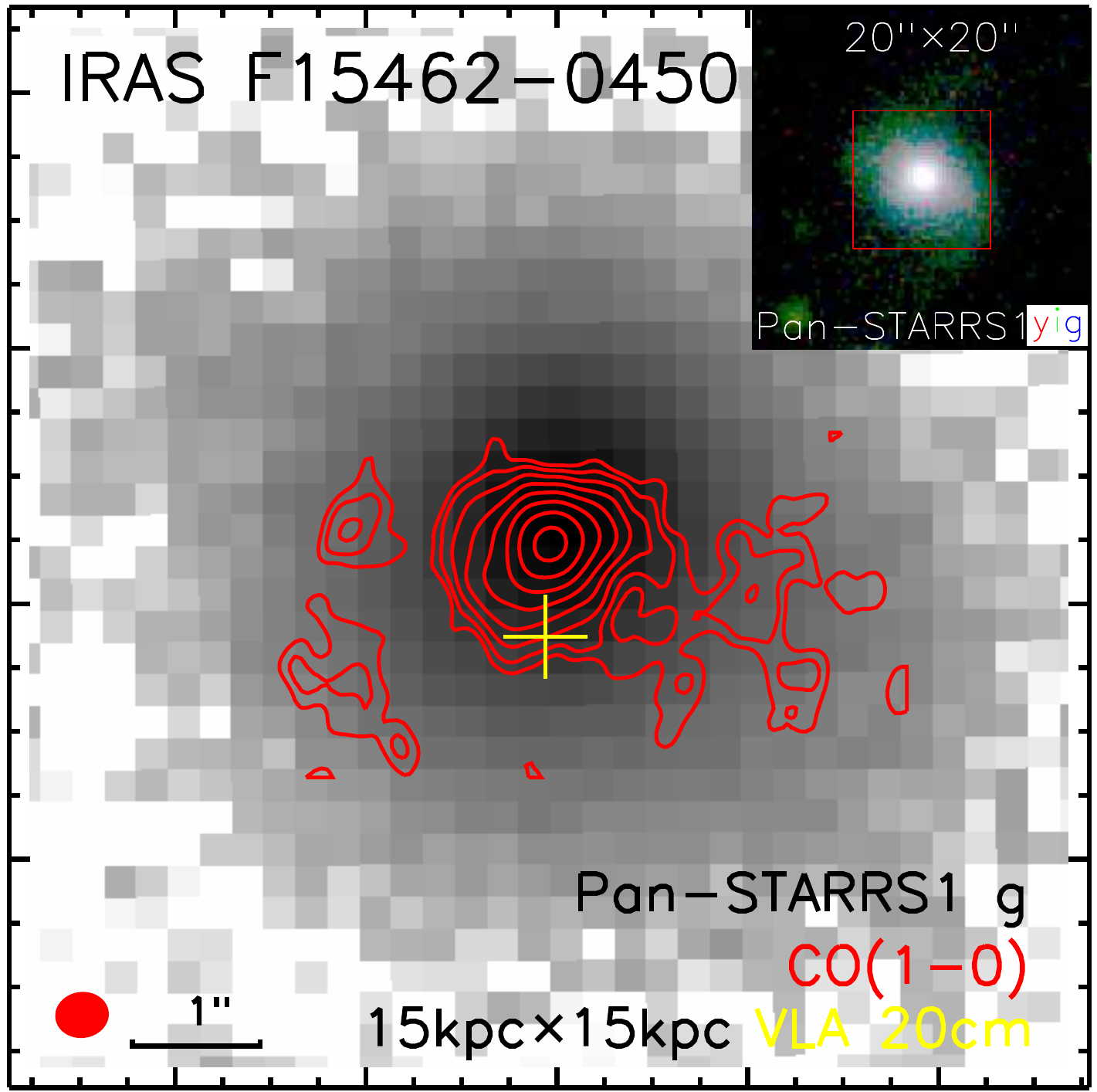}\\
%\vspace{-3pt}
\includegraphics[width=0.25\linewidth]{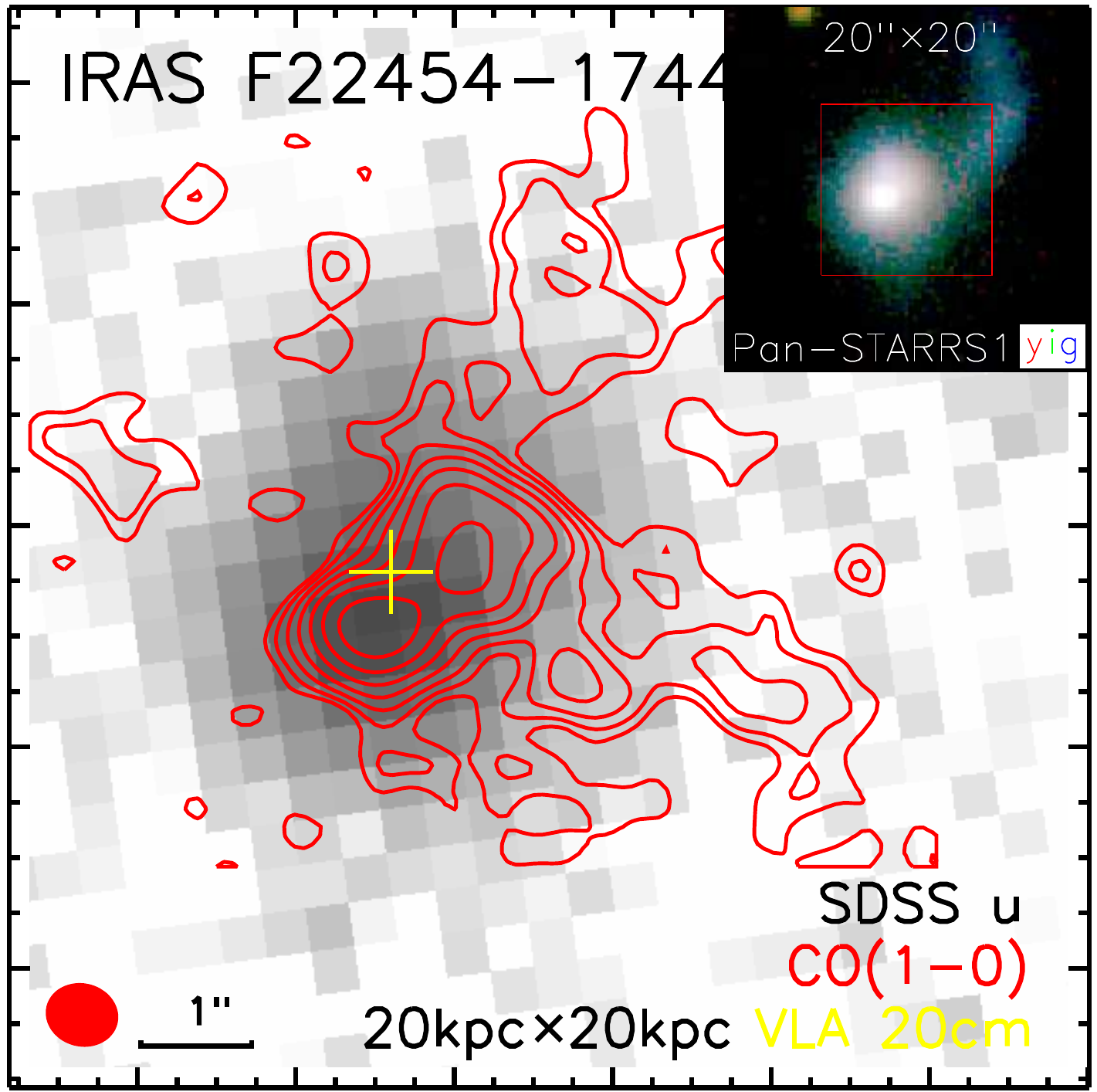}
\hspace{-6pt}
\includegraphics[width=0.25\linewidth]{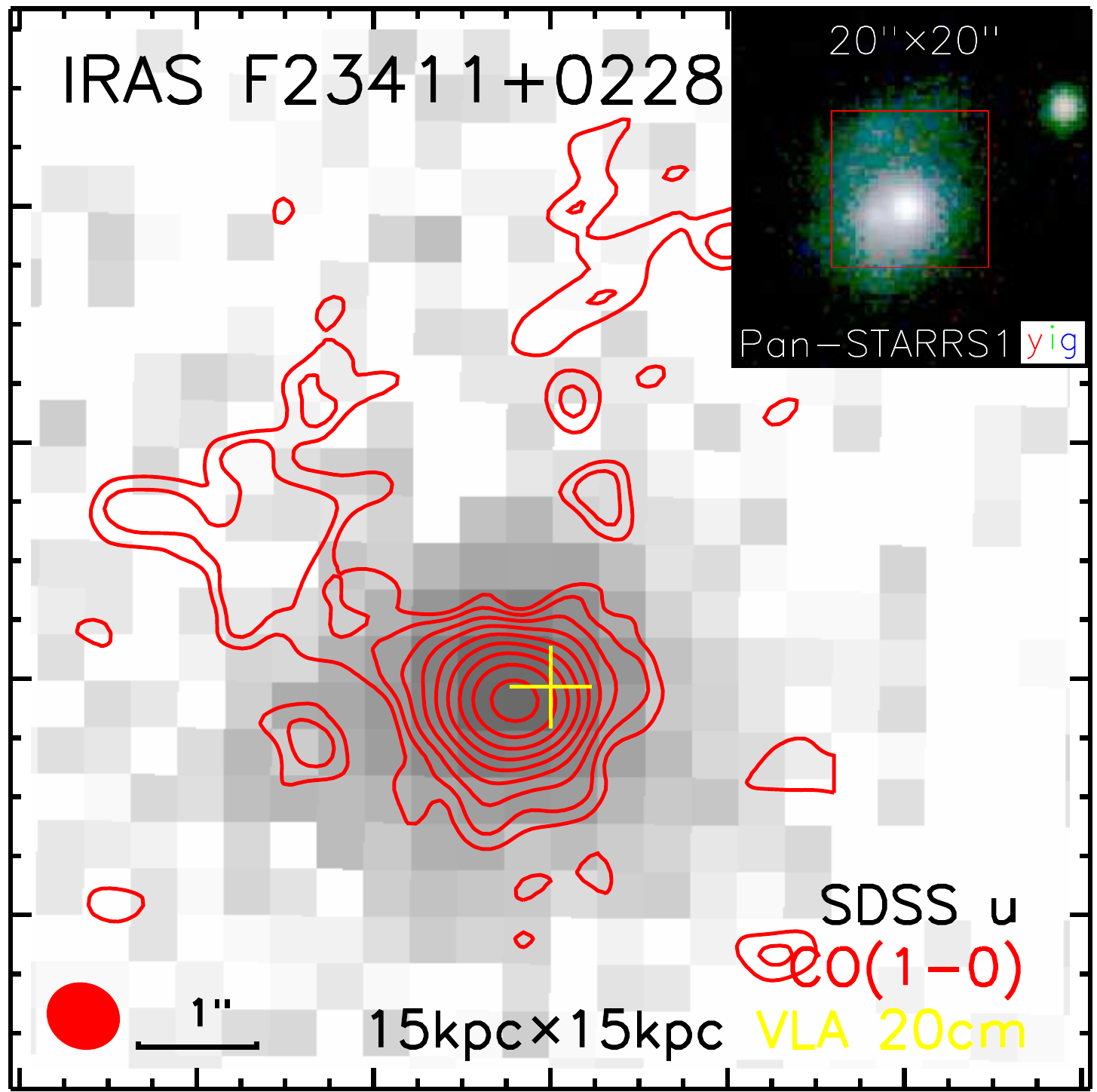}
\hspace{-6pt}
\includegraphics[width=0.251\linewidth]{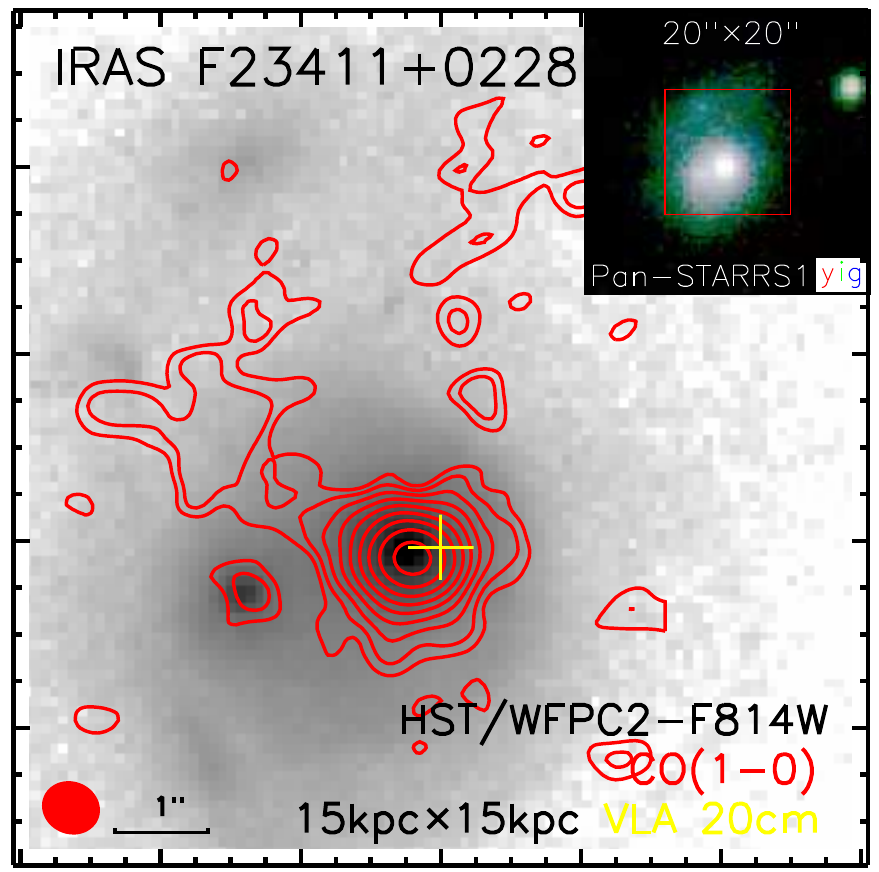}\\
\caption{ALMA CO line contours (red) overlaid on SDSS $u$-band (or Pan-STARRS1 $g$-band if SDSS data are not available) image. For IRAS F23411+0228, we also show the CO(1$-$0) contours overlaid on the HST/WFPC2-F814W image for comparison. The physical scale of each image corresponds to 15~kpc $\times$ 15~kpc (except the two interacting systems, IRAS F15069+1808 and IRAS F22454$-$1744, of which the side length in the image is 20 kpc). Yellow crosses mark the locations of the 20~cm continuum emission observed with the VLA. The CO beam size is shown in the bottom-left corner of each image. Inset panels: False-color Pan-STARRS1 (RGB channels corresponding to $yig$ filters) image with size of $20\arcsec\times 20\arcsec$ (except I~ZW~1, $30\arcsec \times 30\arcsec$). The red box represents the zoomed-in region shown in the background panel. \label{fig:image}}
\end{figure*}

Figure~\ref{fig:image}\footnote{These plots were produced using the software AICER (\url{https://github.com/shbzhang/aicer/}), which is an IDL program for compounding astronomical images.} shows an overlay of the CO(1$-$0) emission and the SDSS $u$-band (or Pan-STARRS1 (PS1) $g$-band if SDSS data are not available) image for the eight IR QSO hosts. The peak of 20~cm radio continuum emission measured from the VLA FIRST survey\footnote{\url{http://sundog.stsci.edu/}} or the VLA 1.4 GHz survey of $ROSAT/IRAS$ galaxy sample \citep{condon98} is marked as a cross symbol in the image. The false-color PS1 $yig$ image is shown in an inset panel. The tidal tails and irregular morphologies shown in the optical images seem to indicate signs of recent and/or ongoing dynamical interactions or mergers in these galaxies. 

%%%%%%%%%%%%%%%- Fig-3 -%%%%%%%%%%%%%%%%

%%%%%%%%%%%%%%%%%%%%%%%%%%%%%%%%%%%%%%%%

Generally, the CO emission peaks roughly coincide with the brightest peaks in the optical images and are consistent with 20~cm continuum peaks within the positional error ($\sim$ 0.\arcsec5 for VLA 20~cm continuum observations). For IRAS~F15069+1808, the 20~cm continuum position lies precisely on the nucleus of F15069+1808SE in CO emission, which also coincides with the UV emission peak. This suggests that the AGN is more likely embedded in the nucleus of F15069+1808SE, while the NW source could be a companion galaxy undergoing an interaction with the QSO. The 20~cm continuum emission peak of IRAS~F22454$-$1744 lies between the two central brightest CO sources but closer to the SE source. Combining with the fact that the UV emission peak coincides with the nucleus of the SE source, we speculate that F22454$-$1744SE is probably the QSO merging with a companion galaxy to the NW which is about to coalesce. The SW molecular tail of IRAS~F22454$-$1744 clearly follows the tidal arm seen in the optical emission of PS1 image, whereas the NW arm has no counterpart in optical wavelength. For IRAS~F23411+0228, the 20~cm continuum emission peaks in the central compact source. In addition, the UV emission is found to peak in the nucleus of central source, indicating that the AGN is likely located in F23411+0228C. It is clear that the NE arm-like structure of CO emission follows the arm seen in the optical image. Higher resolution $HST$/WFPC2 observations in F814W-band identify a bright peak coincident with F23411+0228C, but also a faint peak coincident with the location of the CO peak in F23411+0228E (see the bottom-right panel in Figure~\ref{fig:image}). However, higher resolution observations (e.g., $Chandra$, $JVLA$) of X-ray and radio emission are required to determine the accurate position of the AGN for these merging systems.

\subsection{CO Line Kinematics}\label{subsec:kinematics}

The CO line intensity-weighted velocity fields and velocity dispersion of the eight IR QSO hosts are shown in Figure~\ref{fig:moment}. These were made by smoothing the velocity over the emission region and clipping the intensity using the CASA task, \texttt{immoments}. The cleaned image cube was smoothed in velocity and was clipped at the 3$\sigma$ level per channel. The zero velocity in the velocity maps corresponds to the CO redshifts listed in Table~\ref{tab:meas}. The kinematics of the molecular gas in the eight IR QSO hosts are also quite varied. The maps of the velocity fields reveal smooth and regular rotation velocity for about half of the galaxies in our sample, while the rest are significantly distorted. Maps of the velocity dispersion show increased velocity dispersion in the centers of almost all CO-emitting systems with peak velocity dispersion of about 60$-$130 \kms.

For the five of IR QSOs that are resolved into an isolated compact CO source, four of the host galaxies (I~ZW~1, IRAS 06269$-$0543, IRAS F11119+3257, and IRAS~F15426$-$0450) show velocity gradients in the velocity maps. The velocity dispersion peaks in these galaxies roughly correspond to the peaks of CO emission, except for IRAS~06269$-$0543 where there is a shift of the peak between the CO and 3-mm continuum emission. The velocity dispersion map of IRAS 06269$-$0543 shows an increase in the northwestern direction with respect to the CO peak, with a peak dispersion of $\sim$ 100 \kms\ which is spatially coincident with the location of 3-mm continuum peak. Given the limited spatial resolution of current data, we are not able to infer whether the 3-mm continuum emission is related to AGN activity, although the velocity dispersion is usually found to increase at the AGN location \citep[e.g.,][]{trakhtenbrot17}. The complex velocity distribution revealed in IRAS~Z11598$-$0112 show evidence that the molecular gas is strongly disturbed and has not settled in the galactic plane. For I~ZW~1, the velocity gradient is obvious in the southeast to northwest direction, from about +190 to $-$190 \kms. This is consistent with the velocity map presented in \citet{schinnerer98}. A small velocity gradient is present in the east-west direction of IRAS F11119+3257, while the deviation from regular velocity field seen in the northeast region may be indicative of noncircular motion. The CO velocity map of IRAS F15462$-$0450 shows a clear gradient from the northwest to the southeast in the main body of galaxy, while the CO line of the NE clump is too narrow to show a gradient at our spectral resolution, and the SW clump shows a small velocity gradient along the same direction as that of main body. The SW clump shows a slightly higher velocity dispersion ($\sim$ 85 \kms) than that ($\sim$ 65 \kms) of the peak in main galaxy. 

The CO kinematics for the remaining three galaxies that are resolved into multiple objects are complex. For the merging system IRAS F15069+1808, the SE source (identified as the possible galaxy hosting luminous AGN) shows a rotating gas disk with velocity gradient in the southeast-northwest direction, whereas the molecular gas in the NW galaxy is kinematically disturbed. The overlap region has the highest velocity dispersion ($\sim$ 130 \kms) in CO(1$-$0) emission, possibly marking the new dynamical center of the merger. This is similar to what is seen in IR-bright merger VV114, where the large velocity dispersion detected in the overlap region was interpreted as a turbulence-dominated shock region induced by the interaction between the two colliding galaxies \citep[e.g.,][]{saito15}.

The velocity field in merging system IRAS F22454$-$1744 shows a velocity range of $\sim230$ \kms\ across the galaxy disks with the largest redshifted velocity in the molecular tail SW1 clump (about +130 \kms\ with respect to the systemic velocity of merging system). A regular rotating disk with velocity gradient (ranging from about $-$100 \kms\ to +30 \kms) in the southwest-northeast direction is clearly seen in the NW galaxy, while the velocity is significantly distorted in the SE source which is likely to host the QSO and in the extended CO emission regions. The velocity dispersion map in Figure~\ref{fig:moment} shows a peak velocity dispersion ($\sim$ 80 \kms) in the SE nucleus followed by a large dispersion in the center of NW source and a quiescent structure along the SW filament.

%%%%%%%%%%%%%%- Table-2 - %%%%%%%%%%%%%%

\begin{longrotatetable}
\begin{deluxetable*}{llrrCCCCch}
\tablenum{2}
%\centering
\tablewidth{700pt}
\tabletypesize{\scriptsize}
%\addtolength{\tabcolsep}{-1.pt}
\tablecaption{Summary of the ALMA CO(1-0) and 3-mm continuum measurements}\label{tab:meas}
%\tablewidth{0pt}
\tablehead{
\colhead{Source} & \colhead{Component} & \colhead{RA} & \colhead{DEC} & \colhead{$z_{\rm CO}$} & \colhead{Source Size FWHM} & \colhead{PA} & \colhead{FWHM$_{\rm CO(1-0)}$} & \colhead{Integrated Flux} & \nocolhead{Peak Flux} \\
% &  &  &   &  & \colhead{+FWHM (Circular Gaussian model)} & & & \\
\colhead{} & \colhead{} & \colhead{(J2000)} & \colhead{(J2000)} & \colhead{} & \colhead{(arcsec)} & \colhead{(deg)} & \colhead{(km s$^{-1}$)} & \colhead{} & \nocolhead{(mJy)} \\
\colhead{(1)} &  \colhead{(2)} & \colhead{(3)} & \colhead{(4)} & \colhead{(5)} & \colhead{(6)} & \colhead{(7)} & \colhead{(8)} & \colhead{(9)} & \nocolhead{(10)} 
}
%\colnumbers
\startdata
I~ZW~1 & CO(1$-$0) & 00:53:34.937$\pm$0.004 & 12:41:35.79$\pm$0.08 & 0.06115\pm 0.00003  & (2.75\pm 0.21)\times(2.17\pm 0.29) & 136\pm 14 & 370\pm 16 & 31.97$\pm$1.49 Jy km s$^{-1}$ & \\
 &  &  &   &  & 0.32\pm 0.07 & & & \\
\vspace{0.3cm}
      & Continuum & 00:53:34.934$\pm$0.001 & 12:41:35.91$\pm$0.03 & & - & - & & 0.492$\pm$0.078 mJy & \\      
IRAS 06269$-$0543 & CO(1$-$0) & 06:29:24.782$\pm$0.001 & $-$05:45:26.48$\pm$0.01 & 0.11721\pm 0.00002 & (1.08\pm 0.14)\times(0.68\pm 0.12) & 135\pm 6 & 312\pm 14 &  10.64$\pm$0.68 Jy km s$^{-1}$ & \\
 &  &  &   &  & 0.37\pm 0.09 & & & \\
\vspace{0.3cm}
                & Continuum & 06:29:24.770$\pm$0.001 & $-$05:45:26.42$\pm$0.01 & & 0.27\pm 0.03 & - & & 1.496$\pm$0.063 mJy & \\               
IRAS F11119+3257 & CO(1$-$0) & 11:14:38.901$\pm$0.003 & 32:41:33.51$\pm$0.04 & 0.18988\pm 0.00002 & (1.53\pm 0.24)\times(1.02\pm 0.25) & 10\pm 13 & 226\pm 11 & 6.45$\pm$0.51 Jy km s$^{-1}$ & \\
 &  &  &   &  & 0.32\pm 0.05 & & & \\   
\vspace{0.3cm}           
                 & Continuum & 11:14:38.901$\pm$0.001 & 32:41:33.45$\pm$0.02 &  & 0.31\pm 0.05 & - &  & 0.697$\pm$0.088 mJy & \\
IRAS Z11598$-$0112 & CO(1$-$0) & 12:02:26.768$\pm$0.001 & $-$01:29:15.39$\pm$0.02 & 0.15118\pm 0.00001 & (1.77\pm 0.22)\times(1.41\pm 0.31) & 20\pm 19 & 147\pm 5 &  10.93$\pm$0.62 Jy km s$^{-1}$ & \\
 &  &  &   &  & 0.29\pm 0.04 & & & \\ 
\vspace{0.3cm}
                  & Continuum & 12:02:26.766$\pm$0.001 & $-$01:29:15.35$\pm$0.01 &  & 0.14\pm 0.13 & - &  & 0.556$\pm$0.054 mJy & \\
\vspace{0.3cm}
IRAS F15069+1808NW & CO(1$-$0) & 15:09:13.735$\pm$0.002 & 17:57:12.38$\pm$0.05 & 0.17068\pm 0.00002 & (1.14\pm 0.13)\times(0.77\pm 0.15) & 9\pm 11 & 193\pm 12 & 3.93$\pm$0.22 Jy km s$^{-1}$ & \\
\vspace{0.3cm}
IRAS F15069+1808SE & CO(1$-$0) & 15:09:13.805$\pm$0.004 & 17:57:10.11$\pm$0.11 & 0.17093\pm 0.00006 & (2.41\pm 0.36)\times(1.06\pm 0.27) & 177\pm 7 & 285\pm 30  & 3.07$\pm$0.28 Jy km s$^{-1}$ & \\  
\vspace{0.3cm}          
IRAS F15069+1808 Overlap & CO(1$-$0) & 15:09:13.758$\pm$0.002 & 17:57:11.60$\pm$0.03 & 0.17050\pm 0.00003  & - & - & 142\pm 16 & 1.47$\pm$0.16 Jy km s$^{-1}$ &  \\
IRAS F15462$-$0450C & CO(1$-$0) & 15:48:56.804$\pm$0.001 & $-$04:59:33.56$\pm$0.02 & 0.10039\pm 0.00001 & (1.10\pm 0.16)\times(1.00\pm 0.08) & 45\pm 14 & 196\pm 7 & 7.73$\pm$0.22 Jy km s$^{-1}$ & \\  
 &  &  &   &  & 0.21\pm 0.04 & & & \\ 
\vspace{0.3cm}         
                 & Continuum & 15:48:56.804$\pm$0.001 & $-$04:59:33.53$\pm$0.02 &  & 0.23\pm 0.06 & - &  & 0.645$\pm$0.066 mJy &  \\
\vspace{0.3cm}
IRAS F15462$-$0450NE & CO(1$-$0) & 15:48:56.906$\pm$0.002 & $-$04:59:33.40$\pm$0.04 & 0.10015\pm 0.00002 & - & - & 72\pm 13 & 0.33$\pm$ 0.06 Jy km s$^{-1}$  &   \\  
\vspace{0.3cm} 
IRAS F15462$-$0450SW & CO(1$-$0) & 15:48:56.697$\pm$0.002 & $-$04:59:33.94$\pm$0.05 & 0.10057\pm 0.00007 & - & - & 243\pm 42 & 0.64$\pm$ 0.09 Jy km s$^{-1}$ &  \\    
\vspace{0.3cm}        
IRAS F22454$-$1744NW & CO(1$-$0) & 22:48:04.298$\pm$0.003 & $-$17:28:30.41$\pm$0.05 & 0.11781\pm 0.00001 & (1.38\pm 0.10)\times(1.18\pm 0.19) & 176\pm 17 & 160\pm 6 & 4.65$\pm$0.13 Jy km s$^{-1}$ & \\
IRAS F22454$-$1744SE & CO(1$-$0) & 22:48:04.342$\pm$0.002 & $-$17:28:30.87$\pm$0.02 & 0.11787\pm 0.00002 & (0.55\pm 0.11)\times(0.47\pm 0.07) & 114\pm 14 & 199\pm 9 & 3.67$\pm$0.15 Jy km s$^{-1}$ & \\
\vspace{0.3cm}
                      & Continuum & 22:48:04.346$\pm$0.004 & $-$17:28:30.99$\pm$0.06 &  & - & - &  & 0.219$\pm$0.047 mJy &  \\
\vspace{0.3cm}
IRAS F22454$-$1744SW1 & CO(1$-$0) & 22:48:04.224$\pm$0.003 & $-$17:28:31.35$\pm$0.03 & 0.11813\pm 0.00005 & - & - & 221\pm 26 & 1.31$\pm$0.12 Jy km s$^{-1}$ &  \\
\vspace{0.3cm}
IRAS F22454$-$1744SW2 & CO(1$-$0) & 22:48:04.113$\pm$0.008 & $-$17:28:31.43$\pm$0.07 & 0.11787$\pm$0.00009 & - & - & 328\pm 51 & 1.13$\pm$0.14 Jy km s$^{-1}$ &  \\
\vspace{0.3cm}   
IRAS F22454$-$1744N & CO(1$-$0) & 22:48:04.227$\pm$0.008 & $-$17:28:27.91$\pm$0.18 & 0.11794\pm 0.00008 & - & - & 293\pm 45 & 1.36$\pm$0.16 Jy km s$^{-1}$ &   \\   
IRAS F23411+0228C & CO(1$-$0) & 23:43:39.721$\pm$0.001 & 02:45:03.81$\pm$0.01 & 0.09171\pm 0.00001 & (1.43\pm 0.36)\times(1.26\pm 0.19) & 174\pm 34 & 185\pm 6  & 4.43$\pm$0.15 Jy km s$^{-1}$ &  \\
 &  &  &   &  & 0.22\pm 0.04 & & & \\
\vspace{0.3cm}
                 & Continuum & 23:43:39.719$\pm$0.001 & 02:45:03.86$\pm$0.02 &  & 0.22\pm 0.12 & - &  & 0.433$\pm$0.055 mJy &  \\   
IRAS F23411+0228E & CO(1$-$0) & 23:43:39.837$\pm$0.003 & 02:45:03.45$\pm$0.04 & 0.09173\pm 0.00008 & - & - & 151\pm 47 & 0.12$\pm$0.03 Jy km s$^{-1}$ &  \\
\vspace{0.3cm}
                  & Continuum & 23:43:39.853$\pm$0.023 & 02:45:03.28$\pm$0.14 &  & - & - &  & 0.086$\pm$0.030 mJy & \\
IRAS F23411+0228NE & CO(1$-$0) & 23:43:39.874$\pm$0.017 & 02:45:05.21$\pm$0.29 & 0.09186\pm 0.00004 & - & - & 174\pm 23 & 0.95$\pm$0.10 Jy km s$^{-1}$ &  \\                                                                      
\enddata
\tablecomments{Column (1): galaxy name. Column (2): CO(1-0) emission line and 3-mm continuum. Column (3) and (4): source position derived from a 2D Gaussian fit to the integrated intensity images. Column (5): CO redshift derived from a Gaussian profile fit to the spectrum that extracted from the region with $\geqslant$ 2$\sigma$ signals in the integrated intensity image. Column (6) and (7): FWHM source size and position angle derived from model fitting to the visibilities in the $uv$-plane using the \texttt{UVMULTIFIT} tool \citep{martividal14}. For the CO(1$-$0) line-emitting gas, the source structure is best fitted with a combination of an elliptical Gaussian (extended component) and a circular Gaussian (compact component) profiles, while for the continuum, the source size is inferred from a single circular Gaussian model fit to the visibilities. For the galaxies that resolved into two separate galaxies with comparable gas mass, IRAS~F15069+1808 and IRAS~F22454$-$1744, the line data are best fitted with two elliptical Gaussian models with positions centered on each separate galaxy (see Section~\ref{subsec:size} for details). Column (8): FWHM CO line width derived from a Gaussian profile fit to the spectrum (except for I~ZW~1 and IRAS 06269$-$0543, where the CO spectrum is fitted using a double-Gaussian model). Column (9): CO(1-0) velocity-integrated intensity and 3-mm continuum emission flux. The CO integrated fluxes are measured from the spectra extracted from a circular aperture with diameter shown in Figure~\ref{fig:spec} by integrating the intensity over the line-emitting region in each channel, except for the galaxies spatially resolved into multiple components, IRAS~F15069+1808, IRAS~F15462$-$0450, IRAS~F22454$-$1744, and IRAS~F23411+0228, for which the integrated flux derived for each component is measured from the spectrum extracted from the $\geqslant 2 \sigma$ region in the CO intensity map (see the spectra in Figure~\ref{fig:resolved-spectra}). The continuum fluxes are measured from a 2D Gaussian fit to the images integrated with line-free windows. For the sources that were not resolved in the 3-mm continuum emission (I~ZW~1, IRAS~Z11598$-$0112, IRAS~F22454$-$1744SE, andIRAS~F23411+0228E), the total 3-mm continuum flux corresponds to the peak of the 3-mm continuum. Uncertainties in fluxes quoted in this table are statistical and do not include the absolute flux calibration uncertainty.}
%\tablenotetext{}{}
\end{deluxetable*}
\end{longrotatetable}
%%%%%%%%%%%%%%%%%%%%%%%%%%%%%%%%%%%%%%%

%%%%%%%%%%%%%%%- Fig-4 -%%%%%%%%%%%%%%%%
\begin{figure*}[htbp]
\centering
\includegraphics[width=0.24\linewidth]{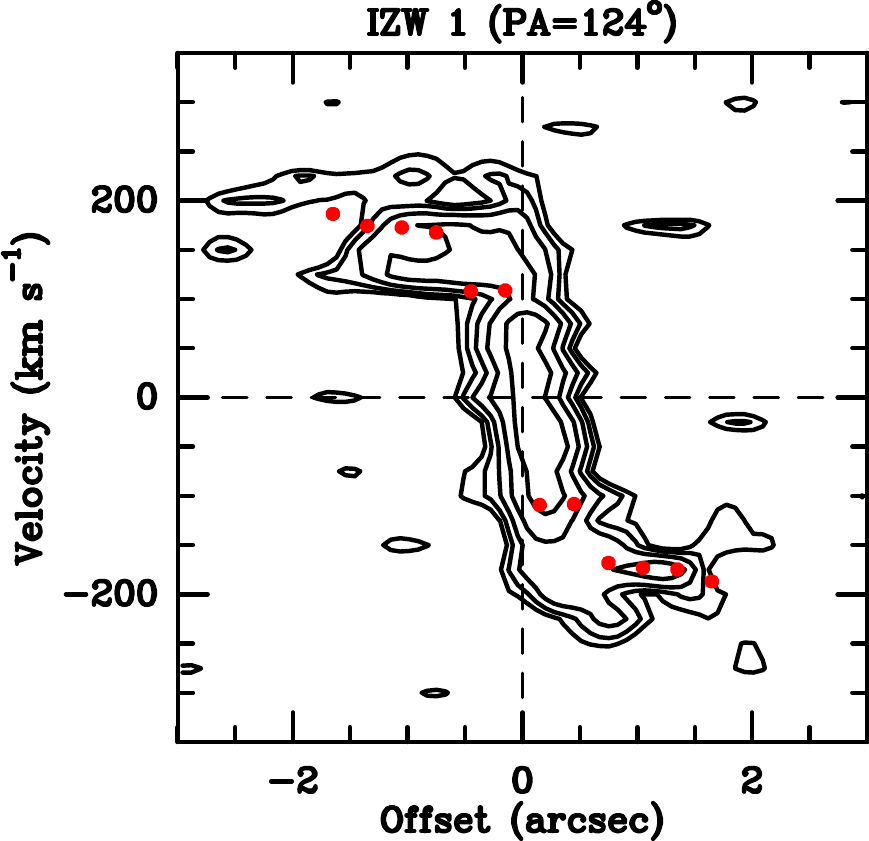}
\includegraphics[width=0.24\linewidth]{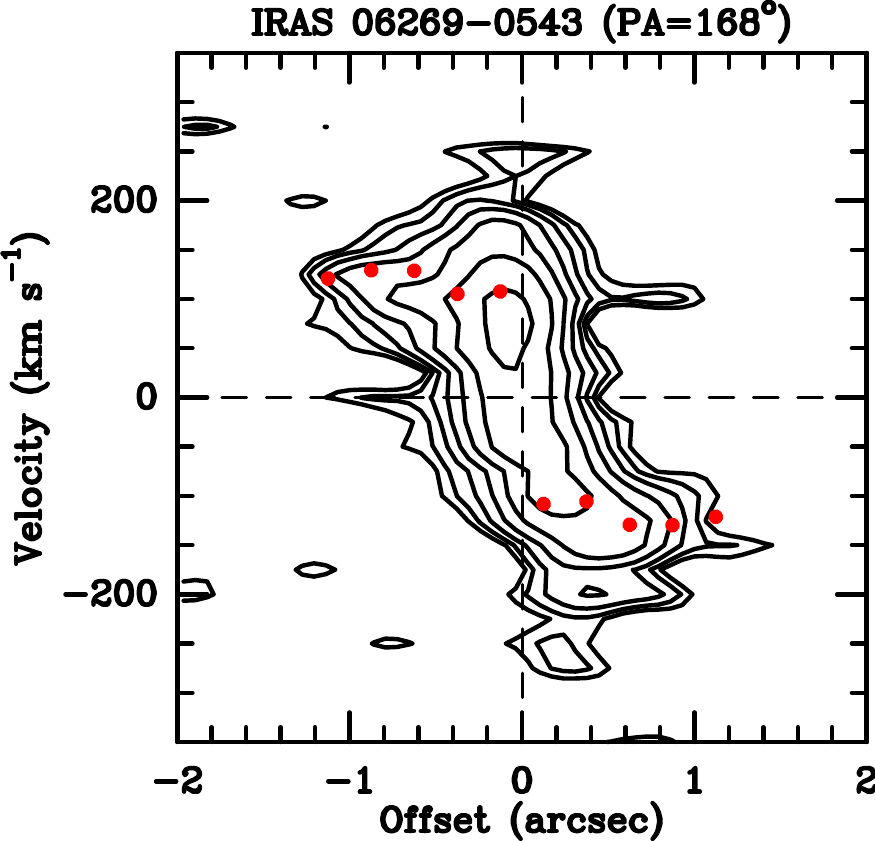}
\includegraphics[width=0.24\linewidth]{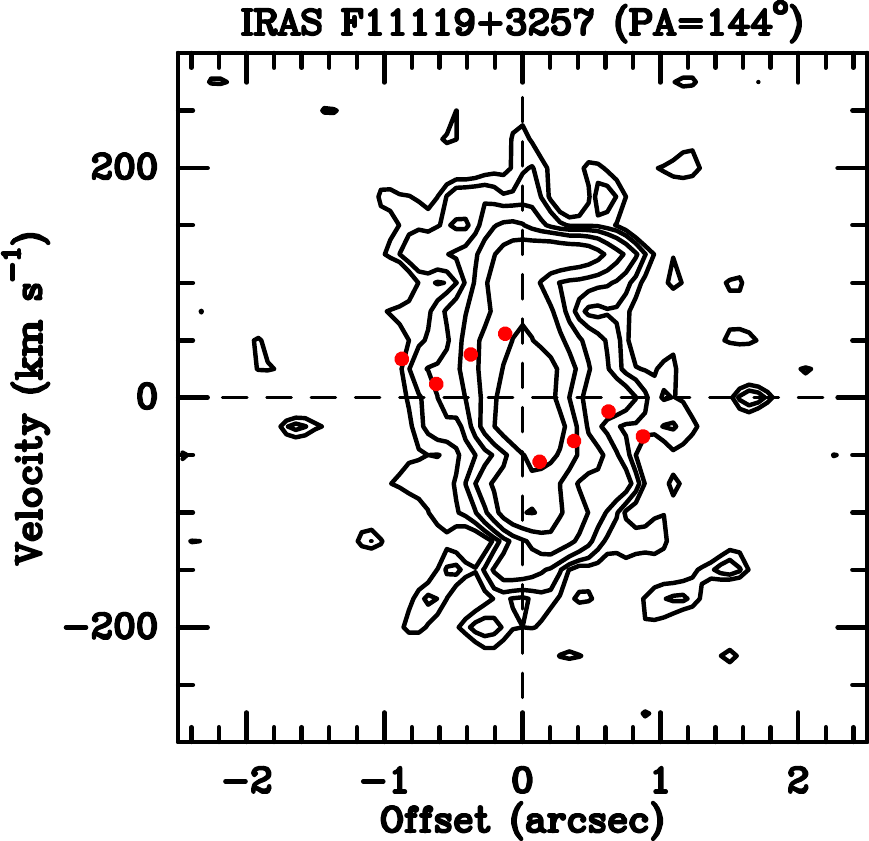}
\includegraphics[width=0.24\linewidth]{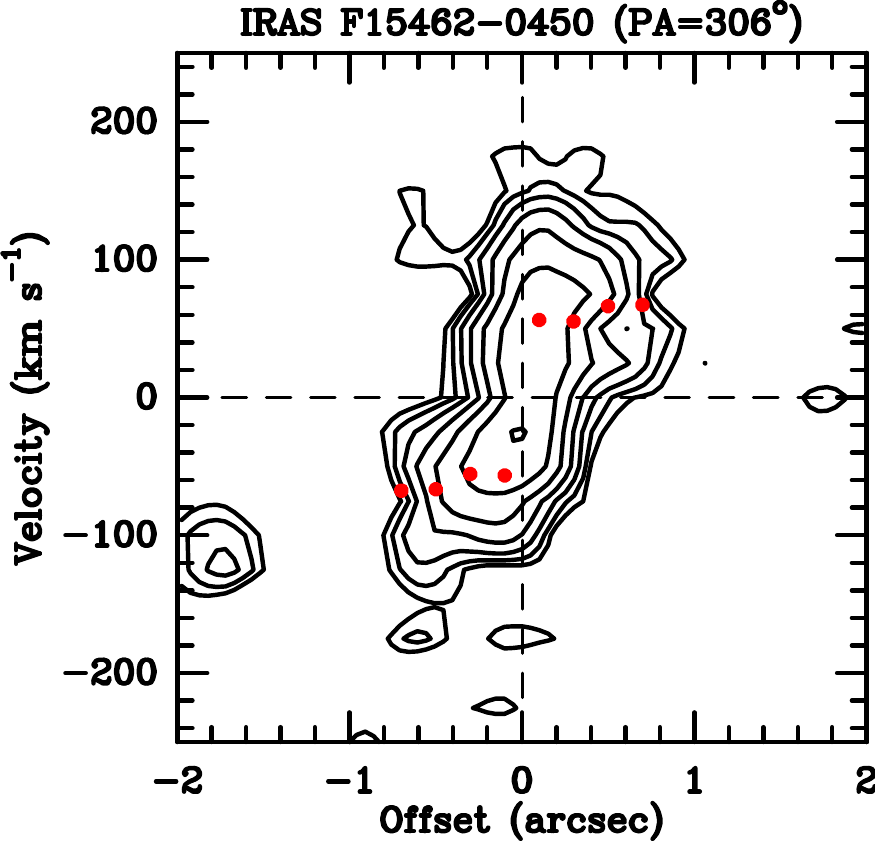}
\caption{Position-Velocity diagrams of CO(1$-$0) emission along the line of rotation axis through the peak CO position for the four galaxies showing velocity gradient in CO emission. Contours start at 2$\sigma$ and increase by factors of 1.5 where $\sigma$ in the spectral velocity resolution of 25 \kms\ for the four galaxies are: 0.90 mJy beam$^{-1}$ (I~ZW~1), 0.46 mJy beam$^{-1}$ (IRAS~06269$-$0543), 0.30 mJy beam$^{-1}$ (IRAS~F11119+3257), and 0.47 mJy beam$^{-1}$ (IRAS~F15462-0450). For IRAS~F11119+3257, we combined our CO data with the archival lower angular resolution ALMA CO data to produce the PV diagram (see text in Sect.~\ref{subsec:kinematics}). The plotted velocity is along the line of sight. The zero offset position corresponds to the peak CO position in all PV diagrams. The red circles denote the best-fit projected rotation velocity derived from $^{3\rm{D}}$BAROLO. \label{fig:pv} }
\end{figure*}
%%%%%%%%%%%%%%%%%%%%%%%%%%%%%%%%%%%%%%%%

For IRAS F23411+0228, the central compact source F23411+0228C and the NE arm show complex and distorted velocity fields, whereas the source E tends to show a small gradient from north to  south. The NE arm has a redshifted velocity of $\sim$ 50$-$150 \kms\ with respect to the main galaxy. The peak of velocity dispersion for F23411+0228C is about 70 \kms , while a relatively low value ($\sim$ 10 \kms) is found in both NE arm and source E. Deeper observations with higher resolution are needed to accurately constrain the kinematics and morphology of the extended CO structures in these galaxies.

Combining the analysis of CO morphology and gas kinematics, we find that the molecular gas in four (I~ZW~1, IRAS 06269$-$0543, IRAS F11119+3257, and IRAS F15462$-$0450) out of the eight IR QSO hosts show signatures of rotating gas disks with ordered velocity field, one (IRAS Z11598$-$0112) is compact CO source with disturbed velocity, and the rest three (IRAS F15069+1808, IRAS F22454$-$1744, and IRAS F23411+0228) resolved into distinct objects showing evidence for strong interactions and merging. 

Figure~\ref{fig:pv} shows position-velocity (PV) diagrams of the CO(1$-$0) emission line for the four isolated QSO host galaxies (I~ZW~1, IRAS~06269$-$0543,IRAS~F11119+3257, and IRAS~F15462$-$0450) with signatures of disk-like rotation in their CO velocity fields, along the velocity gradient through the peak CO position. For IRAS~F11119+3257, we combined our high-resolution CO data with the archival ALMA CO data ($\sim$ 2\arcsec .8 resolution) observed by \citet{veilleux17} to produce the PV diagram. The PV plots show that the rotation curve rises steeply in the inner region, becoming flat at radius $\sim 0\arcsec.5 - 0\arcsec.8$ ($\sim 0.9 - 1.3$ kpc) for I~ZW~1, IRAS~06269$-$0543, and IRAS~F15462$-$0450. However, for IRAS~F11119+3257, the rotation curve is much less prominent. We calculated the inclination angle (between the polar axis of the disk and the line of sight) based on the axis ratio, $i$ = cos$^{-1}$($a_{\rm min}$/$a_{\rm maj}$), where $a_{\rm min}$ and $a_{\rm maj}$ are the semiminor and semimajor axes of the CO emitting regions (assuming a thin-disk geometry; see Sect.~\ref{subsec:size} and Table~\ref{tab:properties}). We then assumed the gas to be in a rotating disk, and studied the kinematic properties by fitting 3D tilted-ring models to emission-line data cubes with $^{3\rm{D}}$BAROLO code \citep{diteodoro15}. The red circles in Figure~\ref{fig:pv} denote the best-fit values for the projected rotation velocity derived from $^{3\rm{D}}$BAROLO. We adopt values of the rotation velocity and velocity dispersion derived at the turnover radius, i.e., the transitional point between the rising and flat parts of the rotation curve, as the intrinsic rotation velocity and local velocity dispersion for each galaxy. Except IRAS~F11119+3257, which is poorly fit by models, the kinematics for the remaining three galaxies can be well-fitted by models. The derived rotation velocity and velocity dispersion range from 157 to 273 \kms\ and 42 to 45 \kms\ (see Table~\ref{tab:pv}), respectively, and the corresponding ratio of the rotation velocity to the velocity dispersion, $V_{\rm rot}/\sigma$, is in the range of 4$-$6 for the three galaxies, somewhat smaller than values of present-day disks \citep{sofue01}.

\subsection{Comparison with Single-Dish CO(1$-$0) Data and the Integrated Intensity}\label{subsec:spect}

We measured the CO fluxes based on the spectra extracted from a circular aperture, of which the CO peak fluxes are consistent with single-dish IRAM 30m measurements (see the spectra in Figure~\ref{fig:spec} and the corresponding apertures used for spectra extraction shown in Figure~\ref{fig:moment}), and found that both the CO integrated fluxes and the line profiles are in good agreement with the IRAM 30m data for the galaxies in our sample.

In Figure~\ref{fig:spec} we also plot the CO spectra extracted from the ALMA cube using a mask that encompasses the $\geqslant 2 \sigma$ region in the velocity-integrated CO(1$-$0) map for comparison. In this work we define a detection if the velocity-integrated line intensity is higher than or equal to 3$\sigma$, but use the 2$\sigma$ cutoff for data analysis as a comparison. It is clear that in most cases, apart from I~ZW~1 and IRAS~06269$-$0543, both the CO peak flux and the line profile of the spectra  extracted from the $\geqslant 2 \sigma$ region agree well with IRAM 30m measurements, suggesting that little emission is filtered out by long baselines and the molecular gas is centrally concentrated.

In the cases of I~ZW~1 and IRAS~06269$-$0543, the spectra extracted from the 2$\sigma$ CO-emitting regions show $\sim$ 30-40\% smaller CO peak flux. The CO spectra of these two galaxies are best described by a double-horned velocity profile, with a separation of 225$\pm$10 \kms\ and 165$\pm$8 \kms\ between red and blue peaks respectively, consistent with the CO spectra measured with the IRAM 30m \citep{evans06,xia12}. In comparison, the 30m spectra (beam size $\approx$ 25\arcsec) have more flux in the horns, indicative of an extended disk component in the host galaxies. For the ALMA data that were extracted from a circular aperture with peak CO fluxes consistent with the single-dish measurements, the diameter of the circular aperture is 10\arcsec\ and 8\arcsec\ for I~ZW~1 and IRAS~06269$-$0543, respectively. This may indicate that the molecular gas is extended to a radius of about 6 kpc and 8 kpc for these two galaxies, respectively. For I~ZW~1, our estimate agrees well with previous PdBI observations which show an extent of about 10 kpc for the molecular disk size of the host galaxy \citep{schinnerer98}. By comparison, the $\geqslant 2 \sigma$ CO-emitting region extend to a size with radius of about 2\arcsec .0 and 1\arcsec .1 (both equivalent to $\sim$2.4 kpc) for I~ZW~1 and IRAS~06269$-$0543, respectively, suggesting that a substantial amount of molecular gas is concentrated in the central few kpc. Given that the flux extracted from a larger aperture is found to be comparable to that measured with the IRAM 30m, it is likely that the flux not recovered within  the 2$\sigma$ region for these two galaxies is due to the emission from extended disk undetected in our ALMA observations. 

For the remaining six galaxies, the CO line profile are well described by a single-Gaussian fit. Similarly, the CO molecular gas is found to be within an extent of about 4 to 8 kpc ($\sim$ 2\arcsec - 3\arcsec) in radius for these galaxies, based on the spectra extracted from a circular aperture with diameter shown in Figure~\ref{fig:spec}). It can be seen from Figure~\ref{fig:moment} that these apertures are typically larger than the 2$\sigma$ CO-emitting regions for the galaxies in our sample. A comparison of the CO(1$-$0) integrated fluxes measured based on the ALMA spectra extracted from the $\geqslant 2 \sigma$ region with the fluxes seen by the IRAM 30m \citep{xia12}, shows that our interferometric data typically recover $\sim$ 80\% of the observed single-dish flux. The consistency of the flux measured from a larger aperture ALMA spectrum with the single-dish observations (Figure~\ref{fig:spec}), is likely to indicate that the flux not recovered in the 2$\sigma$ CO-emitting region is attributed to the insufficient sensitivity of our ALMA observations for the detection of more extended lower surface brightness CO emission. However, it should be noted that the systematic uncertainties in the flux calibration would also affect the comparison of flux measured with different observations.

\begin{figure*}[ht]
\begin{minipage}{0.49\textwidth}
\centering
\includegraphics[width=\textwidth]{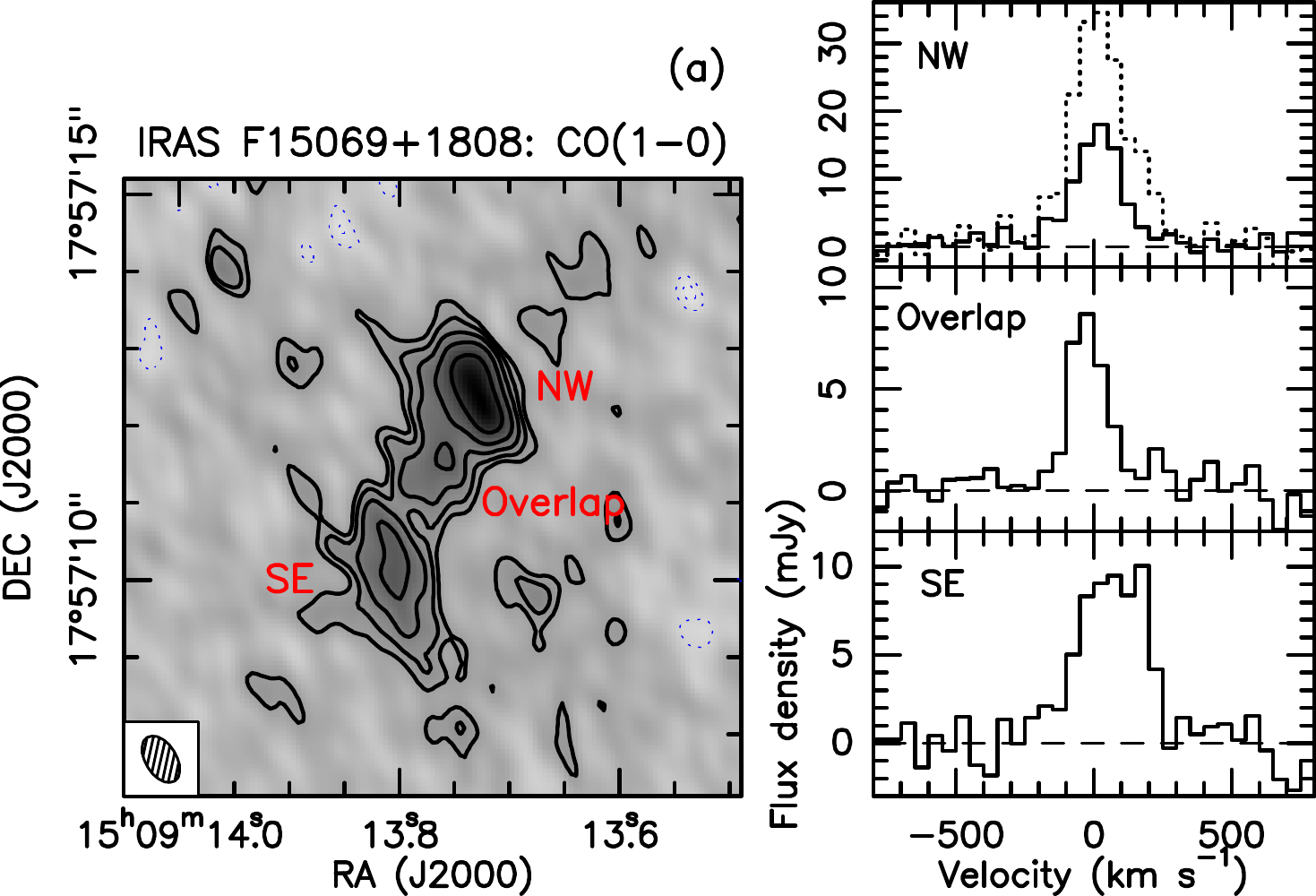}
\end{minipage}
\begin{minipage}{0.49\textwidth}
\centering
\includegraphics[width=\textwidth]{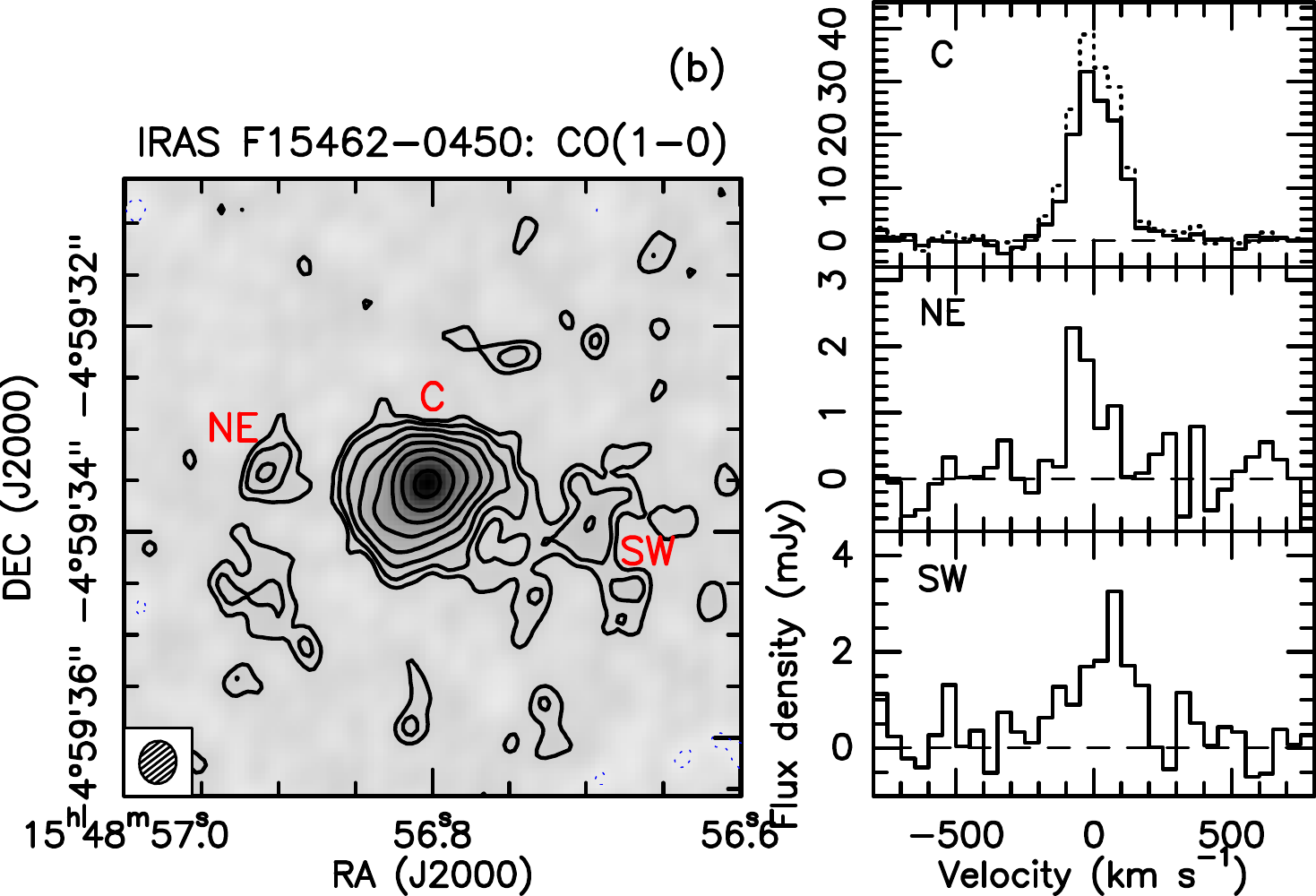}
\end{minipage}\\
\begin{minipage}[b]{0.49\textwidth}
\centering
\includegraphics[width=\textwidth]{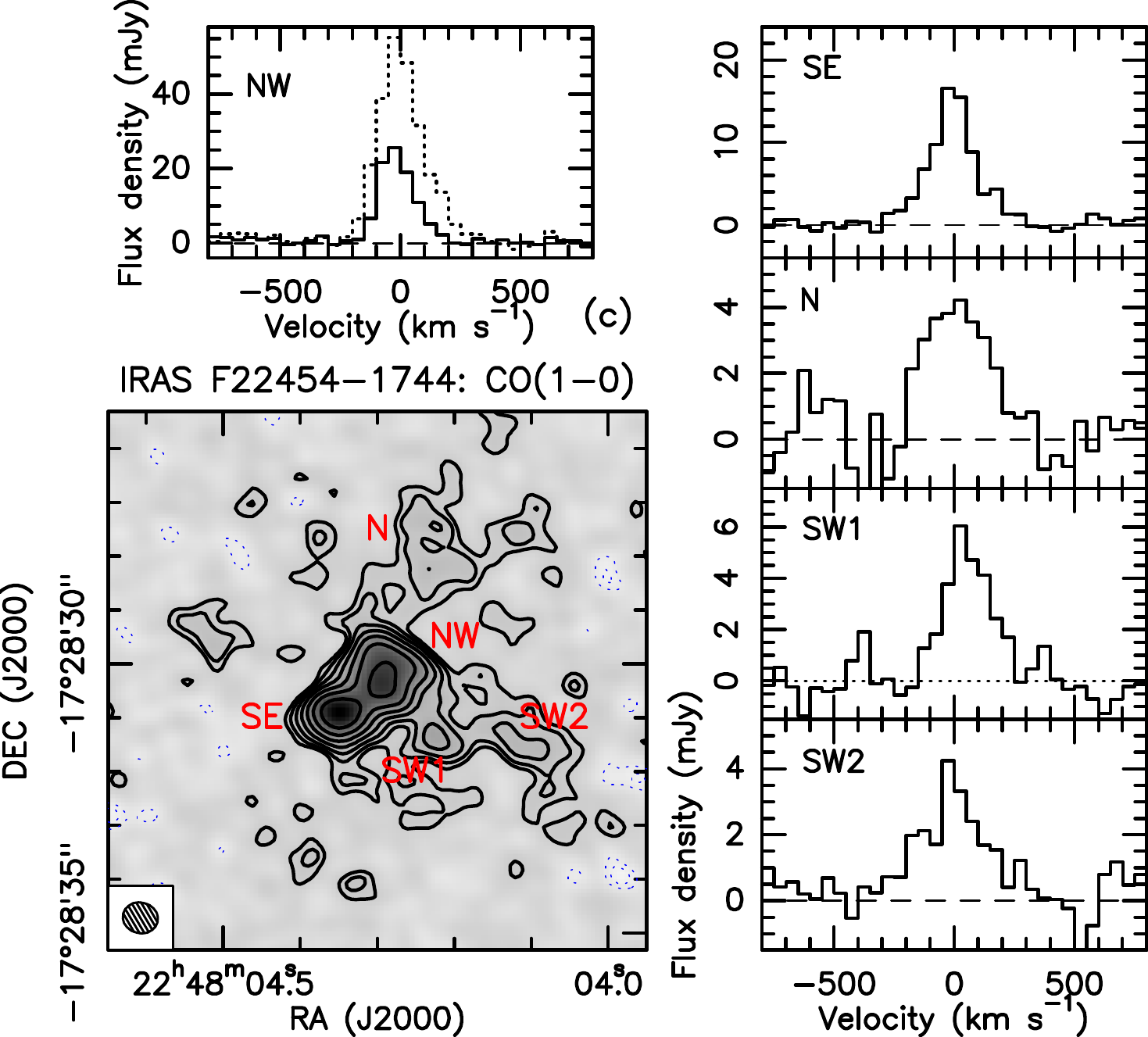}
\end{minipage}
\begin{minipage}[b]{0.49\textwidth}
\centering
\includegraphics[width=\textwidth]{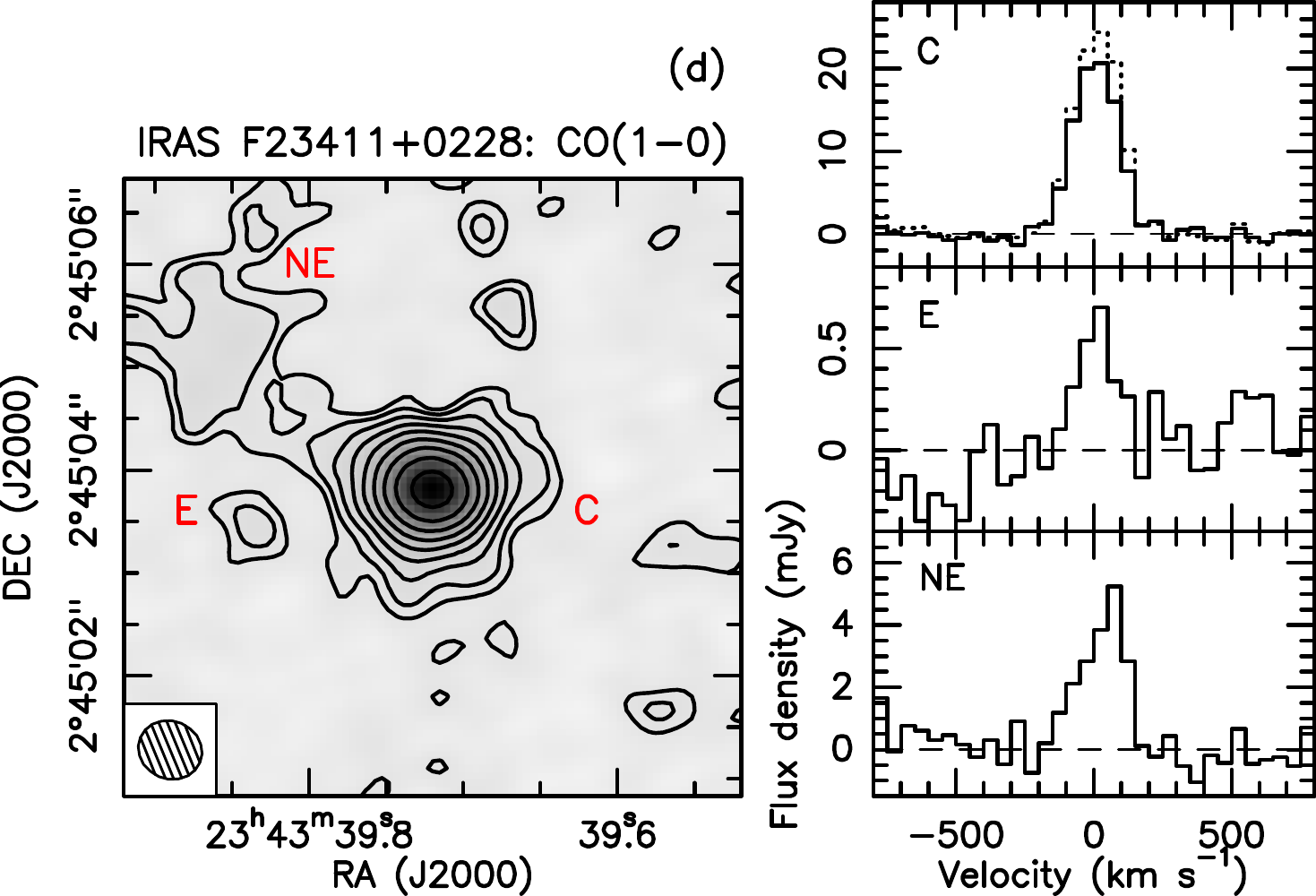}
\end{minipage}

\caption{The CO(1$-$0) velocity-integrated intensity maps (left panel) and the CO(1$-$0) spectra (right panel) extracted from each source component (solid lines) for the galaxies spatially resolved into multiple components overlaid with the global line extracted from the $\geqslant$2$\sigma$ CO-emitting region of IR QSOs (dotted lines, same as in Figure~\ref{fig:spec}): IRAS F15069+1808 (a), IRAS F15462$-$0450 (b), IRAS F22454$-$1744 (c), and IRAS F23411+0228 (d). The CO contours start at 2 $\sigma$ and increase by factors of 1.5, same as in Figure~\ref{fig:moment}. The CO(1$-$0) spectrum of each component is extracted from the $\geqslant$ 2$\sigma$ region in the CO intensity maps shown in the left panel and the spectral resolution is smoothed to $\sim$ 50~\kms. \label{fig:resolved-spectra}}
\end{figure*}

To make a consistent flux measurement with the IRAM 30m data, the total CO integrated flux was measured from the spectrum extracted from a circular aperture (with the diameter shown in Figure~\ref{fig:spec} for each source) by integrating the intensity over the line-emitting region in each channel, except for the galaxies spatially resolved into interacting systems and/or clumpy structures, IRAS~F15069+1808, IRAS~F15462$-$0450, IRAS~F22454$-$1744, and IRAS F23411+0228, for which the integrated flux derived for each component was measured from the spectrum extracted from the $\geqslant 2 \sigma$ region in the CO intensity map (see the spectra in Figure~\ref{fig:resolved-spectra}). The redshift and line width were measured from a Gaussian fit to the CO line. The ALMA CO line properties measured for the galaxies in our sample are summarized in Table~\ref{tab:meas}.

%%%%%%%%%%%%%%%- Fig-5 -%%%%%%%%%%%%%%%%
%%%%%%%%%%%%%%%%%%%%%%%%%%%%%%%%%%%%%%%%

The CO luminosity can be derived from the line integrated intensity listed in Table~\ref{tab:meas}, following \citet{solomon97}:
\begin{equation}\label{eq1}
\begin{split}
L^\prime_{\rm CO}= & 3.25\times10^7 \left(\frac{S_{\rm CO}\Delta v}{{\rm 1\ Jy\ km\ s^{-1}}}\right)\left(\frac{\nu_{\rm obs}}{{\rm 1\ GHz}}\right)^{-2}\\
& \times\left(\frac{D_{\rm L}}{{\rm 1\ Mpc}}\right)^2 \left(1+z\right)^{-3}\ {\rm K\ km\ s^{-1}\ pc^2},
\end{split}
\end{equation}
where $S_{\rm CO}\Delta v$ is the velocity-integrated flux density, $\nu_{\rm obs}$ is the observed line frequency, and $D_{\rm L}$ is the luminosity distance. Similar to the measurements in \citet{xia12}, the CO(1$-$0) luminosity ranges from several times 10$^9$ to 10$^{10}$ \kkmspc\ for our sample of IR QSO hosts (see Table~\ref{tab:properties}). The molecular gas mass derived by $M_{\rm H_2}=\alpha_{\rm CO} L^\prime_{\rm CO}$, corresponds to a few times 10$^9\ M_\odot$ if we adopt an ULIRG-like CO luminosity-to-H$_2$ mass conversion factor, $\alpha_{\rm CO}$, of 0.8 $M_\odot$ pc$^{-2}$ (K km s$^{-1}$)$^{-1}$ \citep[e.g.,][]{downes98}. We note that $\alpha_{\rm CO}$ is uncertain and varies significantly from source to source, and possibly from region to region within a galaxy \citep[e.g.,][]{bolatto13}. For simplicity, we adopt a constant $\alpha_{\rm CO}$ to estimate the molecular gas mass for all the galaxies in our sample. 

For the systems that are resolved into multiple separate objects in CO images, the ALMA data enable us to estimate the molecular gas mass for each component for the first time (see Table~\ref{tab:properties}). The molecular gas mass derived for the two brightest sources NW and SE in the merging system IRAS~F15069+1808 are found to be comparable, implying that these two galaxies are likely undergoing a major merger. Similarly, the approximate equivalent mass derived for the NW and the SE sources in IRAS~F22454$-$1744 is probably indicative of a major merger. In the case of IRAS~F23411+0228, the molecular gas mass estimated for the E source ($\sim 3.8\times10^7\ M_\odot$) is much lower compared to that of central bright galaxy ($\sim 1.4\times10^9\ M_\odot$). Combining with the observed properties discussed above, i.e.,  multi-wavelength identification and gas kinematics, this may suggest that IRAS~F23411+0228 would be a minor merger if the faint E source is indeed an interacting companion.

%%%%%%%%%%%%%%%%%%%%%%%%%%%%%%%%%%%%%%%%%

\subsection{Source Size Measurements} \label{subsec:size}

Our high-resolution CO imaging reveals detailed structure of the molecular gas distribution for IR QSO hosts, which is crucial for placing constraints on the dynamical states of the sources. To avoid possible uncertainties produced by the image reconstruction process that will be introduced into the size measurement, we directly fit models to the interferometric observables (i.e., the visibilities) in the $uv$-plane which is preferable for the cases of simple source structure by using \texttt{UVMULTIFIT} tool\citep{martividal14}.

%%%%%%%%%%%%%%- Table-3- %%%%%%%%%%%%%%

%\begin{longrotatetable}
\begin{deluxetable*}{lCCCCCCCCCC}[htbp]
\tablenum{3}
\centering
\tabletypesize{\scriptsize}
\addtolength{\tabcolsep}{-1.pt}
\tablecaption{Derived properties}\label{tab:properties}
\tablewidth{0pt}
\tablehead{
\colhead{Source} & \colhead{$L'_{\rm CO(1-0)}$} & \colhead{$M_{\rm H_2}$} & \colhead{CO size} & \colhead{Cont. size} & \colhead{Inclination $i$} & \colhead{$M_{\rm dyn}$sin$^2 i$} & \colhead{$M_{\rm dyn}$}  & \colhead{$M_{\rm BH}$} & \colhead{SFR$_{\rm CO}$} & \colhead{$L_{\rm bol}$} \\
\colhead{} & \colhead{(10$^9$ {\bf $L_l$})$\tablenotemark{a}$} & \colhead{(10$^9$ $M_\odot$)} & \colhead{(kpc)} & \colhead{(kpc)} & \colhead{(deg)} & \colhead{(10$^{10}$ $M_\odot$)} & \colhead{(10$^{10}$ $M_\odot$)}  & \colhead{(10$^7$ $M_\odot$)} & \colhead{($M_\odot$ yr$^{-1}$)} & \colhead{(10$^{12}$ $L_\odot$)}\\
\colhead{(1)} &  \colhead{(2)} & \colhead{(3)} & \colhead{(4)} & \colhead{(5)} & \colhead{(6)} & \colhead{(7)} & \colhead{(8)} & \colhead{(9)} & \colhead{(10)} & \colhead{(11)} 
}
%\colnumbers
\startdata
I~ZW~1 & 5.50\pm 0.26 & 4.40\pm 0.21 & 3.2$\pm$0.2 & - & 38\pm 11 & 1.6\pm 0.3 & 4.1\pm 0.7  & 2.1 & 117 & 1.0 \\
IRAS 06269$-$0543 & 6.91\pm 0.44 & 5.53\pm 0.35 & 2.3$\pm$0.3 & 0.6$\pm$0.1 & 51\pm 10 & 0.7\pm 0.1 & 1.1\pm 0.2  & 3.2 & 183 & 1.8 \\
IRAS F11119+3257 & 11.36\pm 0.90 & 9.09\pm 0.72 & 4.8$\pm$0.8 & 1.0$\pm$0.2 & 48\pm 15 & 2.4\pm 0.4 & 4.3\pm 0.8  & 15.7 & 312 & 11.0 \\
IRAS Z11598$-$0112 & 11.99\pm 0.68 & 9.59\pm 0.54 & 4.7$\pm$0.6 & - & 37\pm 19 & - & 1.6\pm 0.2  & 0.4 & 378 & 0.4 \\
IRAS F15069+1808NW & 5.57\pm 0.31 & 4.45\pm 0.25 & 3.3$\pm$0.4 & - & 48\pm 12 & - & 2.0\pm 0.3  & - & 130 & - \\
IRAS F15069+1808SE & 4.35\pm 0.40 & 3.48\pm 0.32 & 7.0$\pm$1.0 & - & 64\pm 8 & 5.5\pm 1.4 & 6.9\pm 1.8  & 1.0 & 96 & 0.5 \\
IRAS F15462$-$0450{\bf C} & 3.64\pm 0.10 & 2.91\pm 0.08 & 2.0$\pm$0.3 & 0.4$\pm$0.1 & 25\pm 21 & 0.15\pm 0.03 & 0.9\pm 0.2  & 0.8 & 70 & 0.2 \\
IRAS F22454$-$1744NW & 3.07\pm 0.09 & 2.46\pm 0.07 & 2.9$\pm$0.2 & - & 31\pm 17 & 0.7\pm 0.1 & 2.8\pm 0.3  & - & 62 & - \\
IRAS F22454$-$1744SE & 2.43\pm 0.10 & 1.94\pm 0.08 & 1.2$\pm$0.2 & - & 31\pm 24 & - & 0.7\pm 0.2  & 0.5 & 47 & 0.6 \\
IRAS F23411+0228{\bf C} & 1.75\pm 0.06 & 1.40\pm 0.05 & 2.4$\pm$0.6 & 0.4$\pm$0.2 & 28\pm 31 & - & 1.3\pm 0.3  & 1.0 & 55 & 1.3 \\
\enddata
\tablecomments{Column (1): galaxy name. Column (2): CO(1-0) luminosity derived with the integrated flux density listed in Table~\ref{tab:meas}. Column (3): molecular gas masses derived with $L'_{\rm CO(1-0)}$ by assuming a CO luminosity-to-molecular gas mass conversion factor of $\alpha_{\rm CO}$ = 0.8 $M_\odot$ (K km s$^{-1}$ pc$^2$)$^{-1}$. Column (4): CO FWHM major axis size of the extended component, derived from visibility fitting (see Table~\ref{tab:meas}). Column (5): 3-mm continuum FWHM size. We take the source as an unresolved point source if the size measurement is only significant at $<2 \sigma$. Column (6): disk inclination angle estimated from the ratio of CO(1-0) minor and major axis as $i$ = cos$^{-1}$($a_{\rm min}$/$a_{\rm maj}$). Column (7): dynamical mass without inclination angle correction for the galaxies with kinematical signatures of rotating molecular gas disks, estimated from the CO(1-0) line width and source size (see Section~\ref{subsec:dynmass} for details). Column (8): inclination angle-corrected dynamical mass within the CO-emitting region. The uncertainty in these dynamical masses does not include the uncertainties in the inclination angle. For the galaxies without a velocity gradient in CO map, we adopt the virial relation to estimate dynamical mass. Column (9): virial black hole masses from \citet{hao05} using H$\beta$ broad emission line and $\lambda L_{\lambda,5100\angstrom}$ measurements. For I~ZW~1, we take the average of the measurements derived in \citet{hao05} and \citet{vestergaard06}. Column (10): star formation rate derived from IR luminosity, which is predicted from CO(1-0) luminosity based on the correlation between $L^\prime_{\rm CO(1-0)}$ and $L_{\rm IR}$ observed for starburst galaxies (see Section~\ref{subsec:growth} for details). Column (11): AGN-associated bolometric luminosities from \citet{hao05} using the bolometric correction of $L_{\rm bol} \approx$ 9$\lambda L_{\lambda,5100\angstrom}$.}
\tablenotetext{a}{$L_l \equiv$ K km s$^{-1}$ pc$^2$}
\end{deluxetable*}
%\end{longrotatetable}
%column(4): For each source, we adopt the median value of inclination derived with the FWHM size measured from the 2D Gaussian fit to the integrated CO map and the fits to the CO visibilities in the $uv$ plane with a circular Gaussian model (see Section~\ref{subsec:distribution})
%Column (9): virial black hole masses from \citet{hao05} using H$\beta$ broad emission line and $\lambda L_{\lambda,5100\angstrom}$ measurements. For I~ZW~1, we take the average of the measurements derived in \citet{hao05} and \citet{vestergaard06}.
%%%%%%%%%%%%%%%%%%%%%%%%%%%%%%%%%%%%%%

%%%%%%%%%%%%%%- Table-4- %%%%%%%%%%%%%%

\begin{deluxetable}{lccc}[htbp]
\tablenum{4}
\centering
\tabletypesize{\scriptsize}
\addtolength{\tabcolsep}{-1.0pt}
\tablecaption{CO gas kinematical properties}\label{tab:pv}
\tablewidth{0pt}
\tablehead{
\colhead{Source} & \colhead{PA} & \colhead{$v_{\rm rot}$} & \colhead{$\sigma_{\rm gas}$} \\
    & \colhead{(deg)} & \colhead{(\kms)} & \colhead{(\kms)}\\
    \colhead{(1)} &  \colhead{(2)} & \colhead{(3)} & \colhead{(4)}
    }
\startdata
I~ZW~1 & 124 & 273 & 44 \\
IRAS 06269$-$0543 & 168 & 166 & 45 \\
%IRAS F11119+3257 & 144 & 51 & 93 \\
IRAS F15462$-$0450 & 306 & 157 & 42 \\
\enddata
\tablecomments{Column (1): galaxy name. Column (2): position angle along the major axis of the rotation disc, measured from the velocity field. Column (3) and Column (4): rotational velocity and velocity dispersion, obtained by fitting 3D tilted-ring models to CO(1$-$0) data cubes using the software $^{3D}$BAROLO.}
\end{deluxetable}
%%%%%%%%%%%%%%%%%%%%%%%%%%%%%%%%%%%%%%

\begin{figure*}[htbp]
\centering
\begin{minipage}[b]{0.99\linewidth}
\includegraphics[width=0.245\linewidth]{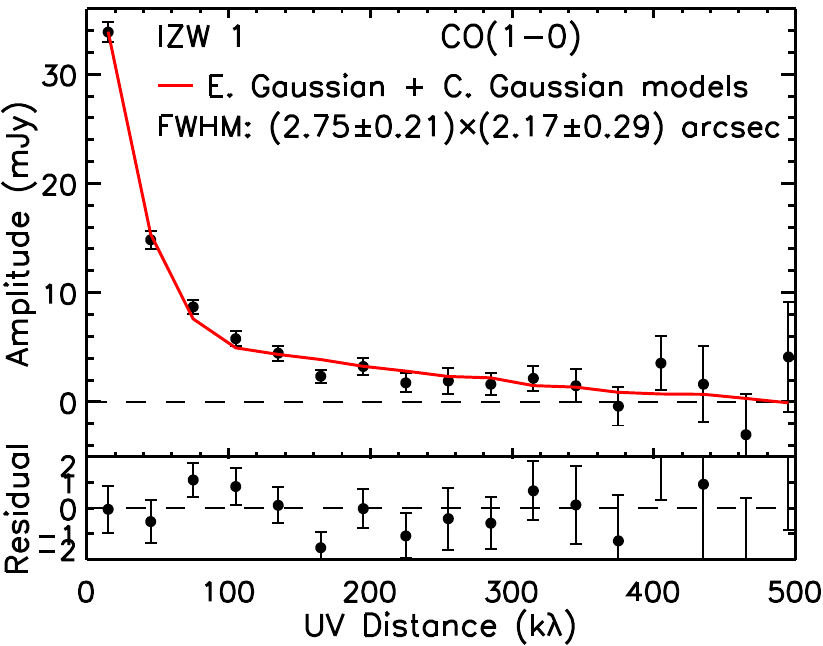}
\includegraphics[width=0.245\linewidth]{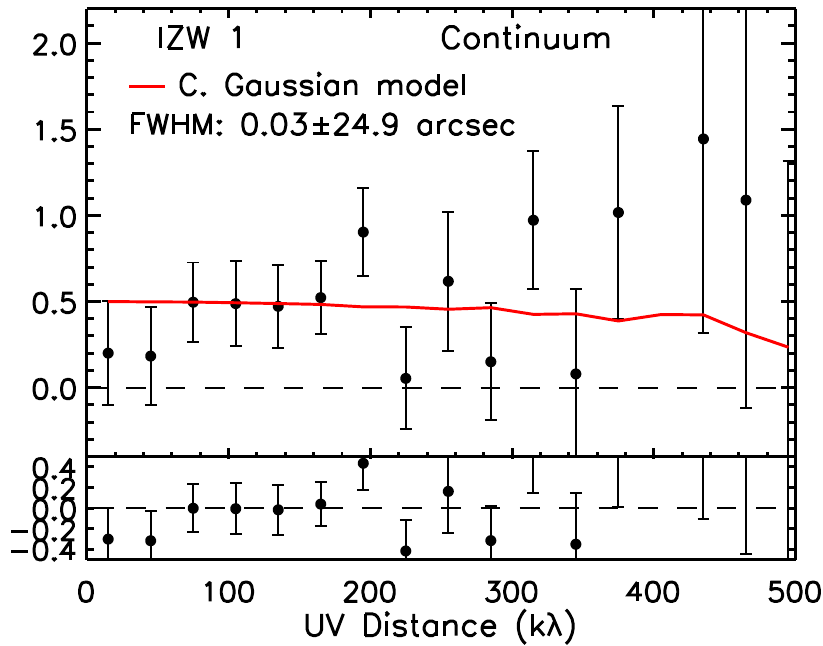}
\includegraphics[width=0.245\linewidth]{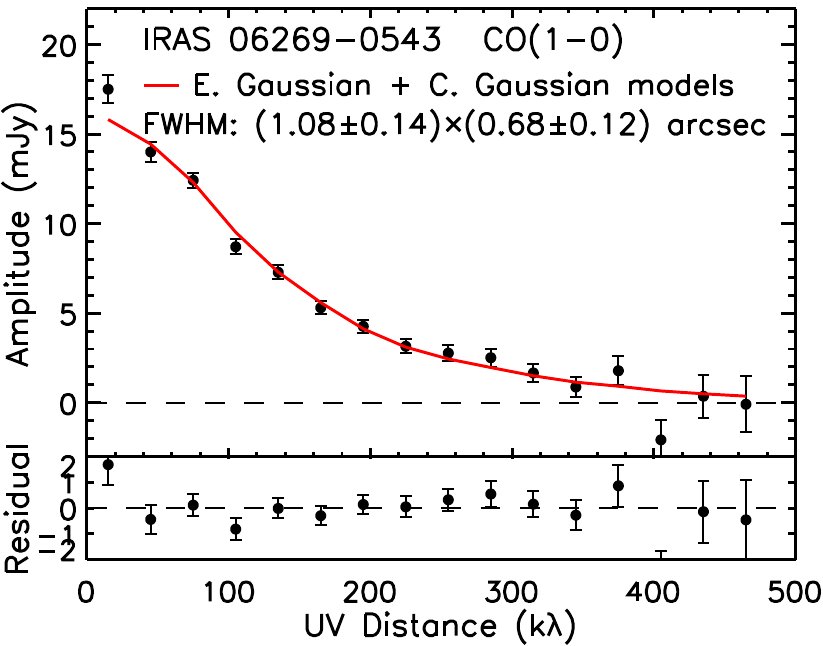}
\includegraphics[width=0.245\linewidth]{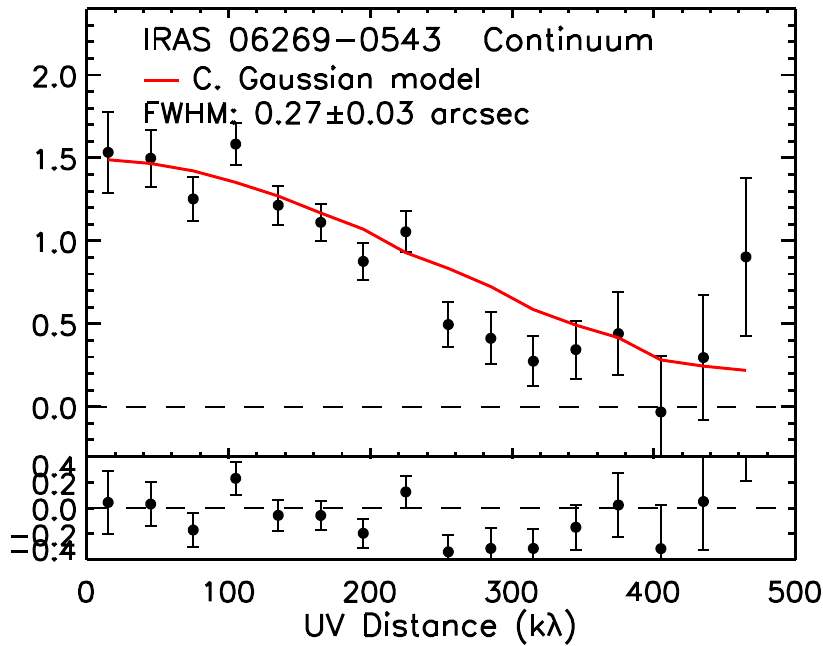}
\end{minipage}
%\vspace{2pt}
\begin{minipage}[b]{0.99\linewidth}
\includegraphics[width=0.245\linewidth]{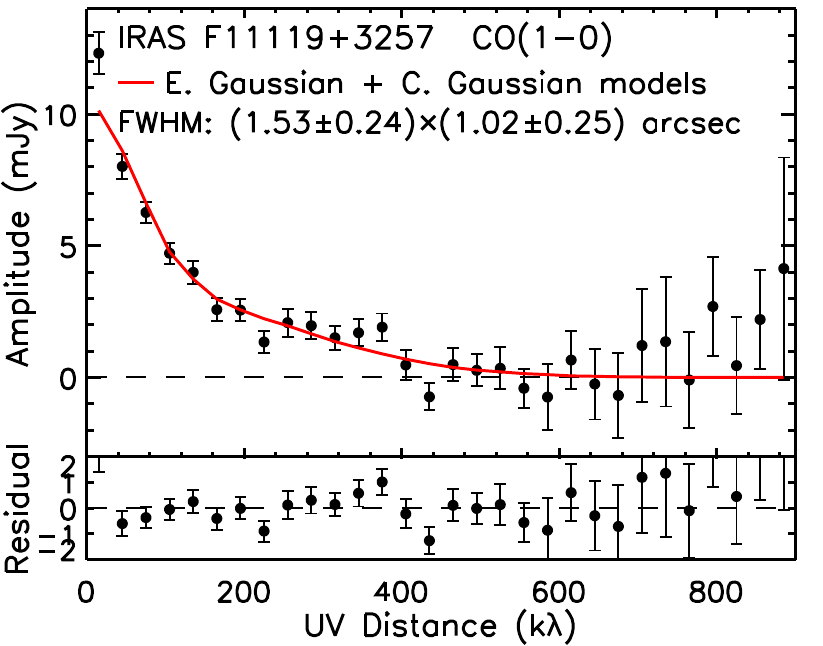}
\includegraphics[width=0.245\linewidth]{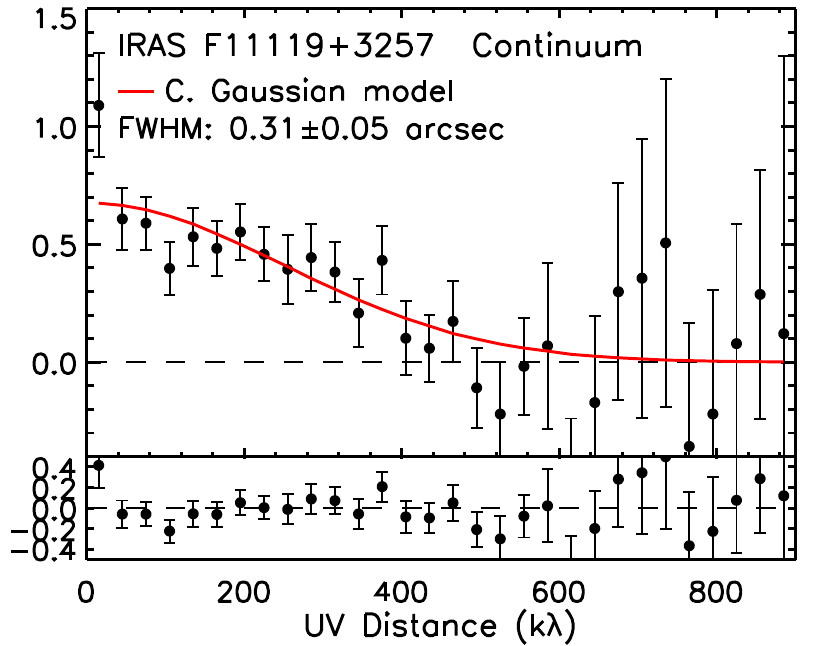}
\includegraphics[width=0.245\linewidth]{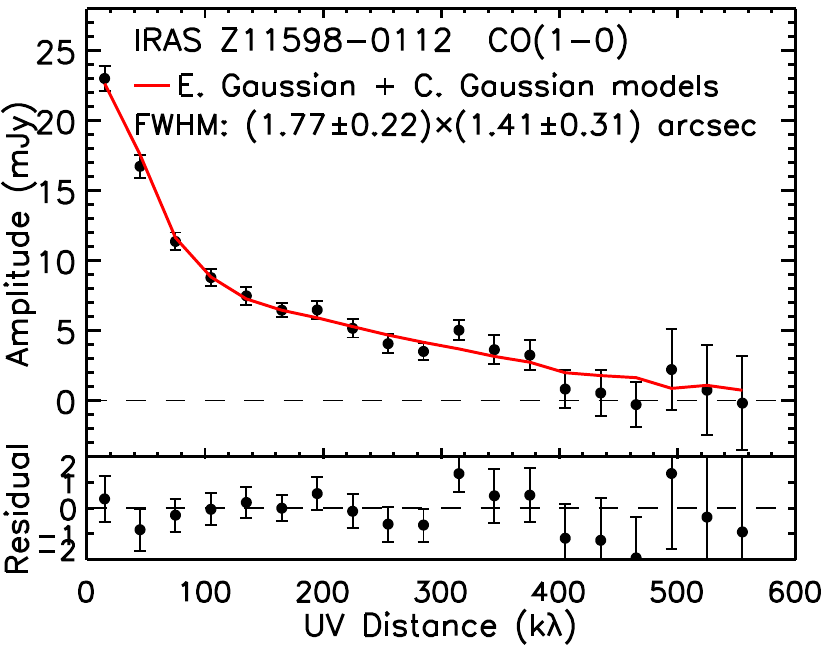}
\includegraphics[width=0.24\linewidth]{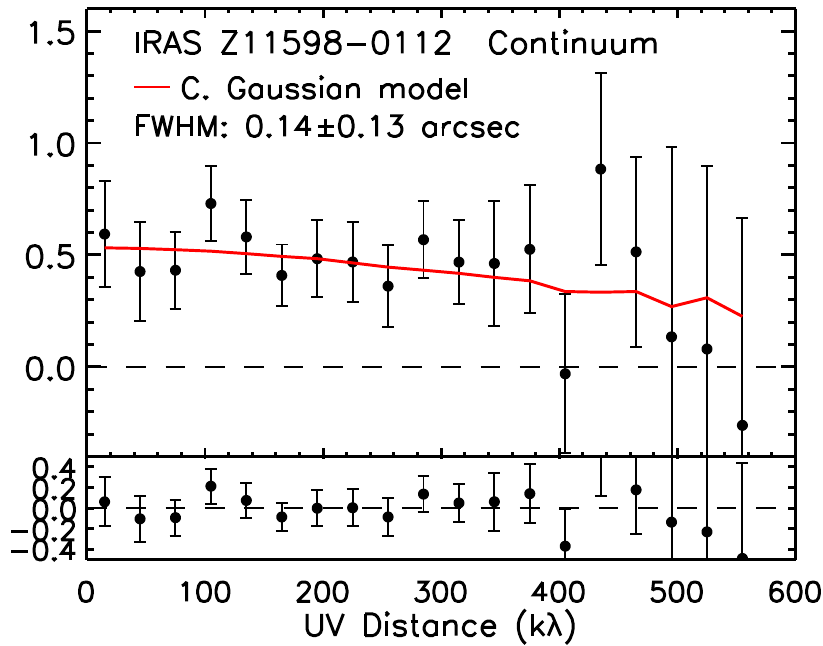}
\end{minipage}
\begin{minipage}[b]{0.99\linewidth}
\includegraphics[width=0.245\linewidth]{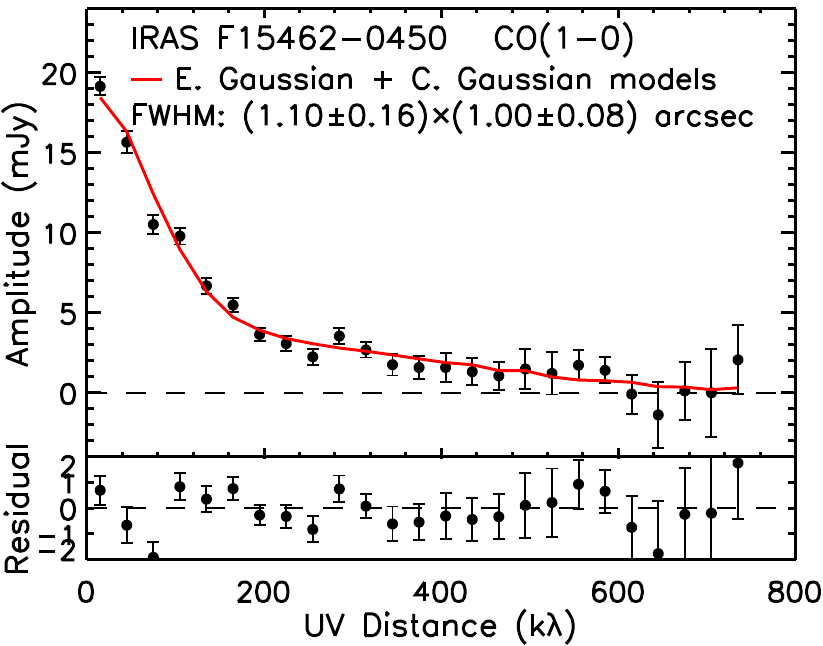}
\includegraphics[width=0.245\linewidth]{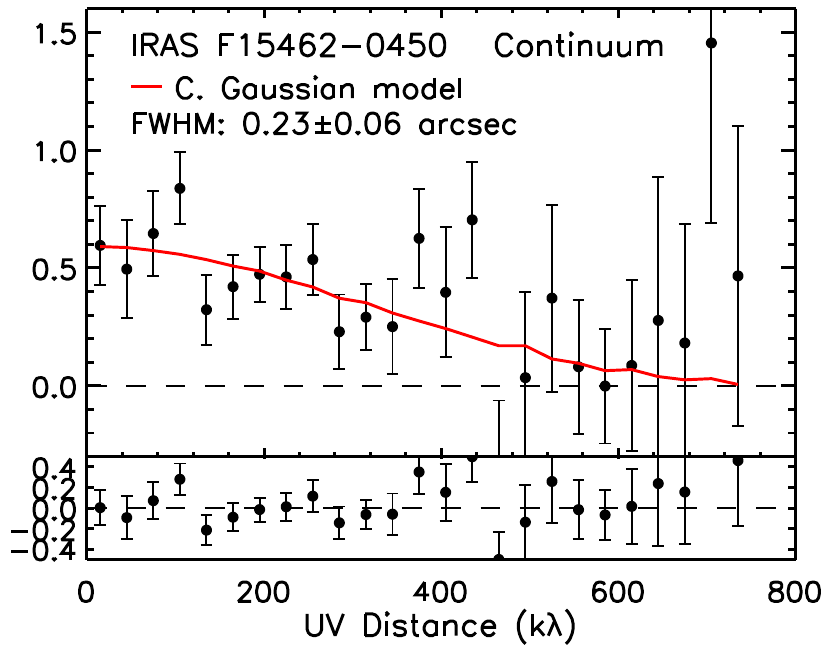}
\includegraphics[width=0.245\linewidth]{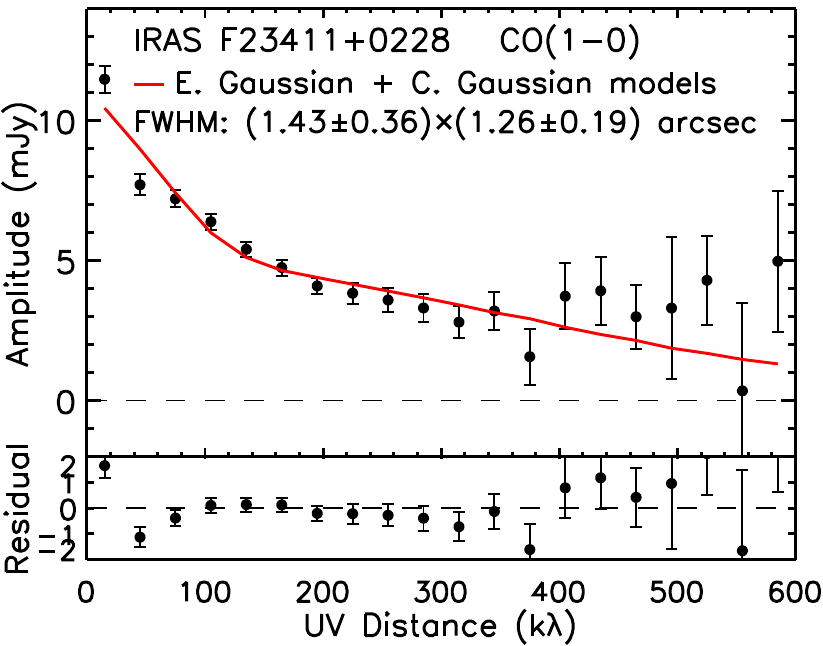}
\includegraphics[width=0.245\linewidth]{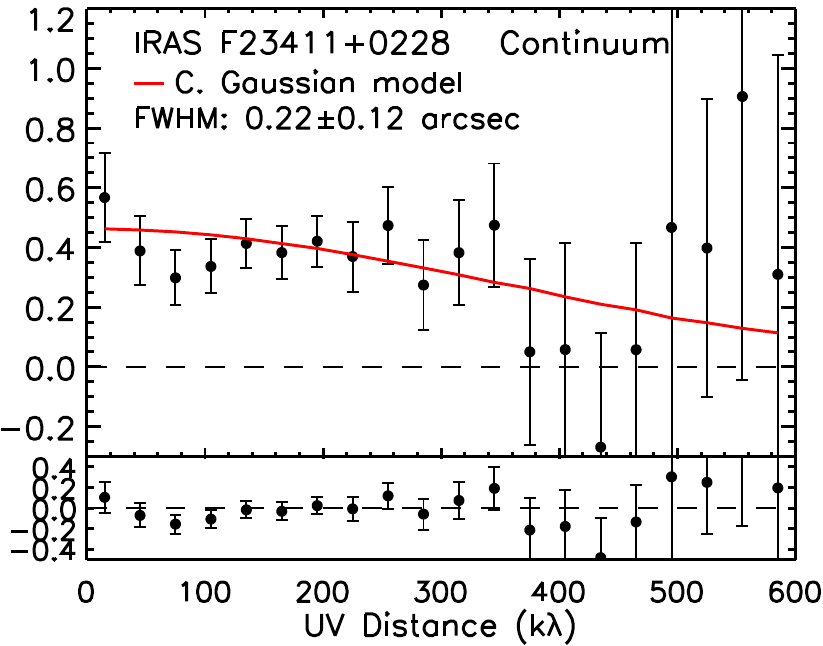}
\end{minipage}
%\vspace{2pt}
\begin{minipage}[b]{0.99\linewidth}
\includegraphics[width=0.245\linewidth]{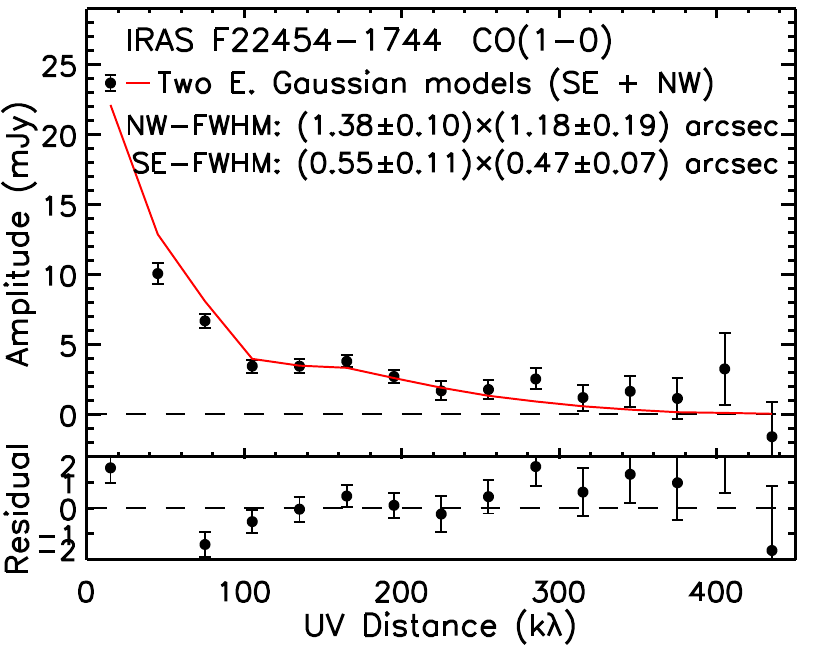}
\includegraphics[width=0.245\linewidth]{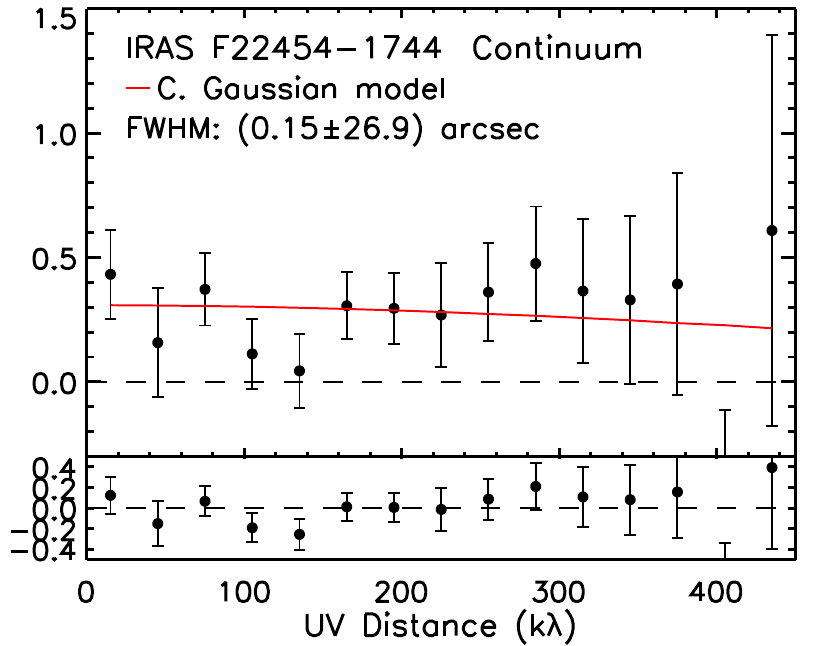}
\includegraphics[width=0.245\linewidth]{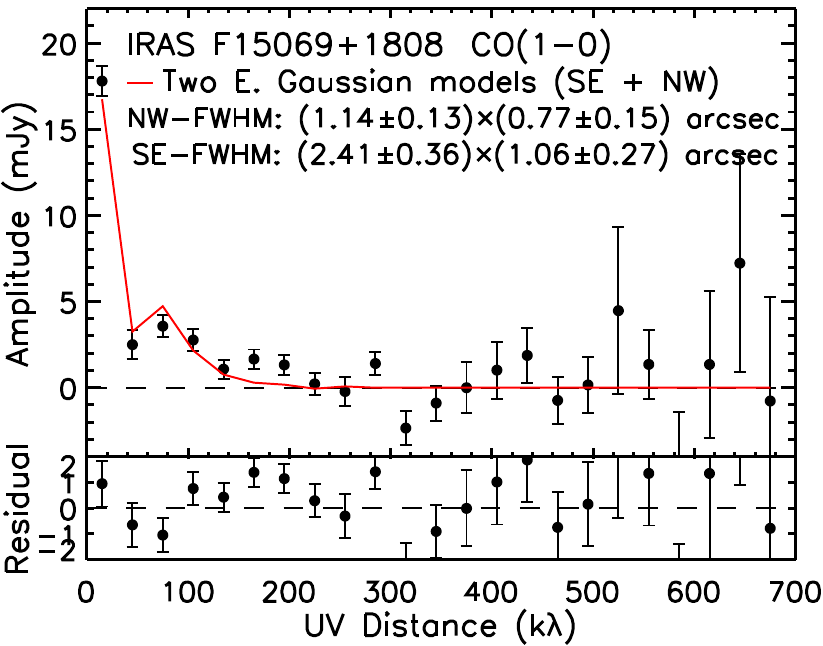}
\includegraphics[width=0.24\linewidth]{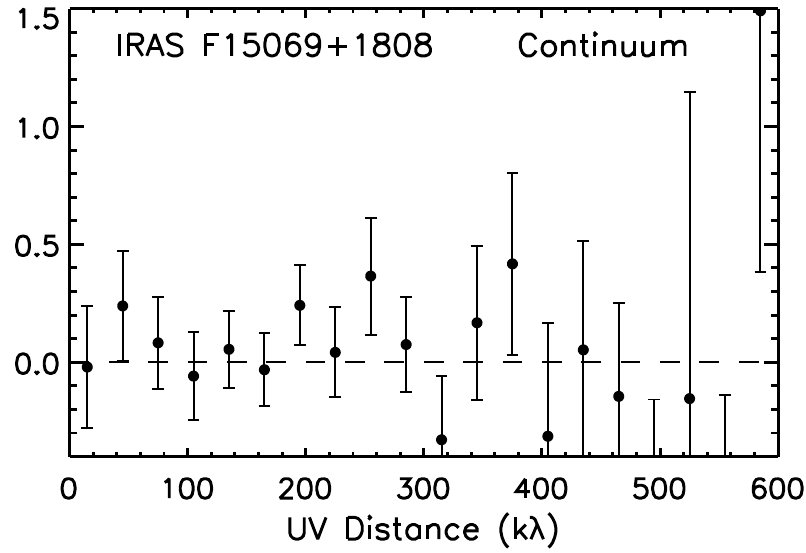}
\end{minipage}
\caption{Visibilitity amplitude as a function of the $uv$-distance for the CO(1$-$0) line (left) and the continuum emission (right). The black circles with error bars are the data binned in $uv$-radius steps of 30 k$\lambda$. Error bars show the statistical photon noise on the average amplitude in each bin. We use the \texttt{UVMULTIFIT} tool to fit the visibility amplitudes to models in the $uv$-plane. For the line data, the red curves are the best-fits with a combination of an elliptical and a circular Gaussian models to the $uv$-data, while for the continuum data the red lines indicate the best-fitting model with a single circular Gaussian function to the data, respectively. For IRAS~F15069+1808 and IRAS~F22454$-$1744, the line data are best fitted with two elliptical Gaussian models with positions centered on each separate galaxy. The residuals between the data and the model are shown in the bottom panels. The FWHM source size shown in the upper-right of each line data panel is from the elliptical Gaussian component and adopted as the CO source size for analysis. The flat distribution of visibilities implies that the source is unresolved in ALMA observations. No model fit is performed for the non-detection of 3-mm continuum emission in IRAS~F15069+1808. \label{fig:size}}
\end{figure*}

Based on the centroid of the surface brightness distribution inferred from 2D Gaussian fitting (Table~\ref{tab:meas}), we shift the phase center of both the line and continuum datasets used for the size measurement to match the centroid of each source. Assuming that the surface brightness distribution of the CO line and continuum follow a symmetric Gaussian profile, we performed fits of the visibilities that were spectrally-averaged with different source models (i.e., a single Gaussian profile, a combination of a point-source model and an elliptical Gaussian model, and a combination of a compact circular Gaussian model and an extended elliptical Gaussian model to investigate the possibility of having compact CO emission embedded in an extended, low surface-brightness component) and found that all the CO sources are well-resolved and the best fit is given by a combination of a circular Gaussian and an elliptical Gaussian components. For the continuum, a single circular Gaussian component model is preferred due to the relatively low SNR of the detection. The best fit is determined by $\chi^2$ minimization. For IRAS~F15069+1808 and IRAS~F22454$-$1744, which are resolved into two separate galaxies with comparable gas mass, the CO sources are best fitted with two elliptical Gaussians with positions centered on the QSO host and the companion galaxy, respectively. The best-fitting values of the double Gaussian models for the line and the one-component model for the continuum are reported in Table~\ref{tab:meas}. Figure~\ref{fig:size} shows the visibility function for the CO and 3-mm continuum, and the corresponding best-fitting profiles. 

For the double-Gaussian models fit to the CO data, the source size of the compact component (FWHM measured from circular Gaussian profile) ranges from 0.\arcsec22 to 0.\arcsec37 ($\sim$ 0.4$-$1.0 kpc with a median of 0.8 kpc) and the size of the extended component (major axis FWHM from elliptical Gaussian profile) ranges from 0.\arcsec55 to 2.\arcsec75 ($\sim$ 1.2$-$7.0 kpc with a median of 3.2 kpc). The largest CO source size (2.\arcsec41, equivalent to a physical scale of 7.0 kpc) is seen in IRAS F15069+1808SE. For the rest galaxies in our sample, the major axis size of the CO source ranges from 1.2 kpc to 4.9 kpc. For the continuum emission, four of the sources are marginally resolved with observed sizes comparable to the synthesized beams, while the remaining detections are point-like sources in our observations. The continuum source sizes are estimated to be 0.\arcsec22$-$0.\arcsec31 ($\sim$ 0.4$-$1.0 kpc with a median of 0.6 kpc) for the four sources with size measurement at $\gtrsim$ 2$\sigma$ significance, which is more than a factor of 3 smaller than their CO sizes, suggesting that the continuum emission is more concentrated than that of molecular gas. Compared to the full CO flux distribution ($\sim$ 4$-$8 kpc in radius) discussed in Section~\ref{subsec:spect}, the 2D Gaussian visibility fitting gives smaller CO size, which represents the distribution of the relatively dominant central component.

\subsection{Dynamical Masses} \label{subsec:dynmass}

As described in the previous section, large-scale systematic velocity gradients are seen for six galaxies (I~Zw~1, IRAS~06269$-$0543, IRAS~F11119+3257, IRAS~F15069+1808SE, IRAS~15462$-$0450, and IRAS F22454$-$1744NW) in our sample , whereas for the remaining four CO sources the molecular gas are found to be highly disturbed with complex velocity structure (see Figure~\ref{fig:moment}). 

%%%%%%%%%%%%%%%- Fig-6 -%%%%%%%%%%%%%%%%

%%%%%%%%%%%%%%%%%%%%%%%%%%%%%%%%%%%%%%

For the six galaxies exhibiting CO velocity gradient across the source, we estimate the dynamical mass within the CO-emitting region by assuming that the bulk of the emission in these sources can be well described by a rotating disk. In this case, the dynamical mass is given by applying the relation, $M_{\rm dyn}\approx 2.3\times 10^5 v_{\rm cir}^2 R$ \citep[e.g.,][]{wang10}, where $R$ is the disk radius in kpc, which we assume equal to 0.75 times the CO FWHM major axis of the extended component measured from visibility fitting \citep[i.e., half of full-width at 20\% of the peak intensity for a Gaussian profile of CO emission;][]{wang13}, and $v_{\rm cir}$ is the maximum circular velocity of the gas disk in \kms. For the three galaxies that are well-fitted by our rotating kinematic models (see Figure~\ref{fig:pv}), the circular velocity is equal to the rotation velocities reported in Table~\ref{tab:pv}. For the remaining three galaxies, one (IRAS~F11119+3257) is not consistent with our model and the other two (IRAS~F15069+1808SE and IRAS~F22454$-$1744NW) are undergoing a major merger interaction. For these galaxies, their circular velocities were estimated by $v_{\rm cir}$ = 0.75 $\Delta v_{\rm FWHM}$/sin$i$ \citep{ho07b,ho07a,wang10}, assuming the gas is distributed in an inclined disk. Adopting the inclination angle listed in Table~\ref{tab:properties}, the derived inclination-corrected dynamical masses range from $9.0\times10^{9}$ to $6.9\times10^{10}$ $M_\odot$. These estimates of $M_{\rm dyn}$ can be significantly affected by the uncertainty of inclination and the assumption of a rotating disk as part of the emitting gas might be associated to non-circular motion, especially in interacting systems with perturbations. For the three galaxies with $v_{\rm cir}$ derived from model fitting, we also calculated the circular velocity using the aforementioned equation, and found the $v_{\rm cir}$ inferred by the FWHM of CO line are roughly twice higher than that derived from model fitting. This is suggestive that the likely presence of non-rotating gas components also contributes to the CO line width, which would affect the determination of circular velocity. If true, this would result in lower $v_{\rm cir}$ and therefore lower $M_{\rm dyn}$ for IRAS~F11119+3257, IRAS~F15069+1808SE, and IRAS~F22454$-$1744NW. 

For the galaxies with irregular velocity fields, we assumed that the galaxy is supported by random motions and estimated the dynamical mass using the virial theorem, assuming that the distribution of molecular gas is spherical with uniform density and has an isotropic velocity dispersion. Under these assumptions, the dynamical mass can be estimated as $M_{\rm dyn} = 5\sigma^2 R/G$ \citep[e.g.,][]{pettini01}, where $R$ is the gas disk radius, $\sigma = \Delta v_{\rm FWHM}$/2.35 is the one-dimensional velocity dispersion, and $G$ is the gravitational constant. The derived dynamical masses range from $7.4\times10^9$ to $2.0\times10^{10}$ $M_\odot$ for these four galaxies. We found that the virial estimator leads to a smaller $M_{\rm dyn}$ compared to that derived from a rotating disk estimator. Note that the virial mass estimates are also affected by large systematic errors, e.g., the unverified assumption that the system is virialized and the source size that is likely underestimated due to the possibly more extended, low surface brightness CO emitting regions missed by our high-resolution ALMA observations. As discussed in Section~\ref{subsec:size}, the CO size inferred from visibility fitting is more likely to represent the distribution of relatively central component. Detailed modeling of spatially resolved gas kinematics are needed to obtain a more accurate estimate of the dynamical mass for the IR QSO hosts in our sample.  

%%%%%%%%%%%%%%%%%%%%%%%%%%%%%%%%%%%%%%

\section{Discussion} \label{sec:discussion}

\subsection{Host Galaxy Properties}\label{subsec:host}

The combined analysis of CO morphology and gas kinematics of the IR QSO hosts in our sample reveals the diversity of host galaxy properties. Three of IR QSO systems in our sample are identified as interacting systems, one galaxy accompanying each of these QSOs. The companion galaxies are located between roughly 2 and 7 kpc in projection from the QSOs, with small velocity offsets ($\Delta v<$ 60 \kms). Two (IRAS~F15069+1808 and IRAS~F22454$-$1744) of these interacting QSO systems are interpreted as major mergers, while the other one (IRAS~F23411+0228) show evidence for a minor merger, although higher resolution multi-wavelength data are needed for further confirmation. Our ALMA data for the first time provide evidence for significant major merger interactions in some of local IR QSO systems. Observations toward high-$z$ have also revealed companion galaxies interacting with the central luminous QSOs, suggesting that merger activity may be a dominant mechanism for the most luminous galaxies \citep{hu96,omont96,decarli17,trakhtenbrot17,diaz-santos18}. The velocity fields of the molecular gas in QSO hosts IRAS~F23411+0228 and IRAS~F22454$-$1744 are significantly disturbed and show complex structure, while the gas in the QSO host of IRAS~F15069+1808 system shows velocity gradient. Simulations of gas-rich mergers have shown that the coalesced nucleus of mergers can rapidly relax into smooth disks \citep[e.g.,][]{springel05b,robertson06,hopkins09}, which has also been shown observationally \citep{ueda14}. We therefore caution that our ALMA data cannot disprove the possibility that the interacting QSO hosts may have already experienced merging process with other galaxies if all the QSOs are triggered by major mergers.

For the rest five IR QSO systems in our sample, of which the QSO hosts are found to be isolated without any interacting companion, most of these sources show evidence for ordered rotation, suggesting that these systems are likely observed in the final coalescence stage. For the three galaxies that can be well-fitted by a rotating kinematic model, I~ZW~1, IRAS~06269$-$0543, and IRAS~F15462$-$0450, the ratio of the rotation velocity to the local velocity dispersion of $V_{\rm rot}/\sigma$ is in the range 4$-$6, indicating that the molecular gas in these galaxies are dominated by rotation. Similar rotating disks with $V_{\rm rot}/\sigma>3$ have also been revealed in massive star-forming galaxies at $z\sim 2$, which were observed at the transition stage from extended disks to compact spheroids with rapid dense cores build-up \citep{tadaki17a,tadaki17b}. In addition, the line-of-sight velocities are found to flatten at $\gtrsim$ 1 kpc for the three galaxies in our sample, implying that these galaxies may do not yet have a significant bulge component, as a dominant bulge would cause the rotation curve to rise steeply at the center and flatten on scales of $<$500 pc \citep[e.g.,][]{sofue99}. However, higher spatially resolved data are necessary to draw a firm conclusion. For IRAS~F11119+3257, the complex velocity field becomes disturbed in the north-east region, probably affected by a turbulent component (see Figure~\ref{fig:moment}). Its kinematic properties reveals the likely presence of a disturbed disk with turbulent molecular gas. This is consistent with the large-scale of a few kpc molecular outflow observed in this galaxy, which is interpreted as likely driven by the strong feedback from the central AGN \citep{veilleux17}. 

High-resolution millimeter observations of Mrk~231, the only well-studied local IR QSO so far, have revealed the molecular gas traced by both CO and dense molecules are dominated by rotation with disk size of $\sim$1 kpc FWHM \citep{downes98,aalto12,feruglio15}. Together with the eight IR QSOs observed in this study, the combined analysis of the CO morphologies and kinematic properties allow us to roughly classify the local IR QSO hosts into three categories, 1) rotating gas disk with ordered velocity gradient, 2) compact CO peak with disturbed velocity indicative of unsettled molecular gas in the galactic plane, and 3) multiple CO distinct sources undergoing a merger between luminous QSO and companion galaxy separated by a few kpc. The diverse host galaxy properties revealed in our observations show evidence that the merging between luminous QSOs and companion galaxies is still ongoing in some of IR QSO systems. This is unexpected since the objects in IR QSO phase are typically characterized by the final coalescence of the galaxies, according to the theoretical models of merger-driven evolutionary sequence \citep[e.g.,][]{sanders88,barnes92,hopkins08,narayanan10}, but consistent with recent discovery of a population of late-stage nuclear mergers with separation $<3$ kpc in obscured luminous black holes by high-resolution near-IR observations \citep{koss18}. We caution, however, given that our classification of IR QSOs is based on a small sample with only nine CO-mapped sources, it is thus a first-order kinematic classification and may be too simplistic. A larger sample of local IR QSOs mapped in gas emission with high resolution is needed to reach a meaningful statistical study for galaxy classification.

Table~\ref{tab:properties} summarizes the size measurements for the IR QSOs spatially resolved in CO(1$-$0) and 3-mm continuum emission. It is clear that the 3-mm continuum sizes are substantially (a factor of $>$3) smaller than the CO sizes, indicative of a more compact structure of continuum emission compared to that of molecular gas which is related to the star formation in host galaxies. I~ZW~1 is spatially resolved in CO emission with a physical size of 3.2$\pm$0.2 kpc, however, unresolved in 3-mm continuum in our observations with an angular resolution of about 0\arcsec .6 ($\sim$ 0.7 kpc). ALMA HCN (3$-$2) and HCO$^+$ (3$-$2) imaging of I~ZW~1 with a comparable angular resolution (0\arcsec .69$\times$0\arcsec . 63) to our CO observations, show that the source is marginally resolved in dense molecular gas emission with a deconvolved size of $\sim $0\arcsec .4 \citep{imanishi16}. Compared to the total molecular gas as traced by CO emission, both HCN and HCO$^+$ lines with high critical densities probe the dense molecular gas (i.e., $n({\rm H_2})\geqslant 10^4\ {\rm cm^{-3}}$), of which the mass is observed to have a tight linear correlation with the SF rate in galaxies, suggesting that the dense gas is likely the direct fuel for star formation \citep[e.g.,][]{gao04}. The significantly smaller size of dense gas observed compared to the CO size may indicate that the star formation is concentrated in sub-kpc scale structure for I~ZW~1. At the same time, ALMA observations performed by \citet{imanishi16} also marginally resolved the continuum emission at 1.2-mm that is usually dominated by dust thermal radiation, with an intrinsic size of about 0\arcsec .6 ($\sim$ 0.8 kpc). This is roughly equal to the dense gas size, but larger than that of 3-mm continuum emission which is unresolved in our ALMA observations with a slightly higher angular resolution (see Table~\ref{tab:obs}).

In addition, VLA snapshot observations (beam size $\approx 0\arcsec .31$) of I~ZW~1 at 8.4 GHz show a point-like structure in the map \citep{kukula95}. Based on this, we collected and processed the A configuration 8.4 GHz radio data (project ID: AB0670) from VLA data archive that was observed much deeper and with a higher resolution of 0\arcsec .27. A 2D Gaussian fit to the 8.4 GHz continuum image obtains a deconvolved size of 0\arcsec .21$\pm$0\arcsec.06. This is obviously smaller than the size of 1.2-mm continuum emission that is mainly related to the star formation in host galaxies. Combining the analysis that the 3-mm continuum size is smaller than that of 1.2-mm, but could be close to the 8.4 GHz radio size, we speculate that the 3-mm continuum emission of I~ZW~1 may be partially related to the compact synchrotron radiation arising from plasma accelerated by the central AGN \citep{panessa19}.

Moreover, the 3-mm flux density we measured for I~ZW~1 is found to exceed the interpolation of the low frequency steep-slope power law \citep[see Figure 11 in ][]{pasetto19}, which may indicate a significant contribution from additional AGN-related synchrotron emission, and thermal free-free emission. In addition, theoretical calculation using Clumpy models together with the mid-IR to X-ray and radio fundamental plane scaling relations, provide evidence for a non-negligible contribution of synchrotron emission to the 3-mm band for bright AGNs \citep{pasetto19}.

According to the above analysis, the different radiation components contributed to the 3-mm band of I~ZW~1, seems to be a plausible explanation for the compact structure of 3-mm continuum emission compared to the CO morphology observed for this source, as well as for the remaining local IR QSOs in our sample. However, deeper observations with higher angular resolution at (sub)millimeter wavelengths together with an SED analysis including radio data, are necessary to explore the different radiation structures and mechanisms, from the galaxy to the nuclear regions and their connections for IR QSOs.

The CO-based dynamical mass we derived for each of the QSO hosts is the sum of all the mass (contributed by various galaxy components, i.e., stars, gas, dust, central black hole, and dark matter) inside the central few kpc of the galaxies. Given that both the dust and BH masses make up a substantially small fraction compared to the total mass, we assumed that the molecular gas fraction can be approximated as the ratio of gas to dynamical mass and found that the gas fraction of $M_{\rm gas}/M_{\rm dyn}$ ranges from 0.05$\pm$0.01 to 0.60$\pm$0.08 with a median value of 0.22$\pm$0.04 for IR QSOs, in line with that of local ULIRGs \citep[e.g.,][]{downes98}. However, it should be noted that the gas fractions derived for the IR QSOs in our sample are largely affected by the uncertainties of dynamical mass, e.g., the relatively smaller CO size compared to the full CO flux distribution that adopted for the dynamical mass estimates would lead to an overestimation of gas fraction (see Section~\ref{subsec:dynmass}). Similar results have been obtained for $z\sim 6$ QSOs detected in CO emission, with a median value of $M_{\rm gas}/M_{dyn}$ = 0.16 \citep[e.g.,][]{wang10,wang16,feruglio18}. For comparison, the gas fraction of molecular gas to total baryonic mass of $M_{\rm gas}/(M_\star+M_{gas})$ is found to have a mean value of about 0.3 in $z=1$ main sequence galaxies, and the fraction increases with look-back time \citep[e.g.,][]{tan13,schinnerer16,magdis17,tacconi18}.

\subsection{SMBH and Host Galaxy Growth}\label{subsec:growth}

As the SMBH grows primarily through mass accretion during the AGN phase and the bulge of the host galaxy builds up from star formation, a comparison of the mass accretion rate of SMBH with the SFR will be useful to understand how the BHs and their host galaxies grow in IR QSO phase, under the assumption that the object is caught in the phase with ongoing processes of galaxy assembly. 

\begin{figure}[tbp!]
\centering
\includegraphics[width=0.98\linewidth]{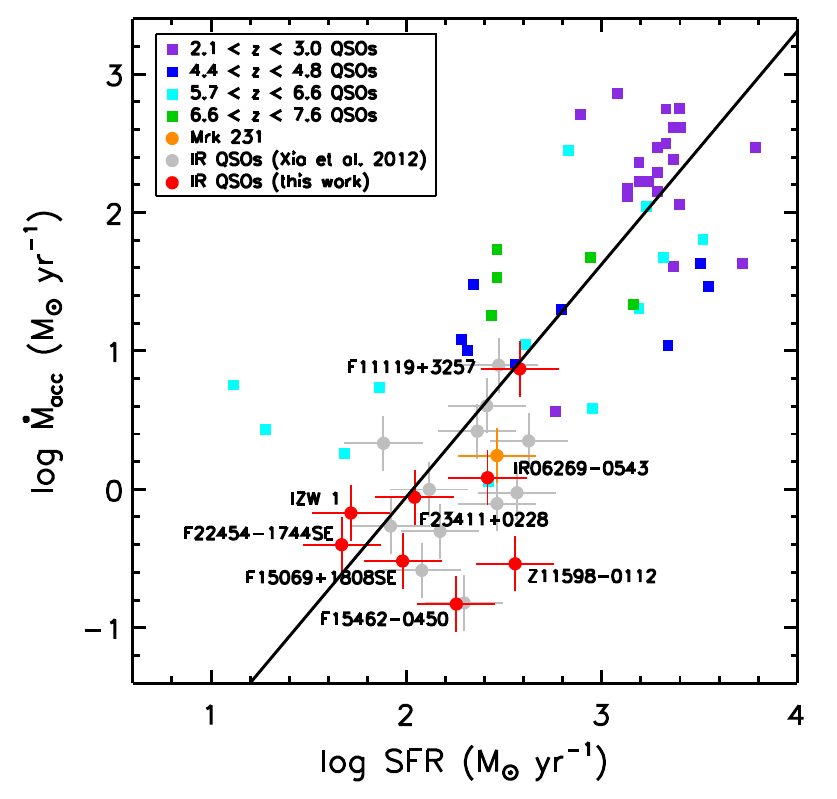}
\caption{Black hole mass accretion rate as a function of the SFR for the local IR QSOs with CO-mapped (colored circles) and the other IR QSOs from \citet{xia12} (grey circles),  compared to the sample of high-$z$ QSOs (squares). Purple squares: \citet{beelen04}, \citet{coppin08}, \citet{feruglio14}, \citet{banerji17}, \citet{hill19}. Blue squares: \citet{carniani13}, \citet{trakhtenbrot17}, \citet{bischetti18}. Cyan squares: \citet{maiolino05},\citet{walter09}, \citet{wang13}, \citet{wang16}, \citet{willott13,willott15,willott17}. Green squares: \citet{venemans12,venemans16,venemans17a,venemans17b}, \citet{banados18}. The solid line indicates the best-fitting relation of log($\dot{M}_{\rm acc}$) = ($1.68\pm0.17$)log(SFR)$-$3.4. \label{fig:maccsfr}}
\end{figure}

Figure~\ref{fig:maccsfr} shows the BH mass accretion rate $\dot{M}_{\rm acc}$ as a function of the SFR for low$-z$ IR QSOs (circles), compared to a compilation of high$-z$ QSOs sample (squares) for which both the far-IR and AGN bolometric luminosities have been measured (see caption of Figure~\ref{fig:maccsfr} for details). The colored and grey circles represent the local IR QSOs mapped in CO emission and the remaining sample in \citet{xia12} without CO images. 

The BH accretion rate is derived as $\dot{M}_{\rm acc}=L_{\rm bol}/\eta c^2$ assuming an accretion efficiency $\eta=0.1$. For the IR~QSOs, the AGN bolometric luminosities $L_{\rm bol}$ were estimated from the extinction-corrected continuum emission at 5100 \angstrom\ and using a bolometric correction, i.e., $L_{\rm bol} \approx$ 9$\lambda L_{\lambda,5100\angstrom}$ \citep{hao05}. For the high$-z$ QSOs, $L_{\rm bol}$ were taken from the literature when available, which generally were derived from the continuum luminosity at 1450 \angstrom, while for the sources with no $L_{\rm bol}$ reported in the literature, we estimated the $\dot{M}_{\rm acc}$ by assuming Eddington accretion, i.e., $\dot{M}_{\rm Edd} \approx 2.2\times10^{-8} M_{\rm BH}$~M$_\odot\ {\rm yr^{-1}}$. For both low and high$-z$ QSOs, we computed the SFR from the far-IR emission which is dominated by the cold dust component tracing the SF-related IR luminosity following the scaling relation found in the local universe: SFR/$M_\odot \ {\rm yr}^{-1} = 1.49 \times 10^{-10} L_{\rm IR}/L_\odot$ \citep{kennicutt12}. Here we converted the far-IR luminosity to the total IR luminosity (in the range of 3-1100 $\mu$m) by adopting $L_{\rm IR} = 1.3 L_{\rm FIR}$ \citep{decarli18}, which is derived for bright QSOs by modeling the dust continuum emission as a modified blackbody.  Note that the SFR measured for the hyper-luminous ($L_{\rm bol}>10^{13}\ L_\odot$) QSOs  may be overestimated due to the possible contribution from AGN to the far-IR emission \citep[][]{dai12,duras17}. 

For the low$-z$ IR QSOs, the far-IR luminosities were calculated based on the flux densities measured at IRAS 60 $\mu$m and 100 $\mu$m \citep[see][]{xia12}. However, for the two IR QSOs that are identified as interacting systems in our resolved CO maps, IRAS F15069+1808 and IRAS F22454-1744, the SFRs were estimated from IR luminosities that are predicted by the CO(1$-$0) emission, based on the correlation between $L^\prime_{\rm CO(1-0)}$ and $L_{\rm IR}$, i.e., log$\left(\frac{L^\prime_{\rm CO(1-0)}}{{\rm K\ km\ s^{-1}\ pc^2}}\right)$ = 0.08$_{-0.08}^{+0.15}$ + 0.81 log$\left(\frac{L_{\rm IR}}{L_\odot}\right)$, observed for starburst galaxies \citep{sargent14}. The CO-based SFRs estimated for IRAS F15069+1808SE and IRAS F22454-1744SE, both of which are likely galaxies hosting the luminous AGN, are found to be lower than the dust-based SFR estimates, which are global measurements for the whole systems. We also calculated the CO-based SFR for the remaining local IR QSOs with CO(1$-$0) luminosities derived from the single-dish data in \citet{xia12} and found a median and a mean value of SFR$_{\rm CO}$ to SFR$_{\rm dust}$ ratio of 0.86 and 1.06 respectively, with a 1$\sigma$ dispersion of 0.57. This suggests that the two estimates are comparable, although with substantial scatter. Recall that the CO distribution of IR QSOs in our sample is found to be composed of a compact component and an extended component by model fitting (see Sect.~\ref{subsec:size}), as well as the much compact 3-mm continuum emission compared to the CO source observed for these galaxies, we speculate that the excitation of a compact component of the CO emission is likely related to the X-ray dominated regions in AGN torus. In this case, the CO-based SFR estimates for IR QSOs might be overestimated. We will investigate this topic by performing SED analysis including multi-wavelength data for more details in a future work.

%%%%%%%%%%%%%%%- Fig-7 -%%%%%%%%%%%%%%%%

%%%%%%%%%%%%%%%%%%%%%%%%%%%%%%%%%%%%%%%%

In Figure~\ref{fig:maccsfr} we find a trend between $\dot{M}_{\rm acc}$ and SFR over about three orders of magnitude in far-IR luminosity/SFR. The solid line shows the best-fitting (a logarithmic linear least-squares bisector fit) correlation for the whole sample of QSOs, with a power law of $\dot{M}_{\rm acc}\propto {\rm SFR}^{(1.68\pm0.17)}$. The Spearman rank correlation coefficient for this correlation is 0.77 with a high significance ($P<10^{-5}$), although with substantial scatter. Our result is in general agreement with the correlation found between AGN activity and star formation, i.e. $L_{\rm AGN} \propto L_{\rm IR,SF}^\alpha$ \citep[$\alpha\sim 1.1-1.7$; see][and references therein]{netzer09,chen13,hickox14,duras17,dai18,izumi19}. As noted in previous works \citep[see e.g.,][]{dai18}, the variation of slope might be attributed to different sample compositions and/or methods of analysis. However, all these studies including our own reveal a significant trend between the SFR and BH accretion spanning a wide range of luminosity, implying that there may be a direct connection between the AGN and star formation activities over galaxy evolution timescales. In addition, there is evidence that SMBHs grow in step with their host galaxies since $z\sim 2$ \citep{daddi07,mullaney12}. The median value of $\dot{M}_{\rm acc}$/SFR is {\bf $6.0\times10^{-3}$} for IR QSOs, consistent with the value derived in \citet{xia12}. In contrast, the $\dot{M}_{\rm acc}$/SFR values for high$-z$ QSOs are found to be about a factor of 10 higher (a median of {\bf $7.3\times10^{-2}$}) than the local IR QSOs. This is consistent with the ``overmassive" black hole revealed for $z>5$ QSOs \citep{walter04,wang10,wang13,wang16}, for which a plausible interpretation is likely related to the different efficiency of gas accretion to the central SMBHs, i.e. faster gas consumption in the most massive SMBHs. Similar results have been reported by \citet{hao08}, who found a trend that the SFR increases with the accretion rate for both low and high-redshift IR-luminous QSOs, and the relative growth of BHs and their host spheroids may depend on the intensity of QSO activities.

Based on the CO-based dynamical mass estimates and the SMBH mass taken from the literature \citep[see Table~\ref{tab:properties};][]{hao05}, we also investigated the possible constraints of the relationship between the black hole and the host galaxy in IR QSO systems, and found that almost all local IR QSOs in our sample are located on or below the local BH-bulge mass relation (equating $M_{\rm dyn}$ to $M_{\rm bulge}$) with mass ratio of $M_{\rm BH}/M_{\rm dyn}$ ranging from 0.02\% to 0.36\% with a median value of 0.08\%, about 3$-$4 times lower than the local BH to bulge mass ratio at $M_{\rm BH}=10^7\ M_\odot$ \citep{kormendy13}. This is similar to IBISCO hard X-ray selected AGN sample galaxies at $z<0.05$, most of which are found to show large offsets with BH mass below the local relationship \citep{feruglio18}, and low-luminosity QSOs at $z>6$ as well \citep{izumi19}. As hypothesized by \citet{xia12}, the low ratio of $M_{\rm BH}/M_{\rm dyn}$ is likely to imply that the star formation and the AGN activities are not necessarily synchronized and the SMBH may grow faster after the short IR QSO phase, even though they are intimately connected. It should be noted, however, there are significant uncertainties in the estimates of both $M_{\rm dyn}$ and $M_{\rm BH}$, although the high spatial resolution and sensitivity CO observations toward low-$z$ IR QSOs allow us to better constrain the gas geometry and the dynamical properties of host galaxies compared to those of high$-z$ QSOs. Recent reverberation mapping observations show evidence that, the empirical scaling relation between the size of broad-line region and the continuum luminosity of AGN for $M_{\rm BH}$ estimates, may depend tightly on accretion rates, i.e., AGNs with extremely high accretion rates tend to have shorter H$\beta$ time lags and hence to be over-estimated in their BH mass, if adopting correlations inferred based on the low accretion rate sources \citep[see][and references therein]{du16,wang17}. In addition, it remains unclear whether a significant bulge component has formed in the IR QSO phase. It has been found that bulges are less common at high redshift \citep[e.g.,][]{cassata11}. Therefore, we may not expect the growth of bulge and central SMBH follow the local BH-bulge mass relationship for IR-luminous QSOs.

\subsection{The Evolutionary Status of IR QSOs}\label{subsec:evolutionary}

According to the major-merger evolutionary scenario first proposed by \citet{sanders88}, IR QSOs are likely short-lived ULIRG-to-QSO transition objects. A comparison of local CO-detected IR QSOs with ULIRGs revealed similar cold molecular gas properties, e.g., molecular gas mass, CO line width, and star formation efficiency (SFE; $L_{\rm FIR}/L^\prime_{\rm CO}$), between these two populations \citep{xia12}. Moreover, the CO source sizes we measured for IR QSOs are found to be comparable to ULIRGs \citep[radius $\lesssim 2$ kpc; e.g.,][]{downes98,iono09,ueda14}, but slightly smaller than those of less IR-bright luminous infrared galaxies (LIRGs; $10^{11}\ L_\odot \leqslant L_{\rm IR} < 10^{12}\ L_\odot$), indicative of more compact molecular gas reservoir and star formation region in ULIRGs and IR QSOs. \citet{gao04} showed that the SFE indicated by $L_{\rm FIR}/L^\prime_{\rm CO}$ reflects the dense molecular gas fraction and the starburst activity in galaxies, since a tight linear correlation is observed between the far-IR and the HCN molecule (a tracer of dense molecular gas) luminosities. The comparable SFE found for ULIRGs and IR QSOs and a smaller value for PG QSOs \citep{xia12}, suggesting that there are much stronger and massive starburst activities in ULIRGs and IR QSOs than in PG QSOs. Similarly, the SFE is lower in LIRGs compared to ULIRGs \citep[e.g.,][]{iono09}. These are consistent with the gas-rich merger models \citep{dimatteo05,hopkins08}, which predict that these objects represent different phases in an evolutionary sequence, where the most intense star formation activity is seen in stages of ULIRGs and IR QSOs due to massive gas inflows to galaxy center. As molecular gas is consumed by starburst as well as the central black hole, a buried AGN continues to grow a SMBH and evolves to an IR luminous QSO during the final stages of the coalescence. 

Overall, the data of local IR QSOs in this study are broadly consistent with the AGN-galaxy coevolution scenario, although some of IR QSOs are found to undergo mergers with companion galaxies by our resolved CO images. However, we must emphasize that, combined with extensive multi-wavelength data, a systematic analysis of the molecular ISM properties, such as the gas content, morphology, excitation, dynamics, and dust properties for a complete sample of local IR luminous galaxies/QSOs is necessary for testing the evolutionary hypothesis comprehensively. We will focus on these issues in future work.

\section{Summary}\label{sec:summary}

We have presented ALMA band 3 observations of the CO(1$-$0) line and 3-mm continuum emission in eight IR QSO hosts at $z\leqslant 0.19$. The typical FWHM  synthesized beam size of our observations is about 0.\arcsec45 and the achieved 1$\sigma$ rms noise levels are 0.4$-$0.9 \mjybeam\ at the velocity resolution of 25 \kms\ . The data allow us to investigate spatially resolved properties of the ISM on scales of $\sim$1 kpc. Our main findings are as follows.

1. All eight IR QSO hosts are clearly detected and resolved in the CO(1$-$0) line emission. The CO molecular gas in these sources is found to be extended over FWHM $\sim$ 1.2$-$7.0 kpc with a median of 3.2 kpc. Seven out of eight IR QSOs are detected in 3-mm continuum emission, and four of which are marginally resolved with 3-mm continuum source size estimate of 0.4$-$1.0 kpc. 

2. The combined analysis of the morphology and kinematics of IR QSO hosts reveals a wide variety of host galaxy properties. The IR QSO hosts can be roughly classified into three categories, rotating gas disk with ordered velocity gradient, compact CO peak with disturbed velocity, and multiple CO distinct sources undergoing a merger with complex velocity structure. Three IR QSO hosts are identified as interacting systems with companion galaxies located $\sim$ 2-7 kpc in projection from the QSOs and within radial velocities $<$ 60 km s$^{-1}$. Two of these interacting merger QSO systems are interpreted as major mergers with complex velocity structure while the other one show evidence for a minor merger. The remaining five QSO hosts are found to be isolated without an interacting companion in the CO images, four of which have velocity gradients suggestive of rotation whereas the last one is a compact CO source with disordered velocity structure. 

3. We investigate the dynamics of the molecular gas in our sources and model the four isolated QSO host galaxies (Mrk~231-alike) with velocity gradients in their CO kinematics. The model fitting shows that three of these sources are rotation-dominated with $V_{\rm rot}/\sigma=4-6$, whereas the other one shows evidence for a disturbed disk with turbulent molecular gas, consistent with the molecular outflow observed in this galaxy previously.

4. Compared the CO(1$-$0) integrated fluxes recovered by the ALMA with the fluxes measured by the IRAM 30m, our interferometric data typically recover $\sim$ 80\% of the total single-dish flux. For the three interacting QSO systems, the molecular gas masses are found to be comparable between the QSO hosts and companion galaxies for IRAS~F15069+1808 and IRAS~F22454$-$1744, while for IRAS~F23411+0228, the molecular gas mass of the companion galaxy is only $\sim$ 3\% that of the QSO host. 

5. Using rotating disk and isotropic virial estimators, we estimate CO-based dynamical masses of $7.4\times10^9-6.9\times10^{10}\ M_\odot$ within the CO-emitting regions of the sources in our sample. The molecular gas fraction of $M_{\rm gas}/M_{\rm dyn}$ is found to have a median of 0.22$\pm$0.04, similar to that found for local ULIRGs.

6. We find a trend between BH accretion rate of mass and SFR over three orders of magnitude, consistent with the correlation between AGN bolometric luminosity and star formation activity. 

The diversity of molecular gas morphology and kinematics revealed in the host galaxies of IR QSOs, clearly indicates more complicated evolutionary stages from merging (U)LIRGs to QSOs, even during the IR QSO phase. Our studies certainly exemplify the importance of classifying the IR QSOs by further meaningful statistical study. Further high spatial resolution mapping of molecular gas, in terms of assembling larger samples of IR QSOs and obtaining deeper sensitivity, will help shed light on the exact nature of the host galaxies of IR QSOs and the understanding of coeval black hole-galaxy growth. Such high-resolution ALMA studies of IR QSOs will provide better local analogues and valuable information for fair comparison and understanding of the high-$z$ population.

%% Note that the \setcounter and \renewcommand are needed here because
%% this example is using a mix of deluxetable and tabular.  Here the
%% deluxetable counters are set with \tablenum but the situation is a bit
%% more complex for tabular.  Use the first command to set the Table number
%% to ONE LESS than it should be.  The next command will auto increment it
%% to the desired number.

\acknowledgments

%We thank ...

We thank the anonymous referee for the constructive comments which helped improve the paper. QT would like to thank Ran Wang, Tao Wang, and Zhi-Yu Zhang for helpful discussions. This paper makes use of the following ALMA data: ADS/JAO.ALMA\#2015.1.01147.S. ALMA is a partnership of ESO (representing its member states), NSF (USA) and NINS (Japan), together with NRC (Canada), MOST and ASIAA (Taiwan), and KASI (Republic of Korea), in cooperation with the Republic of Chile. The Joint ALMA Observatory is operated by ESO, AUI/NRAO and NAOJ.

This work was supported by National Key Basic Research and Development (R\&D) Program of China (Grant No. 2017YFA0402704), NSFC Grant No. 11803090, 11861131007, and 11420101002, and Chinese Academy of Sciences (CAS) Key Research Program of Frontier Sciences (Grant No. QYZDJ-SSW-SLH008). X.Y.X and C.N.H. acknowledge the support from the NSFC (Grant No. 11733002). The work by SM and YS is partly supported by the National Key R\&D Program of China (No. 2018YFA0404501), by the NSFC (Grant No. 11821303 and 11761131004). ED acknowledges support from CAS President's International Fellowship Initiative (Grant No. 2018VMA0014).

%% To help institutions obtain information on the effectiveness of their 
%% telescopes the AAS Journals has created a group of keywords for telescope 
%% facilities.
%
%% Following the acknowledgments section, use the following syntax and the
%% \facility{} or \facilities{} macros to list the keywords of facilities used 
%% in the research for the paper.  Each keyword is check against the master 
%% list during copy editing.  Individual instruments can be provided in 
%% parentheses, after the keyword, but they are not verified.

\vspace{5mm}
\facilities{ALMA, HST, IRAM 30m, Pan-STARRS1, SDSS, VLA}

%% Similar to \facility{}, there is the optional \software command to allow 
%% authors a place to specify which programs were used during the creation of 
%% the manusscript. Authors should list each code and include either a
%% citation or url to the code inside ()s when available.

\software{AICER \url{(https://github.com/shbzhang/aicer/)},
          $^{\rm 3D}$BAROLO \citep{diteodoro15},
          CASA \citep{mcmullin07},  
          GILDAS \url{(http://www.iram.fr/IRAMFR/GILDAS/)},        
          UVMULTIFIT \citep{martividal14}
          }

\bibliographystyle{aasjournal.bst}
\bibliography{reference}{}

%\begin{thebibliography}{}

%\bibitem[Astropy Collaboration et al.(2013)]{2013A&A...558A..33A} Astropy Collaboration, Robitaille, T.~P., Tollerud, E.~J., et al.\ 2013, \aap, 558, A33 
%\bibitem[Bertin \& Arnouts(1996)]{1996A&AS..117..393B} Bertin, E., \& Arnouts, S.\ 1996, \aaps, 117, 393 

%\end{thebibliography}

%% This command is needed to show the entire author+affilation list when
%% the collaboration and author truncation commands are used.  It has to
%% go at the end of the manuscript.
%\allauthors

%% Include this line if you are using the \added, \replaced, \deleted
%% commands to see a summary list of all changes at the end of the article.
%\listofchanges

\end{document}